\documentclass[prd,twocolumn,reprint,preprintnumbers,nofootinbib,longbibliography]{revtex4-1}

\usepackage{soul}
\usepackage{hyperref}
\usepackage{tabularx}
\usepackage{multirow}
\usepackage{tikz}
\usetikzlibrary{tikzmark, calc}
\usepackage{amsmath}
\usepackage{amsthm}
\usepackage{amssymb}
\usepackage{graphicx}
\makeatletter
\makeatletter \renewcommand{\@dotsep}{10000} \makeatother
\usepackage[utf8]{inputenc}
\usepackage{lmodern}
\hypersetup{
	unicode,
	colorlinks,
	breaklinks,
	urlcolor=cyan, 
    linkcolor=black, 
	pdfauthor={Author One, Author Two, Author Three},
	pdftitle={A simple article template},
	pdfsubject={A simple article template},
	pdfkeywords={article, template, simple},
	pdfproducer={LaTeX},
	pdfcreator={pdflatex}
}
\usepackage[T1]{fontenc} 
\usepackage{amsmath}
\usepackage{graphicx, color}
\usepackage{physics}

\usepackage{bm}
\usepackage{float}
\usepackage{xcolor}
\usepackage{adjustbox}
\usepackage{tikz}
\usetikzlibrary{quantikz2, positioning, calc}

\definecolor{gate}{RGB}{198, 216, 211}
\tikzset{
    operator/.append style={fill=gate!20},
    meter/.append style={fill=gate!20},
}

\newcommand{\be}{\begin{eqnarray}}
\newcommand{\ee}{\end{eqnarray}}
\def\be{\begin{equation}}
\def\ee{\end{equation}}
\def\bea{\begin{eqnarray}}
\def\eea{\end{eqnarray}}

\newcommand{\gsim}{\;\raisebox{-0.9ex}{$\textstyle\stackrel{\textstyle >}{\sim}$}\;}
\newcommand{\lsim}{\;\raisebox{-0.9ex}{$\textstyle\stackrel{\textstyle<}{\sim}$}\;}
\def\lsim{\raise0.3ex\hbox{$\;<$\kern-0.75em\raise-1.1ex\hbox{$\sim\;$}}}
\def\gsim{\raise0.3ex\hbox{$\;>$\kern-0.75em\raise-1.1ex\hbox{$\sim\;$}}}
\def\ie{{\it i.e.\ }}

\usepackage{graphics}
\usepackage{epsfig}
\usepackage{slashed}
\usepackage[utf8]{inputenc}
\usepackage{slashed}

\usepackage{multirow}
 \usepackage{pstricks}
\usepackage{dcolumn}

\theoremstyle{plain}

\theoremstyle{definition}

\DeclareMathOperator*{\argmax}{argmax}

\makeatletter
\newcommand{\fmslash}[2][0mu]{%
  \mathchoice
    {\fmsl@sh\displaystyle{#1}{#2}}%
    {\fmsl@sh\textstyle{#1}{#2}}%
    {\fmsl@sh\scriptstyle{#1}{#2}}%
    {\fmsl@sh\scriptscriptstyle{#1}{#2}}}
\newcommand{\fmsl@sh}[3]{%
  \m@th\ooalign{$\hfil#1\mkern#2/\hfil$\crcr$#1#3$}}
\makeatother

\usepackage[capitalise]{cleveref}
\newcommand{\btheta}{\boldsymbol{ \theta}}
 
\begin{document}

\author{Zhongtian Dong$^{1,2}$}

\author{Taejoon Kim$^{3}$}

\author{Kyoungchul Kong$^{2}$}

\author{Myeonghun Park$^{4,5,6}$}

\author{Jacob L. Scott$^{2}$}

\author{Heechan Yi$^{7}$}

\affiliation{$^{1}$Center for High Energy Physics, Peking University, Beijing 100871, China \\
$^{2}$Department of Physics and Astronomy, University of Kansas, Lawrence, KS 66045, USA
\\$^{3}$School of Electrical, Computer and Energy Engineering, Arizona State University, Tempe, AZ 85281, USA\\
$^{4}$School of Physics, Korea Institute for Advanced Study, Seoul 02455, KOREA\\
$^{5}$Institute of Convergence Fundamental Studies, Seoultech, Seoul 01811, KOREA\\
$^{6}$School of Natural Sciences, Seoultech,  Seoul 01811, KOREA\\
$^{7}$Department of Physics, Yonsei University, Seoul 03722, Republic of Korea}

\author{Willie Aboumrad}
\affiliation{IonQ Inc., 4505 Campus Dr, College Park, MD 20740, USA}

\author{Ananth Kaushik}
\affiliation{IonQ Inc., 4505 Campus Dr, College Park, MD 20740, USA}

\author{Martin Roetteler}
\affiliation{IonQ Inc., 4505 Campus Dr, College Park, MD 20740, USA}

\title{Hybrid quantum-classical approach for combinatorial problems at hadron colliders}

\begin{abstract} 
 In recent years, quantum computing has drawn significant interest within the field of high-energy physics. We explore the potential of quantum algorithms to resolve the combinatorial problems in particle physics experiments. As a concrete example, we consider top quark pair production in the fully hadronic channel at the Large Hadron Collider. We investigate the performance of various quantum algorithms such as the Quantum Approximation Optimization Algorithm (QAOA), a feedback-based algorithm (FALQON) and a variational quantum imaginary time evolution algorithm (VarQITE). We demonstrate that the efficiency for selecting the correct pairing is greatly improved by utilizing quantum algorithms over conventional kinematic methods. Furthermore, we observe that gate-based universal quantum algorithms perform on par with machine learning techniques and either surpass or match the effectiveness of quantum annealers. Our findings reveal that quantum algorithms not only provide a substantial increase in matching efficiency but also exhibit adaptability and the potential for scalability, making them promising candidates for a variety of high-energy physics applications, as quantum hardware technology matures. Moreover, quantum algorithms eliminate the extensive training processes needed by classical machine learning methods, enabling real-time adjustments based on individual event data. 
\end{abstract}

\maketitle

\section{Introduction}
\label{sec:intro}

Understanding the mysteries of Nature at the smallest possible scales requires extraordinarily large and complex particle physics experiments. 
With the ongoing success of the Large Hadron Collider (LHC) at CERN, we can control $\mathcal{O}(10^{13})\,\rm eV$ energy scale experiment to probe the physics at the length scale of $\mathcal{O}(10^{-19})\,\rm m$. While there are sound theoretical reasons to believe that new physics beyond the Standard Model is going to be revealed in these experiments, formidable theoretical and experimental challenges lie ahead \cite{P5:2023wyd}. The future upgrade to the High-Luminosity LHC (HL-LHC) will boost the data delivery rate by 100-fold, reaching approximately 1 exabyte per year. This massive increase in event size, data volume, and complexity will introduce both quantitative and qualitative challenges, significantly stretching the limits of available computational resources.
In this era of big data, a discovery in particle physics experiments will only be achievable through innovative methods for data collection, processing, and analysis \cite{Franceschini:2022vck}. 

Among many others, the combinatorial problems have become one of the central challenges in hadron collider physics analyses. Unlike lepton colliders, the environment at hadron colliders is vastly more complex, making these tasks even more difficult especially with events involving jets. In typical hadron collider events, numerous jets are present - some originating from the decays of heavy particles, while others may result from the dynamics of quantum chromodynamics including initial/final state radiation (ISR/FSR). This complicates the event reconstruction process. The problem is exacerbated by finite detector resolution, pile-up, multi-parton interactions, the finite detector acceptance, and the inability to determine properties of a parton which acts as the seed of a jet. These factors contribute to a significant combinatorial challenge: determining which of the many jets in an event is the correct one to associate with a specific particle decay.

Solving the combinatorial problem enhances the precision in the measurement of particle properties and expedites the discovery of new physics \cite{Barr:2010zj}. 
Many kinematic methods as well as machine learning (ML) methods have been proposed. Only recently have studies begun to explore specific machine learning methods - particularly those that leverage the permutation-invariant structure of attention-based neural networks - to address the combinatorial problems in the fully hadronic channel \cite{Shmakov:2021qdz,Fenton:2020woz,Lee:2020qil} and dilepton \cite{Alhazmi:2022qbf} channel for $t\bar t$ production.
See \cite{Franceschini:2022vck,Alhazmi:2022qbf} for references on kinematics and feature engineering for collider phenomenology. 

In this paper, we go beyond the classical machine learning methods and explore various universal quantum algorithms to resolve the combinatorial problems. Taking the fully hadronic $t\bar t$ production as a concrete example, we examine the performance of hybrid quantum-classical algorithms, including the Quantum Approximation Optimization Algorithm (QAOA) \cite{Farhi:2014ych}, its variants \cite{Blekos:2023nil,Kim:2024jte}, the feedback-based algorithm FALQON \cite{Magann:2021evo} and variational quantum imaginary time evolution (VarQITE) \cite{Motta:2019yya,Yuan2019-oj,McArdle2019-zp,Morris2024-gj}. We compare their results to the performance of quantum annealing \cite{Kim:2021wrr} and kinematic approaches like the hemisphere method \cite{Matsumoto:2006ws,CMS:2007sch}. 

There are a few ways to assess the performance of algorithms utilizing quantum circuits\,\cite{Daley2022PracticalQA,herrmann2023quantum}: the number of adjustable parameters, convergence of accuracy and loss, the number of gates used, the computational complexity, etc. In this paper, we will focus on the efficiency of various algorithms, \ie how well we can identify the correct combination. 
In other words, we focus on the fraction of events for which the output state of a given algorithm has as its most probable eigenstate the bitstring that correctly assigns the outgoing particles. Note that this bitstring does not necessarily correspond to the minimum eigenstate. Although, the more boosted the event, the more likely it is for this correspondence to be true.  For example, let the most probable eigenstate of the circuit output be $\ket{\psi_{\rm max}}=\ket{001011}$ and let the bitstring that correctly assigns the outgoing particles be $111000$ (\ie the first 3 particles are assigned to $t$ and the last 3 are assigned to $\bar{t}$). Since $\ket{\psi_{\rm max}}\neq\ket{111000}$, the algorithm is considered to have failed at identifying the correct assignments.  
We find that this efficiency (or the matching accuracy) in selecting the correct pairing is enhanced when using quantum algorithms compared to conventional kinematic methods.

This paper is structured as follows.
In section \ref{sec:qa_intro}, we review quantum optimization briefly and introduce the problem Hamiltonian that we study in this article. In section \ref{sec:dataset}, we describe preparation of the dataset. 
We provide a short review on quantum algorithms that we use to resolve the combinatorial problem and show results for each method in section \ref{sec:algorithms}. 
In section \ref{sec:comparison}, we compare the performance of different algorithms altogether. 
Section \ref{sec:summary} is reserved for summary and outlook.
In the appendices, we provide schematic diagrams on quantum circuits and more information on their construction (appendix \ref{app:circuits}) and discuss effects of quantum noise (appendix \ref{sec:noise}).
We make short comments on detector effects (appendix \ref{sec:smearing}), performance of quantum algorithms for different processes with the same final state (appendix \ref{sec:diff_processes}) and quantum algorithms based on a non-adiabatic path (appendix \ref{sec:adapt}), and present proof-of-principle results obtained on real quantum hardware (Appendix \ref{sec:quantum_hardware}).

\section{Quantum Optimization and Collider Analyses}
\label{sec:qa_intro}

Optimization problems are ubiquitous, appearing across a wide range of fields and applications. Often classical optimization methods struggle due to the complexity or size of the problem. These are often problems that are computationally intensive, difficult to solve exactly, or have exponentially growing solution spaces as the problem size increases. This is where quantum optimization can be particularly useful \cite{Abbas:2023agz}.  
Recently, the exploration of various quantum algorithms in high energy physics research has gained attention \cite{DiMeglio:2023nsa}. 
Examples include QCD/event generation \cite{Bravo-Prieto:2021ehz,Kiss:2022pjw,Tuysuz:2024hyl,Afik:2025ejh}, parton showers \cite{Bauer:2019qxa,Bauer:2021gup,Bepari_2021,Gustafson:2022dsq}, jet clustering \cite{Okawa:2024goh,Zhu:2024own,deLejarza:2022bwc,Pires:2021fka,Pires:2020urc,Wei:2019rqy,Delgado:2022snu}, reconstruction of particle tracks \cite{Tuysuz:2020ocw,Nicotra:2023rmn,Crippa:2023ieq,Bapst:2019llh}, application in lattice field theory \cite{Funcke_2023}, new physics searches \cite{Yang:2024bqw}, variational quantum eigensolver for loop Feynman diagrams \cite{Clemente:2022nll} etc.
We refer to Refs. \cite{2012NatPh...8..264C,Georgescu_2014,Delgado:2022tpc,humble2022snowmass,Bauer:2022hpo,Catterall:2022wjq,Alam:2022crs} for more details on quantum computing in high energy physics phenomenology.
 
To practically implement quantum optimization techniques, hybrid quantum-classical approaches are frequently utilized \cite{Ge:2022vmm,Callison:2022zfe,Campos:2024wvc}. A prominent example is the Variational Quantum Algorithm (VQA), which employs a quantum circuit with adjustable parameters that are fine-tuned via classical optimization procedures. These adjustable parameters are optimized on a strict event-by-event basis, fundamentally distinguishing them from the globally trained, learnable parameters used in traditional machine learning approaches. 

Because hybrid quantum-classical algorithms optimize parameters for each input without the need for a training process, they do not face generalization issues like those seen in modern classical machine learning (ML) algorithms. This is particularly useful in high energy physics analysis, since classical ML algorithms rely on Monte Carlo (MC) simulations at various stages (parton-level, hadron-level, and detector-level) during training, leading to generalization challenges because the test samples come from collider data and MC simulations are imperfect. In contrast, algorithms based on quantum circuits adjust parameters for each individual event, eliminating the need for training.

While the optimization of a VQA is similar to the training of a classical ML model through the minimization of a cost function, the two approaches differ fundamentally in their scope. Whereas ML training aims to converge on a single, global set of parameters to describe an entire dataset, VQAs find optimal parameters on an event-by-event basis. This implies that, in principle, a study may produce as many unique optimal circuits as there are individual events. This high degree of instance-specific specialization provides a framework for the superior expressivity of VQAs compared to classical neural networks \cite{PhysRevX.4.021041,Sim_2019,Du_2020,Abbas_2021}. Consequently, numerical simulations in Refs. \cite{chen2020QCNN,chen2020quantum,chen2021hybrid,Cara:2024spj,Unlu:2024nvo,Forestano:2023lnb,Dong:2023oqb,Hammad:2023wme} have demonstrated that these hybrid architectures can reach higher accuracies than conventional neural networks, even when constrained to a similar number of parameters. 

In particular, quadratic unconstrained binary optimization (QUBO) problems are studied in depth for many reasons \cite{QUBO-survey,Glover:2018ikr,Lucas:2013ahy}. For one, QUBO appears ubiquitously across science - in social networks, engineering, and physics, among other fields. In physics, it is known as the Ising model. Additionally, there are specialized quantum devices, \ie the quantum annealer (QA), designed to solve QUBO problems \cite{Rajak:2022tgo,Yarkoni:2021zvu}.
However, it is widely recognized that the quantum annealer performs poorly in cases with a degenerate spectrum \cite{Zhou:2018fwi}, leading to the development of several algorithms for universal quantum computers \cite{Farhi:2014ych,Blekos:2023nil}.

Interestingly, the combinatorial problems at particle physics experiments can be recast into a QUBO problem, and the complex environment at hadron colliders is an excellent playground to test such proposals \cite{Kim:2021wrr}.
Combinatorial problems at colliders are notorious and forbid full reconstruction of final states, which is essential to understand the underlying physics associated with the events.

To introduce the problem Hamiltonian, let us consider the production of two particles $A$ and $B$ at the LHC via a $2\to 2$ process with their subsequent decays: $pp \to A B \to (j_1^A \cdots j_{n_A}^A)(j_1^B \cdots j_{n_B}^B)$, where $A$ ($B$) decays into $n_A$ ($n_B$) jets and $n = n_A + n_B$ is the total number of observed jets in the final state. We assume no initial state radiation (ISR) for this discussion\footnote{Ref. \cite{Kim:2021wrr} chooses the first six hardest jets after full reconstruction with parton shower and hadronization.}. 
Then, the combinatorial problem (assigning each daughter particle to either $A$ or $B$) becomes a binary classification, whose computational complexity increases exponentially as $\mathcal{O}(2^n)$. Note that the computational complexity of the quantum annealer is known to be $\mathcal{O}(n^2)$ for the same problem \cite{Kim:2021wrr}. 

\begin{figure*}[t!]
    \centering
    \includegraphics[width=0.75\textwidth]{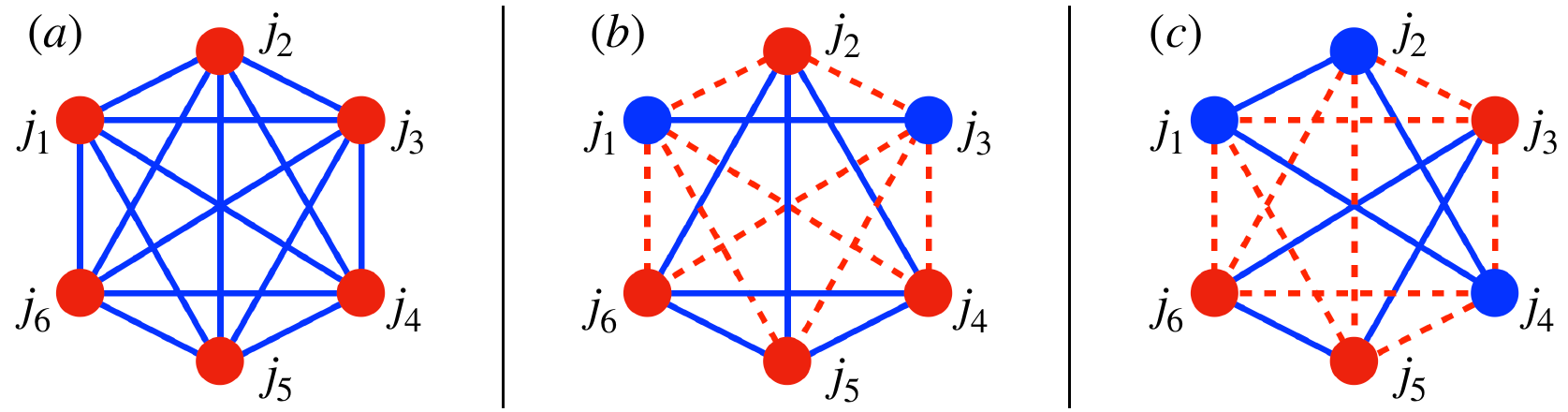}
    \caption{Illustration of the QUBO problem considered in this paper as a fully connected graph. Each node represents a particle originating from the decay process of a specific parent particle. Red nodes represent particles from particle $A$, while blue nodes are from particle $B$. In our example of the $t\bar t$ process, there would be six hard jets resulting from the decay of the two top quarks. For example: (a) all particles originate from $A $, (b) $\{j_2, j_4, j_5, j_6\}\subset A$, $\{j_1, j_3\}\subset B$, and (c) $\{j_3,j_5,j_6\} \subset A$, $\{j_1,j_2,j_4\} \subset B$. 
    Considering $H_2$, the value of the edge connecting nodes $i$ and $j$ is given by $J_{ij}+(1/2)\lambda P_{ij}$. Edges connecting nodes from the same parent particle contribute positively to the problem Hamiltonian $H_P$, and are shown as solid (blue) lines, while edges connecting nodes from different parent particles contribute negatively to $H_P$ and are represented by dashed (red) lines. The task of finding the correct combinatorics is equivalent to identifying the graph configuration that minimizes $H_P$.}
\label{fig:qubo}
\end{figure*}

Given that no further assumptions are made, we need to define a guiding criterion to assign observed particles to the decay products of either $A$ or $B$. Inspired by the general `minimum energy principle' in physics, one might first consider minimizing the total invariant mass. However, this approach leads to a trivial partonic total energy for all visible particles and does not resolve the combinatorial issue \cite{Kim:2021wrr}. A more viable alternative is to focus on the mass difference between $A$ and $B$ \cite{Kim:2021wrr}, 
\begin{equation}
    \label{eq:cost}
    H_0=\left(P_1^2-P_2^2\right)^2 \, .
\end{equation}
Here the four momenta of two mother particles $A$ and $B$ are defined as 
\begin{eqnarray}
P_1 &=& \sum_{i=1}^n \, p_i \, x_i  \, , \\ 
P_2 &=& \sum_{i=1}^n \, p_i \, (1-x_i) \, ,
\label{eq:P1P2}
\end{eqnarray}
where $x_i \in \{ 0, 1 \}$ is an indicator variable: $x_i=1$ if particle $i$ with momentum $p_i$ is from the particle $A$, otherwise it is 0. Using the substitution $x_i=(1+s_i)/2$ with $s_i=\pm1$, the problem Hamiltonian in Eq. (\ref{eq:cost}) can be written as 
\begin{align}
    H_0&=\left(\sum_{ij}P_{ij}x_ix_j-\sum_{ij}P_{ij}(1-x_i)(1-x_j)\right)^2 \nonumber \\
    &=\left(\frac{1}{4}\sum_{ij}P_{ij}\big[(1+s_i)(1+s_j)-(1-s_i)(1-s_j)\big]\right)^2 \nonumber \\
    &=\left(\sum_{ij}P_{ij}s_i\right)^2 = \sum_{ij}J_{ij}s_is_j \, ,
    \label{eq:H0}
\end{align}
where $P_{ij}=p_i\cdot p_j$ and $J_{ij}=\sum_{k\ell}P_{ik}P_{j\ell}$ \cite{Kim:2021wrr}.
%

%

\begin{figure*}[t!]
    \centering
\hspace*{-1cm}
    \includegraphics[width=0.94\textwidth]{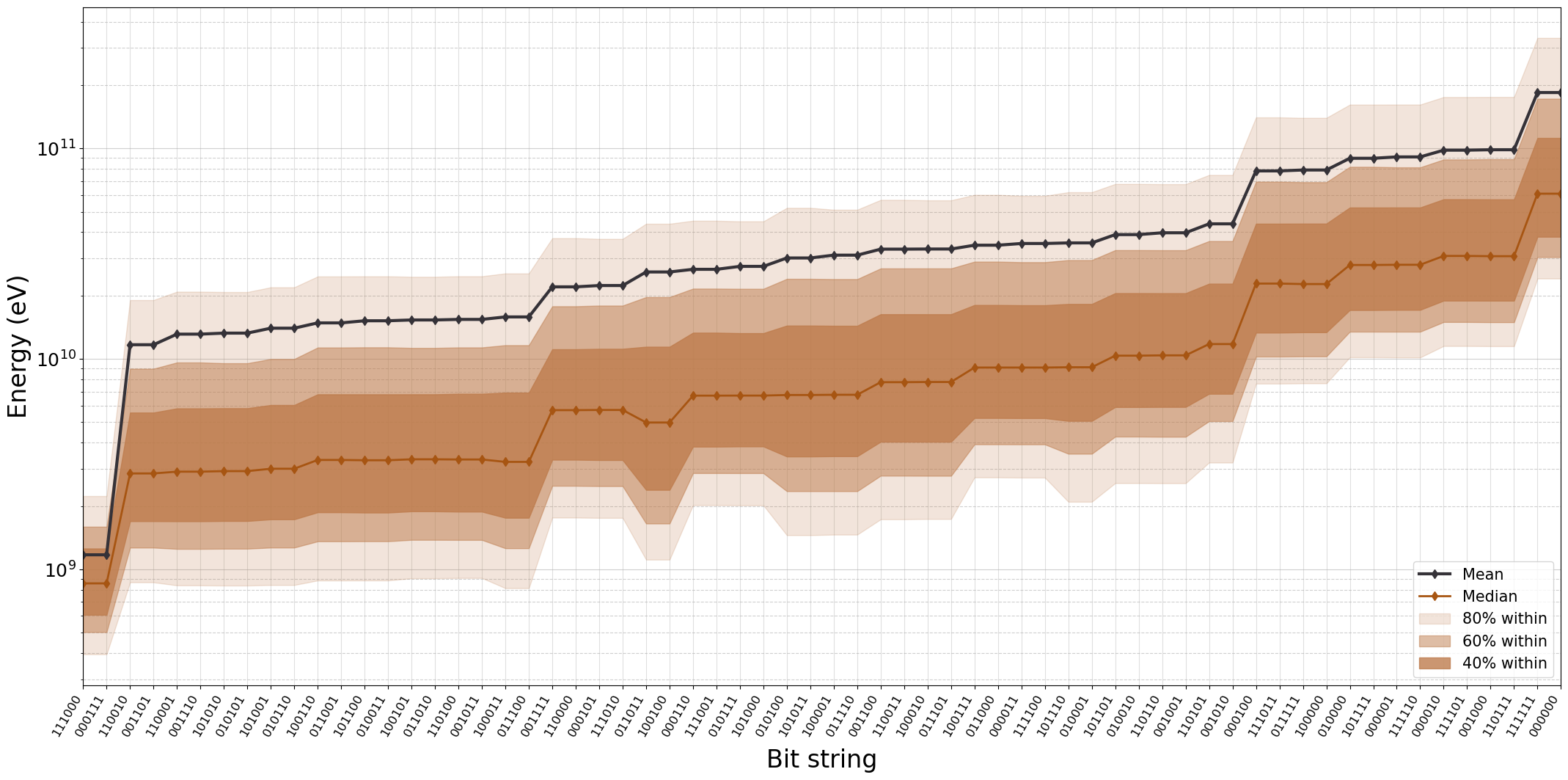}
    \caption{The energy spectrum of the full Hamiltonian $H_2$ in Eq. (\ref{eq:fullH}) for parton-level events.  
    The shaded regions represent the range in which that percentage of events fall in, e.g. 40\% of events fall within the dark orange region.
    }
\label{fig:energy_jets_brute_force_full_H}
\end{figure*}
The target cost function $H_0$ in Eq. (\ref{eq:cost}) is optimized for $m_A = m_B$, which corresponds to the many conventional new physics searches at the LHC. While $H_0$ can serve as a useful starting point, it must be generalized to accommodate more complex scenarios, such as asymmetric production and off-shell effects. This can be achieved by introducing an additional term \cite{Kim:2021wrr}
\begin{eqnarray}
    H_1 &=&(P_1^2+P_2^2) \nonumber \\
    &=& \frac{1}{4}\sum_{ij}P_{ij}\big[(1+s_i)(1+s_j)+(1-s_i)(1-s_j)\big] \, \nonumber \\
    &\to& \frac{1}{2}\sum_{ij}P_{ij}s_is_j \, ,
        \label{eq:h1}
\end{eqnarray}
with the dropping of a constant term $(1/2)(P_1+P_2)^2$.
The full cost function or full (problem) Hamiltonian becomes
\begin{align}
H_2 = H_0 + \lambda H_1 \, ,
        \label{eq:fullH}
\end{align}
where $\lambda$ is a hyperparameter to be chosen during the optimization procedure. Ref. \cite{Kim:2021wrr} suggests $\lambda = \frac{\min (J_{ij}) }{\max (P_{ij})} $ to ensure similar weights for the two terms.
Unless noted otherwise, we will treat the full Hamiltonian $H_2$ as the problem Hamiltonian $H_P$ throughout most of the remaining discussion.

Then the optimization procedure will return a specific assignment of particles which minimizes the cost function or the problem Hamiltonian, therefore resolving the combinatorial problem. We will apply this idea for the fully hadronic channel of the top quark pair production, as shown in Fig. \ref{fig:qubo}. 
Here, each circle (node) represents a particle, with particles of the same color belonging to the same group. The lines connecting two nodes indicate non-zero $J_{ij}+(1/2)\,\lambda\,P_{ij}$ ($J_{ij}$) coefficients in the QUBO model for $H_2$ ($H_0$), illustrating a dense network. Here, the self-loop is ignored as its contribution to $H_2$ remains constant. The objective is to correctly identify two groups, each containing three nodes—one group in red and the other in blue. A significant advantage of the proposed Hamiltonian framework is that it operates without prior assumptions regarding event topology or the specific masses of the particles involved. While traditional reconstruction techniques are often tailored to a fixed decay chain (such as the two-step, two-body decay of the top quark), our formulations, $H_0$ and $H_2$, rely solely on the kinematic relations between final-state objects. This flexibility allows the algorithm to be applied to a wide range of physics processes beyond $t\bar t$ production, including those where the underlying mass scales or decay structures are unknown or modified by new physics.

\section{Data preparation}
\label{sec:dataset}

We generate parton-level events for top quark pairs in the fully hadronic channel using \texttt{MadGraph5\_aMC@NLO} \cite{Alwall:2014hca} at the 13 TeV LHC. Offshell effects of the top quark and $W$ boson are taken into account properly. 
Basic cuts of $p_T > 25$ GeV, rapidity $|\eta |< 2.5$ and $\Delta R(j,j) > 0.4 $ are imposed for the parton-level events. 
Here $p_T = \sqrt{p_x^2 + p_y^2}$ is the transverse momentum and $\Delta R(a,b) = \sqrt{ (\eta_a -\eta_b)^2 + (\phi_a - \phi_b)^2}$ is the angular separation of two particles $a$ and $b$ in the $(\eta, \phi)$ plane, where $\eta=-\ln\tan\left(\theta/2\right )$ is the pseudo-rapidity, $\phi$ is the azimuthal angle, and $\theta$ is the polar angle (see \cite{Barr:2011xt,Barr:2010zj,Franceschini:2022vck} for more information on kinematic variables).
We will use these parton-level events to present the main results and compare against the kinematic methods. Effects of ISR/FSR, parton shower and hadronization are neglected to isolate the performance of the quantum algorithms and allow for an unbiased comparison with Monte Carlo truth. Results with detector effects and more realistic simulation are discussed in Appendix \ref{sec:smearing}. In the boosted regime, where decay products (three jets) may merge into a fat jet, the combinatorial ambiguity between top quarks is effectively removed. Consequently, the problem shifts away from the setup explored here as in this study we focus on the fully resolved case.

Note that there are 6 jets ($n=6$) in this study and therefore a total of 64 possible states. Since there are always two degenerate solutions (swapping $A=t$ and $B=\bar t$ with $n_A=3$ and $n_B=3$), the computational complexity in this case will be $2^{n-1} = 2^5$. For a given event, one can compute the values of the cost function for all possibilities, and declare the combination which leads to the smallest possible value (ground state) as the state that minimizes the Hamiltonian. As stated above, this state may not correspond to the solution to the combinatorial problem but this correspondence becomes more likely for the boosted regimes.  

\begin{figure*}[th]
    \centering
    \includegraphics[height=25em]{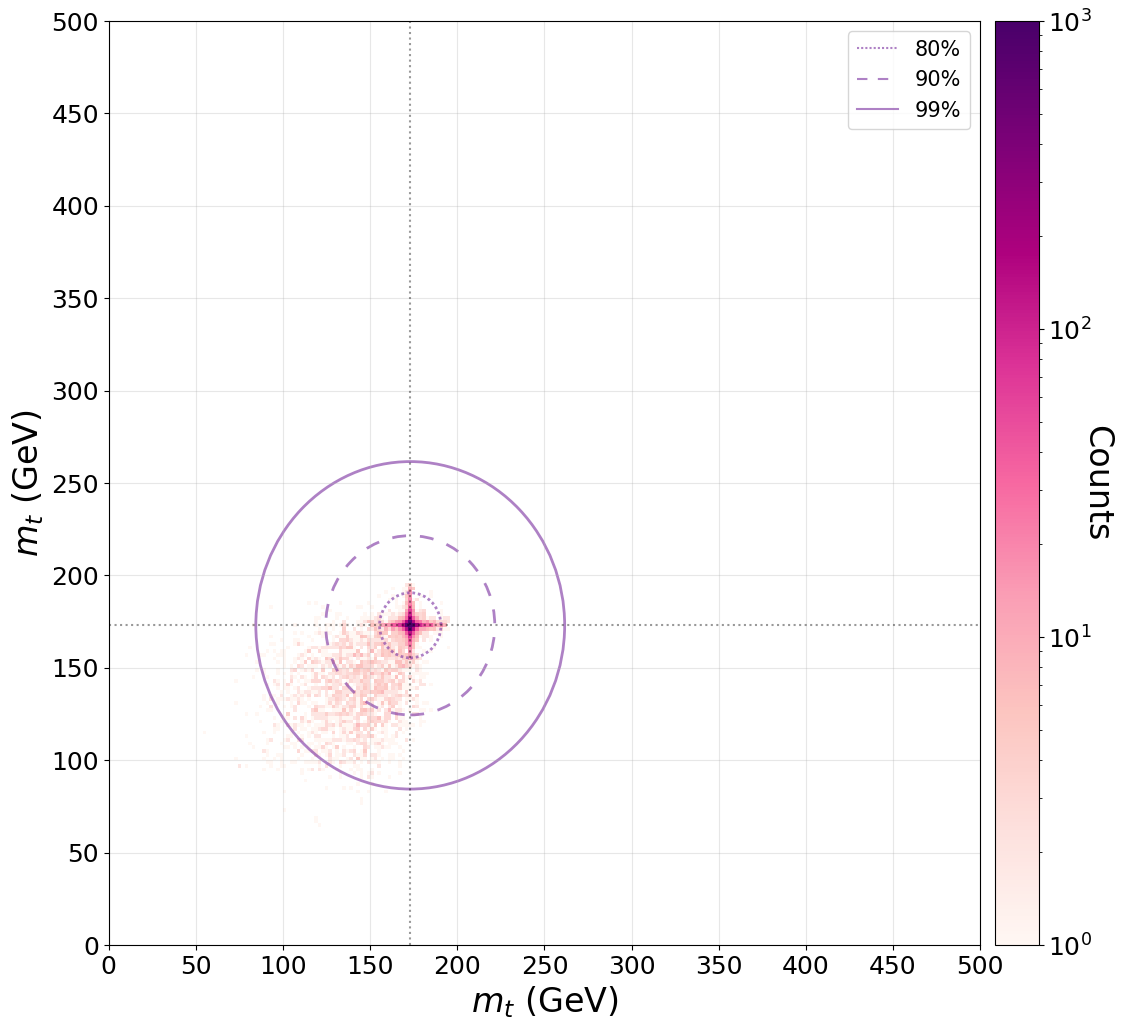}
    \hspace{0.03\textwidth}
    \includegraphics[height=25em]{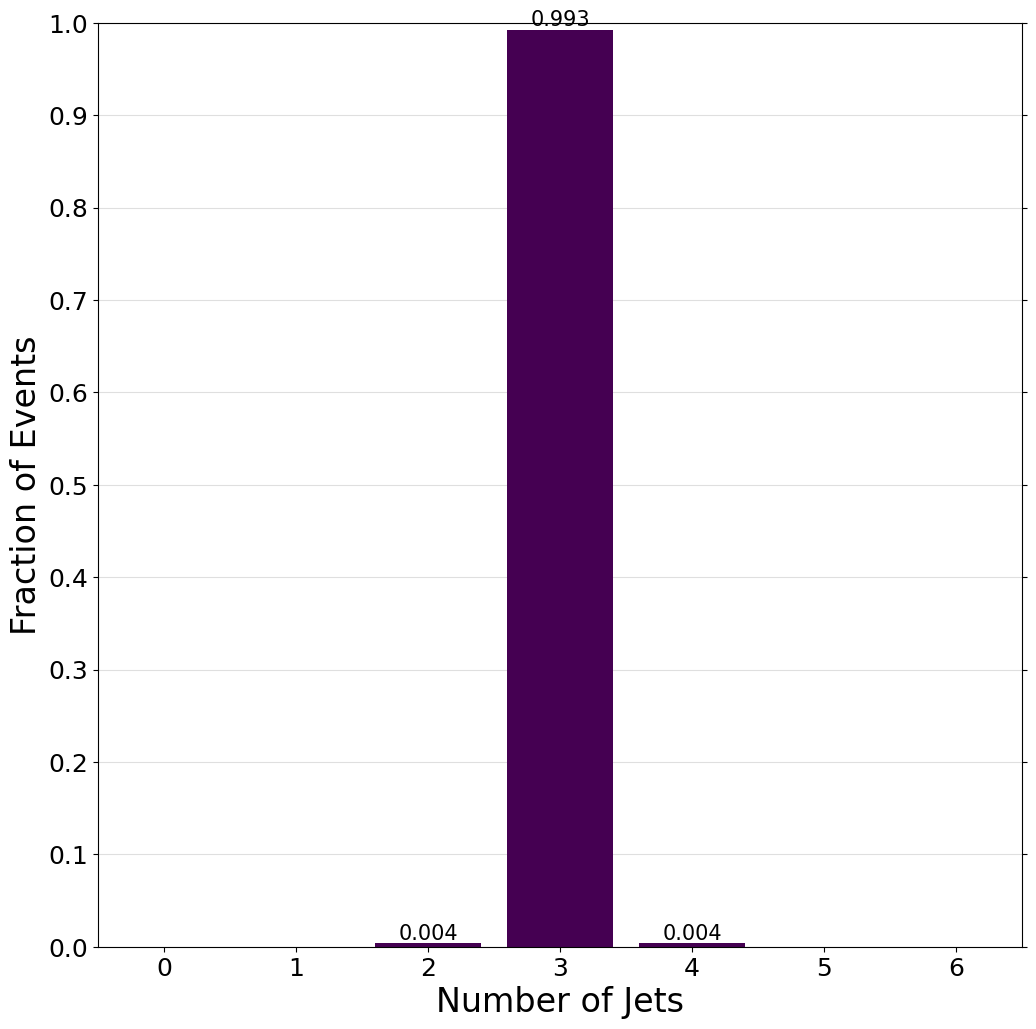}
    \caption{The reconstructed masses (left) and jet-assignment (right) for parton-level events using the true ground state of the $H_2$. The circles contain the specified percentage of events within their circumference. The circles contain the specified percentage of events within their circumference (solid for 99\%, dashed for 90\% and dotted for 80\%). The overall efficiency of resolving the combinatorial problem is about 79\%.
    }
\label{fig:mass_brute_force_full_H}
\end{figure*}
\begin{figure*}[t!]
    \centering
    \includegraphics[height=25em]{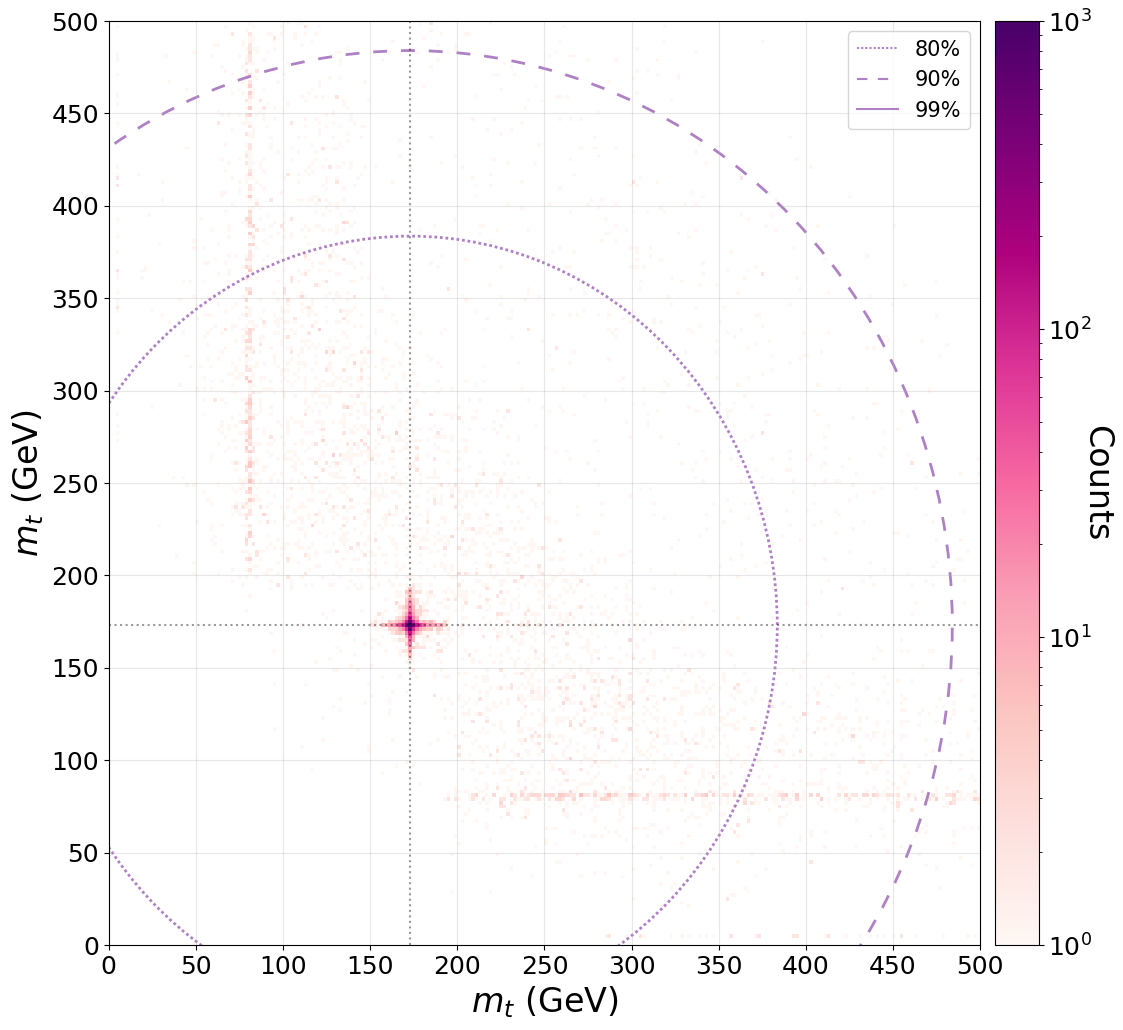}
    \hspace{0.03\textwidth}
    \includegraphics[height=25em]{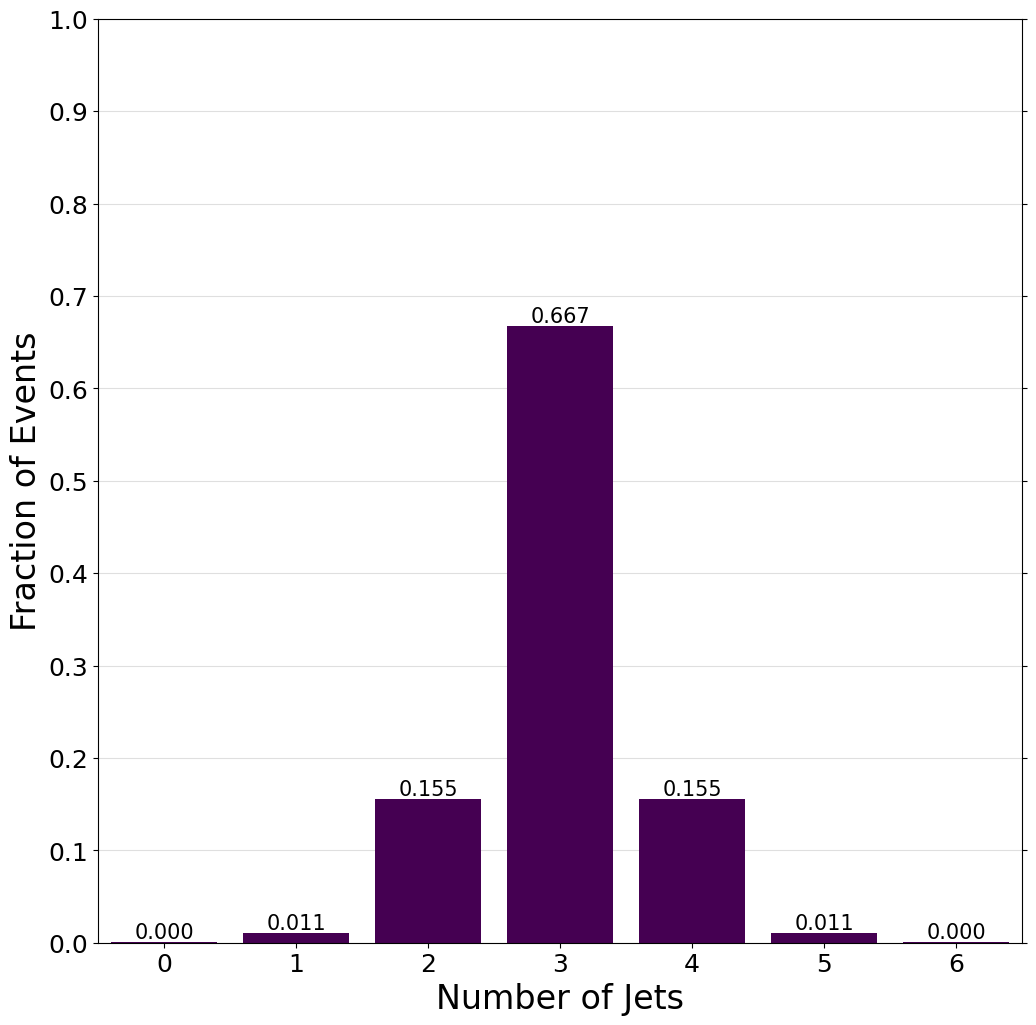}
    \caption{The same as Fig. \ref{fig:mass_brute_force_full_H} but for the hemisphere method. The overall efficiency (matching accuracy) is about 50\%.}
\label{fig:hemisphere}
\end{figure*}

Fig. \ref{fig:energy_jets_brute_force_full_H} illustrates the energy value of the full Hamiltonian in Eq. (\ref{eq:fullH}) for 250k parton-level events. The states ($x_1 x_2 x_3 x_4 x_5 x_6$) are ordered by their energy spectrums such that the far-left corresponds to the average ground state and the far-right state gives the largest average energy values. In this case, the ground state is given by $(x_1 x_2 x_3 x_4 x_5 x_6) = (000111)$, which implies that $P_1 = p_4+p_5+p_6$ for top quark $t = (j_4 j_5 j_6)$ and $P_2 = p_1 + p_2 + p_3$ for anti-top quark $\bar t = (j_1 j_2 j_3)$. The energy spectrum clearly shows the substantial energy difference between the two lowest states. Note the $Z_2$ symmetry in the bitstring, \ie the statistics for $(000111)$ is identical to $(111000)$.

Taking the ground state, one can compute the masses of $t$ and $\bar t$, which are shown in the left panel of Fig. \ref{fig:mass_brute_force_full_H}. 
The circles contain the specified percentage of events within their circumference (solid for 99\%, dashed for 90\% and dotted for 80\%). As expected, most events are populated at the correct mass with very little spread due to the offshell effects of top quark and $W$ boson. 
Our study shows that 99.2\% of the time the ground state is given by the symmetric configuration (3 for top and 3 for anti-top), while 0.8\% of the time, 2 or 4 particles make up top quark (4 or 2 for anti-top quark) due to the finite width effects, as shown in the right panel of Fig. \ref{fig:mass_brute_force_full_H}\footnote{A note about the right panel in \Cref{fig:mass_brute_force_full_H,fig:hemisphere,fig:parton_qaoa8,fig:parton_maqaoa8,fig:parton_xqaoa8,fig:parton_falqon,fig:parton_varqite}: first, the numbers above the bars are rounded to their third digit so they may not add up exactly to 1; second, the histograms take into account \textit{all} jets, so there will be an equal count of $n$ jets as there are of $6-n$ jets. As an example, if an event resolves to 2 jets associated to $t$ and, thus, 4 jets associated to $\overline{t}$, then both are included in the final histogram. Ergo the histograms must be symmetric.}.
The overall efficiency of resolving the combinatorial problem is about 79\%. Explicitly, the Hamiltonian $H_2$ has, as its ground state, the bitstring that resolves the combinatorial problem 79\% of the time.

We also show results in Fig. \ref{fig:hemisphere} for the standard hemisphere method \cite{Matsumoto:2006ws,CMS:2007sch}, when applied to our event topology. In the hemisphere method, one clusters the visible particles into two groups trying to keep the invariant mass of each cluster to a minimum. It is the only kinematic method which does not require prior assumption regarding the event topology or particle masses. While alternative methods such as the Kinematic Likelihood Fitter (KLFitter) \cite{Erdmann:2013rxa} exist, they typically rely on predefined values for the top quark and W boson masses \cite{Franceschini:2022vck}. In our example with 6 jets from the top quark production, we begin with two hardest jets assigned in $A$ and $B$. The third jet is assigned to the side which leads to the smaller invariant mass. One repeats the same process for all remaining jets. We find that the overall efficiency of resolving the combinatorial problem with the hemisphere is about 50\%.
Comparing against results in Fig. \ref{fig:mass_brute_force_full_H}, how can we improve the mass resolution in the left panel and the correct jet assignment in the right panel of Fig. \ref{fig:hemisphere}?
One can attempt to further improve on the hemisphere algorithm with suitable cuts on the invariant mass or the jet $p_T$ by excluding certain reconstructed objects from the clustering algorithm \cite{Debnath:2017ktz}. In this paper we would like to investigate a different approach using quantum algorithms.

Whether or not the ground state resolves the combinatorial problem, Figs. \ref{fig:energy_jets_brute_force_full_H} and \ref{fig:mass_brute_force_full_H} are the best results with the given choice of the Hamiltonian. Now we will investigate various algorithms to find this ground state in the next sections. We will present our results in terms of reconstructed mass, the jet-assignment and the success rate.

\section{Brief review on algorithms used in this paper}
\label{sec:algorithms}

Variational quantum algorithms (VQAs) are a hybrid quantum-classical approach in which the parameters are adjusted using classical optimization techniques. For parameters $\bm\theta$, the circuit produces the state $\ket{\bm\theta}$ with the goal of minimizing $\expval{C}{\bm\theta}$ where $C$ is the cost function. As an example, if $C$ is a Hamiltonian, then we are parameterizing the circuit to find the minimum energy eigenstate of the Hamiltonian. 
In this section, we briefly introduce three classes of algorithms: the hybrid variational quantum algorithms based on QAOA \cite{Farhi:2014ych}, and the purely quantum algorithms FALQON \cite{Magann:2021evo} and VarQITE \cite{Motta:2019yya,Yuan2019-oj,McArdle2019-zp,Morris2024-gj}. All of our quantum algorithms are simulated with PennyLane \cite{Bergholm:2018cyq} and all code is available on \href{https://github.com/crumpstrr33/collider_combinatorics_with_QAs}{GitHub}\footnote{\url{https://github.com/crumpstrr33/collider_combinatorics_with_QAs}}. For the variational algorithms, we use the Adam optimizer \cite{Kingma:2014vow} for the classical optimization of the parameters\footnote{We employed the Adam optimizer for its reliability and availability within PennyLane, though we recognize the NFT optimizer \cite{Nakanishi:2019rrm} as a strong alternative with complementary strengths. For a six-qubit state-vector simulation of this scale, the specific choice of a stable optimizer is unlikely to alter our core qualitative findings.}. We run all algorithms over 12k events split equally into invariant mass bins for the results presented in sections \ref{sec:algorithms} and \ref{sec:comparison}, and in the Appendices unless otherwise stated. That is, we test the algorithms on 2k events per invariant mass bin. These events were chosen randomly once from the full 250k events used in Figure \ref{fig:energy_jets_brute_force_full_H} uniformly. These events are available in the GitHub repository.  In our analysis, we use the state vector simulations, which corresponds to infinite shots. We discuss effects of finite shots in Appendix \ref{sec:finiteshots}. 

\subsection{Quantum Approximation Optimization Algorithm (QAOA)}
\label{sec:qaoa}

\begin{figure*}[t!]
    \centering
    \includegraphics[height=25em]{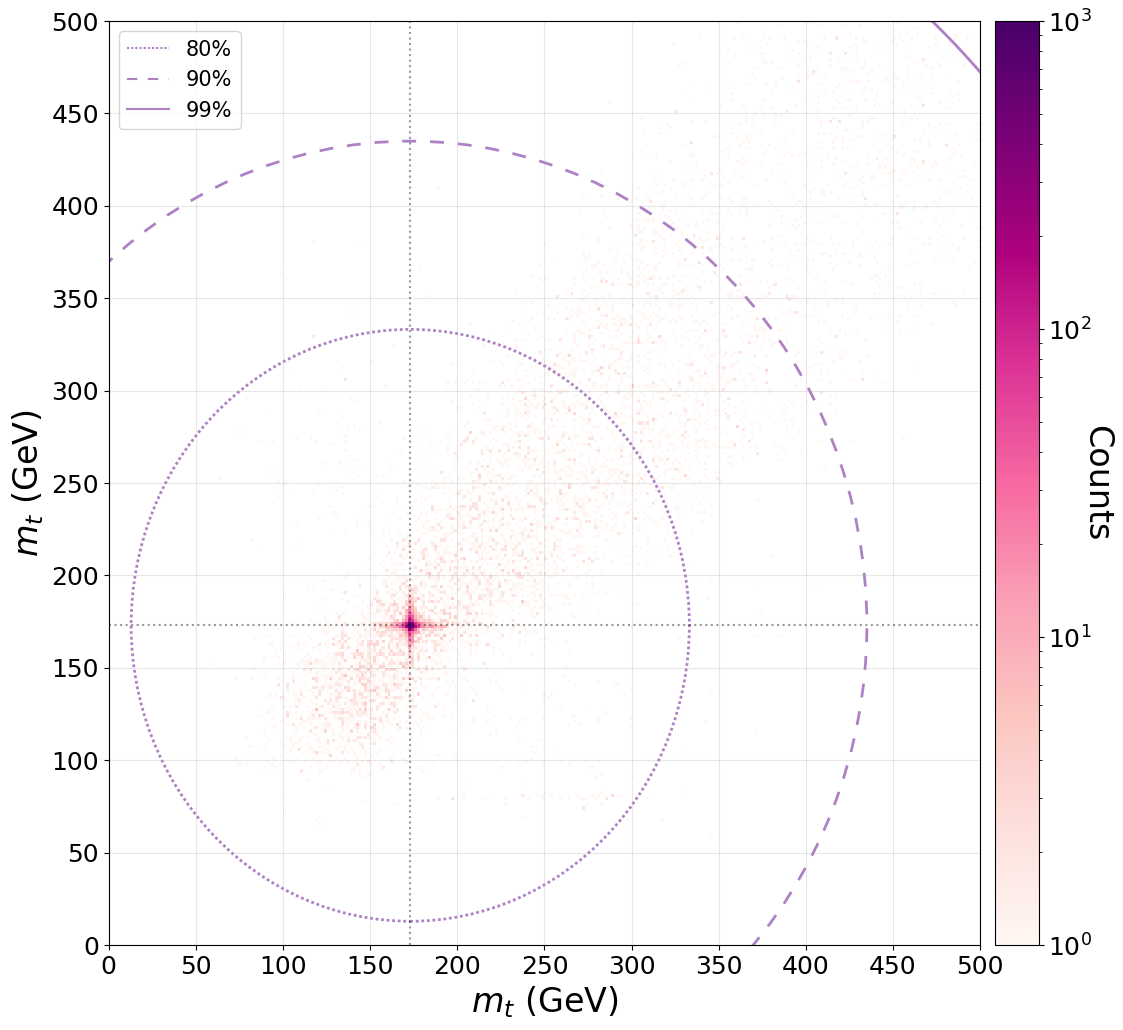}
    \hspace{0.03\textwidth}
    \includegraphics[height=25em]{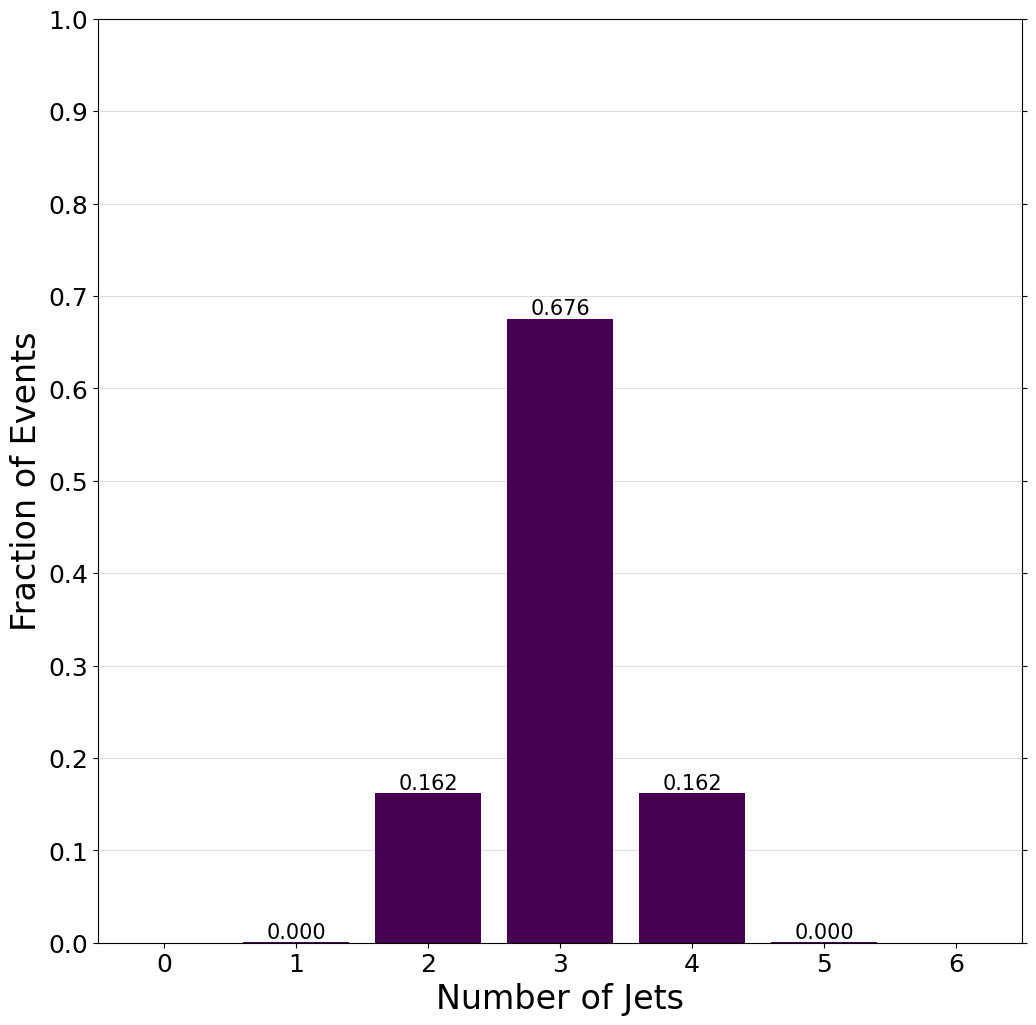}
    \caption{The reconstructed masses (left) and jet-assignment (right) for parton-level events using QAOA with depth $p=8$. The ground state is found for the full Hamiltonian. 
    The overall efficiency (matching accuracy) is 43\%. 
 } \label{fig:parton_qaoa8}
\end{figure*}

The Quantum Approximation Optimization Algorithm (QAOA) \cite{Farhi:2014ych} is one of the most widely known VQAs. 
We consider the Hamiltonian of the form 
\begin{equation}
    H(t)=\big(1-a(t)\big)H_M+a(t)H_P
    \label{eq:timedep-ham}
\end{equation}
where $a(0)=0$ and $a(T)=1$ for the final time $T$. 
Therefore, the full Hamiltonian transitions from being $H_M$ initially to $H_P$ where $H_M$ is called the \textit{mixer} Hamiltonian and $H_P$ is the \textit{problem} Hamiltonian. The former is a system that is easily solvable and preparable. We use 
\begin{equation}
H_M= \sum_{i=1}^{n}\sigma^X_i \, ,    \label{eq:mixer}
\end{equation}
where $\sigma^X_i = X_i$ is the Pauli-X gate operating on the $i$th qubit. 
The latter is a system with an energy eigenstate, specifically the ground state, which represents the solution to the given problem. Therefore, minimizing the cost function corresponds to finding the lowest eigenvalue.

We assume that the system begins with the Hamiltonian $H_M$ and evolves into the Hamiltonian $H_P$ slowly enough such that the adiabatic theorem is applicable. The time-evolution operator can be broken up into steps:
\begin{align}
    U(T,0)&=U(T,T-\Delta t)U(T-\Delta t,T-2\Delta t)\cdots U(\Delta t,0) \nonumber \\
    &=\prod_{j=1}^pU(j\Delta t,(j-1)\Delta t)
    \intertext{where $\Delta t=T/p$. If we consider $\Delta t$ to be small, \ie $p$ is large, then we can approximate the time-evolution operator as follows with constant $H$ over the short time interval $\Delta t$: }
    U(T,0)&\approx\prod_{j=1}^pe^{-i\Delta tH(j\Delta t)}\\
    &=\prod_{j=1}^p\exp\Big[-i\Delta t\big[(1-a(j\Delta t))H_M+a(j\Delta t)H_P\big]\Big] \, .\nonumber
    \intertext{Using the approximation $e^{\varepsilon(A+B)}=e^{\varepsilon A}e^{\varepsilon B }+\mathcal{O}(\varepsilon^2)$ for non-commuting operators $A$ and $B$, The time-evolution operator can be rewritten as }
    U(T,0)&
    \approx\prod_{j=1}^p\exp\Big[-i\Delta t(1-a(j\Delta t))H_M\Big] \nonumber \\
    & \hspace*{2cm} \times \exp\Big[-i\Delta ta(j\Delta t)H_P\Big] \, .
    \intertext{Note that for $\Delta t\to0$ we get back $U(T,0)$ exactly. Introducing $\beta_j\equiv\Delta t(1-a(j\Delta t))$ and $\gamma_j\equiv\Delta t \, a (j\Delta t)$, we obtain }
    U(T,0)&=\prod_{j=1}^p\exp\Big[-i\beta_jH_M\Big]\exp\Big[-i\gamma_jH_P\Big] \, .
    \intertext{Defining the unitary operator $U(\alpha,H)\equiv\exp\left[-i\alpha H\right]$, the time-evolution operator can be written as }
    U(T,0)&=\prod_{j=1}^pU(\beta_j,H_M)U(\gamma_j,H_P) 
\end{align}
which is exact in the limit $p\to\infty$. 
Then the quantum state initialized in the eigenstate of $H_M$, $\ket{+}^{\otimes n}$, evolves into the following final state 
\begin{equation}
\ket{\bm\beta,\bm\gamma}=\sum_{j=1}^pU(\beta_j,H_M)U(\gamma_j,H_P)\ket{+}^{\otimes n} \, ,
\end{equation}
where $\bm\beta = (\beta_1, \cdots \, , \beta_p)$ and $\bm\gamma = (\gamma_1, \cdots \, , \gamma_p)$.
Minimizing the expectation value of $H_P$
\begin{equation}
\langle  H_P \rangle = \expval{H_P}{\bm\beta,\bm\gamma} \, ,
\end{equation}
is then the work of a classical optimizer. Having returned updated values for $\bm\beta$ and $\bm\gamma$, the circuit is rerun and the process repeated.
The schematic QAOA diagram of depth $p$ for $n$ qubits is shown in Fig. \ref{fig:qaoa} of Appendix \ref{app:circuits}, where $U(\beta, H_M)$ and $U(\gamma_i, H_P)$ are shown in Fig. \ref{fig:mixer_layer_qaoa} and Fig. \ref{fig:cost_layer}, respectively. 
\begin{figure*}[hbt!]
    \centering
    \includegraphics[height=25em]{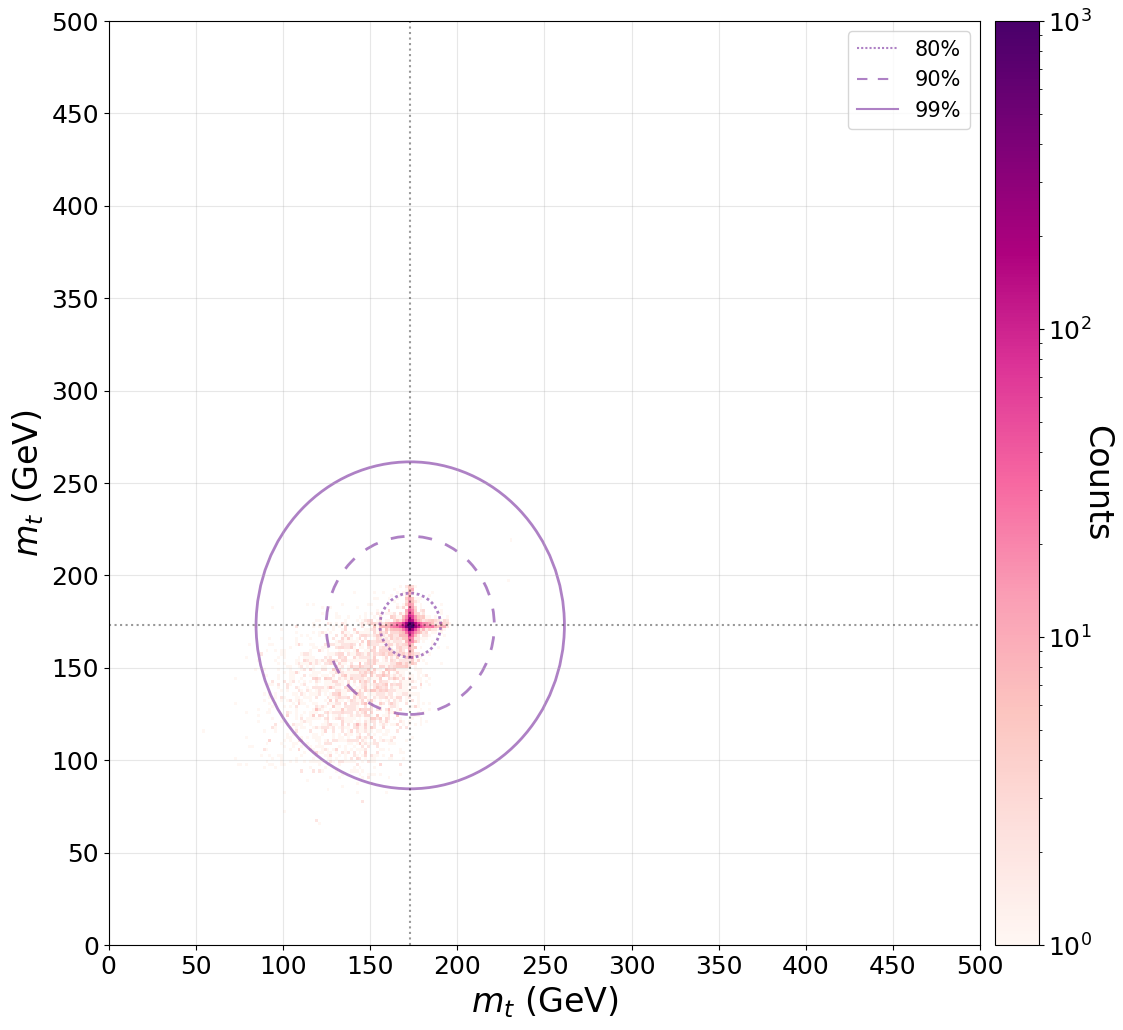}
    \hspace{0.03\textwidth}
    \includegraphics[height=25em]{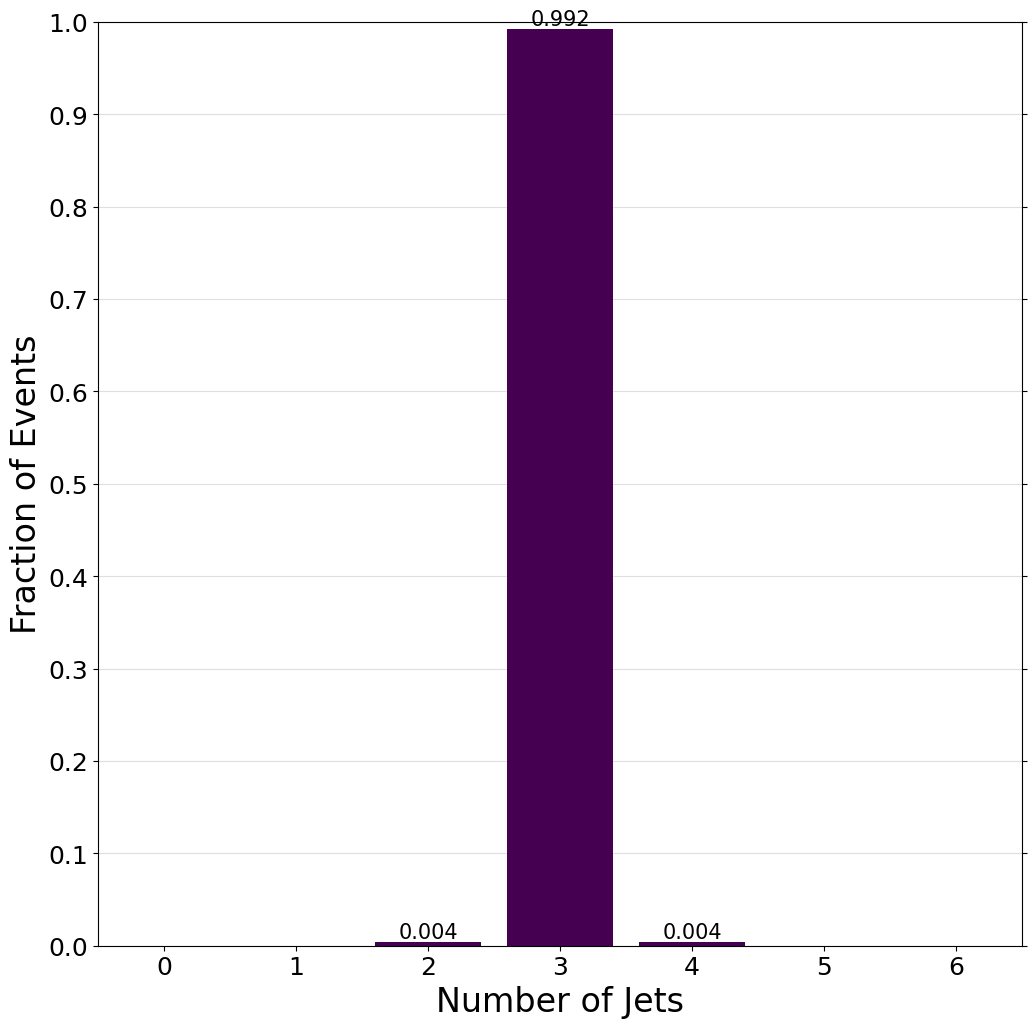}
    \caption{The same as Fig. \ref{fig:parton_qaoa8} but for ma-QAOA with $p=8$. The overall efficiency (matching accuracy) is 79\%. }
    \label{fig:parton_maqaoa8}
\end{figure*}

Fig. \ref{fig:parton_qaoa8} shows results on the reconstructed masses (left) and jet-assignment (right) for parton-level events using QAOA with depth $p=8$ and initial parameters set to 0.5. The ground state is found for the full Hamiltonian. 
Our model for the combinatorial problem is somewhat dense, since all coefficients are non-zero, $J_{ij} \neq 0$, although this particular problem requires 6 qubits. In general, a large number of layers is needed for dense models, while one can find fast and exact solution for sparse models \cite{rehfeldt2022fasterexactsolutionsparse}.

There are many studies which extend this framework to allow alternation between more general families of operators \cite{Blekos:2023nil,Crooks:2018vud,Zhou_2020}. The essence of this extension, the quantum alternating operator ansatz, is the consideration of general parameterized families of unitary transformations rather than only those corresponding to the time evolution under a fixed local Hamiltonian for a time specified by the parameter \cite{Hadfield_2019}. We will introduce a few variations which are relevant for our study in next subsections.

\subsection{Multi-Angle QAOA}
\label{sec:maqaoa}

Multi-angle QAOA (ma-QAOA) \cite{herrman2021multiangle} addresses an issue with the large depth (high $p$) usually needed for good results when using QAOA by increasing the number of adjustable parameters effectively shifting the burden from the quantum computer to the classical optimizer. In QAOA, there are $2p$ parameters with one parameter per each unitary operation per layer. Instead, in ma-QAOA, we allow each qubit to have its own free parameter, increasing the number from $2p$ to $(n+m)p$ where $n$ is the number of qubits (the number of jets in this paper) and $m$ depends on the form of $H_P$. The unitary for $H_M$ for layer $j$ then becomes
\begin{equation}
    U(\bm\beta_j,H_M)=\exp\left[-i\sum_{i=1}^n\beta_{ji}\sigma^X_i\right] \, .
\end{equation}
See Fig. \ref{fig:mixer_layer-maqaoa} in Appendix \ref{app:circuits} for the mixer layer of layer $i$ of ma-QAOA for $n$ qubits. 
Similarly for $U(\bm\gamma_j,H_P)$ we use 
\begin{equation}
    U(\bm\gamma_j,H_P)=\exp\left[-i\sum_{k<\ell=1}^n\gamma_{jk\ell}\sigma_k^Z\sigma_\ell^Z\right] \, .
\end{equation}
Here, $\beta_{ji}$ and $\gamma_{jk\ell}$ are the adjustable parameters. 
For the Ising model Hamiltonian (QUBO type), we expect $\displaystyle m=\binom{n}{2}$ as every pair of qubits will have a 2-qubit gate. The expressibility of each qubit is enhanced by ma-QAOA, enabling greater flexibility to explore its Bloch sphere, which in turn allows the circuit to cover a larger portion of the Hilbert space.
Fig. \ref{fig:parton_maqaoa8} shows results which are greatly improved (with initial parameters set to 0.5). The jet assignment changed from 80\% (QAOA) to $\sim$99\% (ma-QAOA), thus improving the resolution of the reconstructed masses.

\subsection{Expressive QAOA}
\label{sec:xqaoa}

\begin{figure*}[hbt!]
    \centering
    \includegraphics[height=25em]{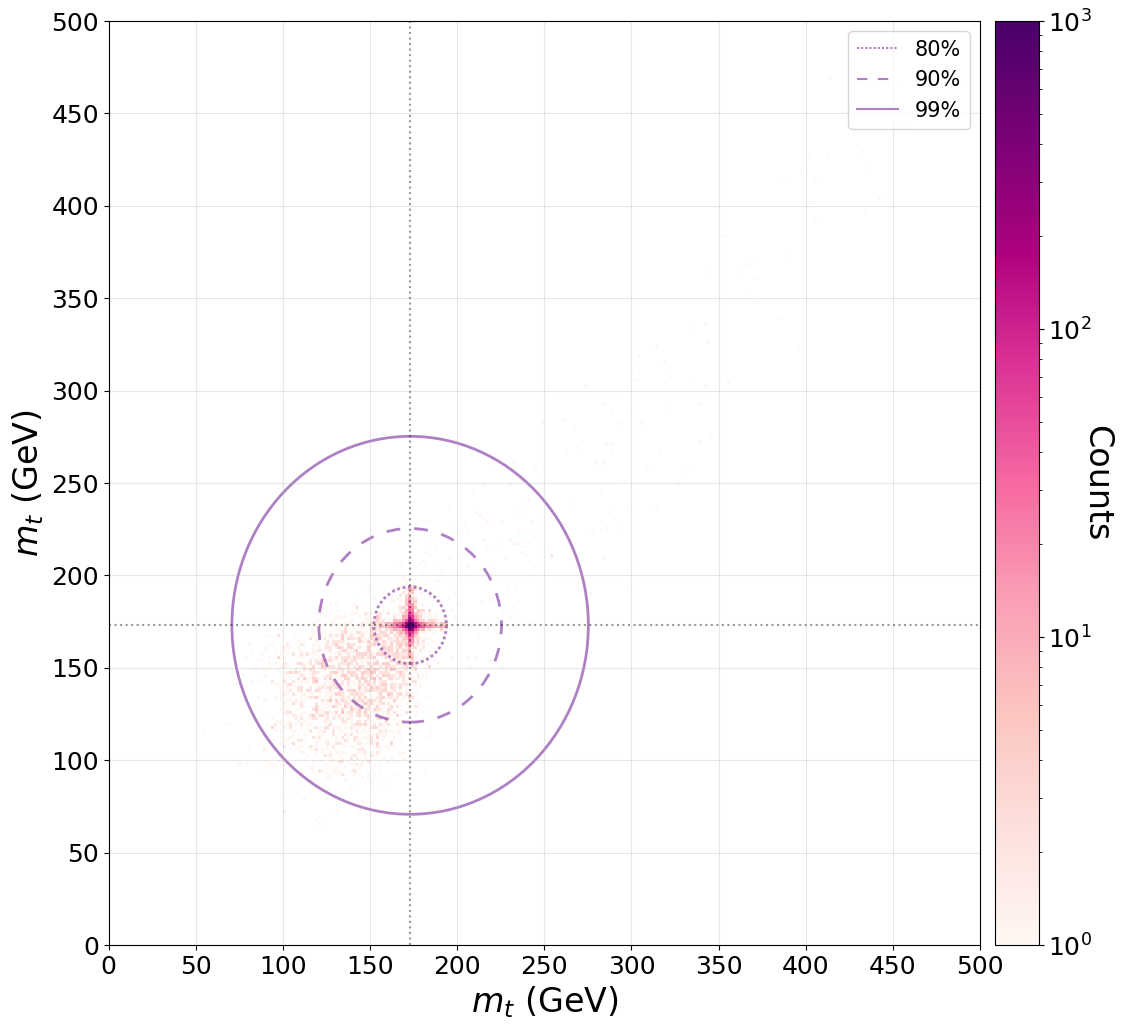}
    \hspace{0.03\textwidth}
    \includegraphics[height=25em]{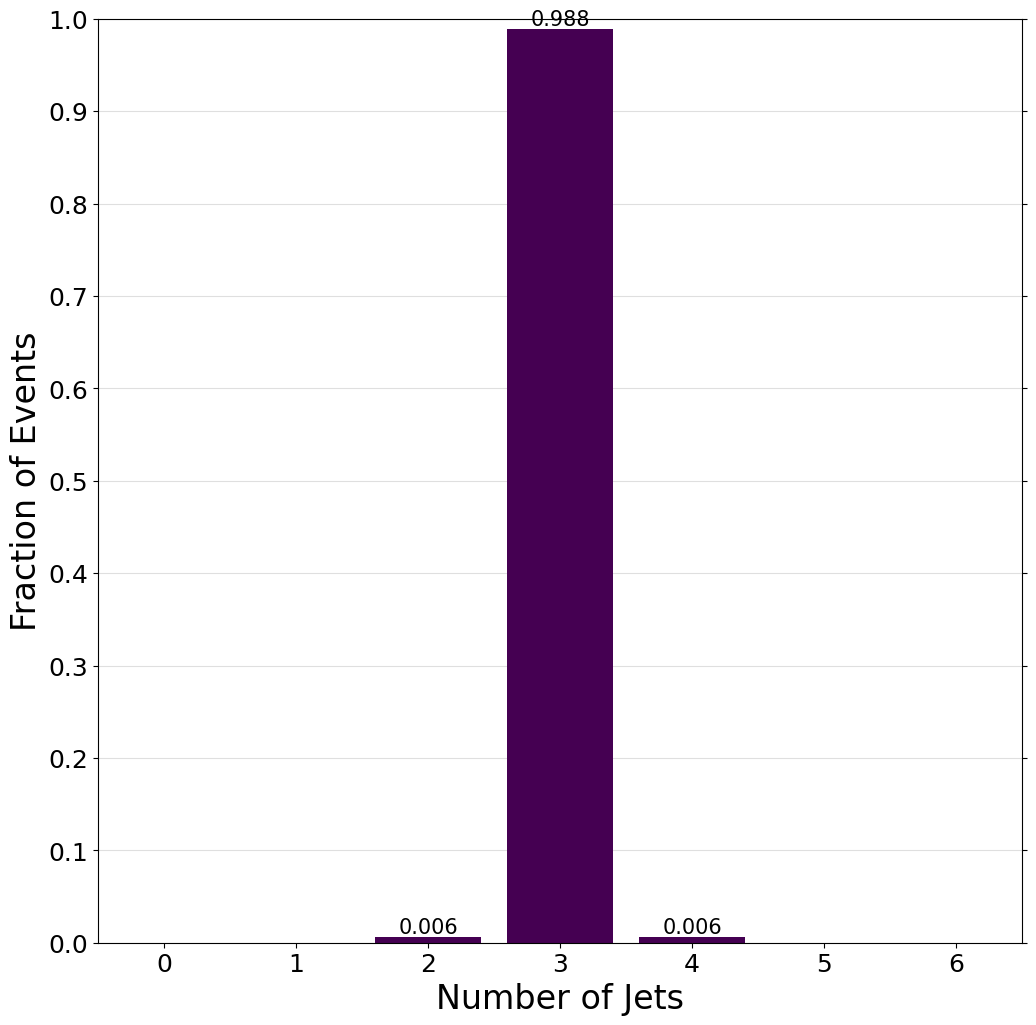}
    \caption{The same as Fig. \ref{fig:parton_qaoa8} but for XQAOA with $p=8$. The overall efficiency (matching accuracy) is 77\%. }
    \label{fig:parton_xqaoa8}
\end{figure*}

Expressive QAOA (XQAOA) \cite{vijendran2023expressive} builds upon ma-QAOA by adding another mixer Hamiltonian of $Y$ gates such that
\begin{equation}
\ket{\bm\alpha,\bm\beta,\bm\gamma}=\sum_{j=1}^pU(\bm\alpha_j,H_X)U(\bm\beta_j,H_M)U(\bm\gamma_j,H_P)
\end{equation}
where
\begin{equation}
    U(\bm\alpha_j,H_X)=\exp\left[-i\sum_{i=1}^n\alpha_{ji}\sigma^Y_i\right].
\end{equation}
Here $\sigma^Y_i = Y_i$ is the Pauli-Y gate operating on the $i$th qubit. This increases the number of adjustable parameters to $(2n+m)p$. The motivation for XQAOA is that it offers a higher expressibility still than ma-QAOA due to the addition of a $Y$ gate rotation on top of the $X$ gate with ma-QAOA being recovered when $\bm\alpha=0$.

The original QAOA is equivariant under the parity, $P = \sum_{i=1}^{n} X_i$, making the model learn the data better than non-equivariant models \cite{Skolik:2022qwn,Forestano:2023lnb,Dong:2023oqb}.
Inclusion of Y gates breaks equivariance of the quantum circuit but increases the expressivity \cite{Meyer:2022fjx}.
As a result, a smaller depth (with initial parameters set to 0.5 for $\bm\beta$ and $\bm\gamma$ and 0.01 for $\bm\alpha$) leads to a better optimization for $p \leq n$. 
We observe that XQAOA exhibits performance comparable to that of ma-QAOA, with differences remaining within a few percent and largely consistent with statistical fluctuations. The overall optimization behavior of XQAOA closely mirrors that of ma-QAOA for the problem considered here, suggesting that the two ansatzes possess similar expressive power in this regime. We further examined the confidence of XQAOA, defined as the average probability assigned to the minimum-energy (correct) bitstrings, and found it to be similarly consistent with the performance of ma-QAOA. This behavior is consistent with the well-known trade-off between trainability and expressibility in variational quantum algorithms \cite{Holmes:2021qjw, Barzen:2025nfo}. These results indicate that both approaches capture the relevant structure of the target states with comparable effectiveness, and that more challenging problem instances may be required to clearly distinguish their respective performance characteristics. 

\subsection{Feedback-based ALgorithm for Quantum OptimizatioN (FALQON)}
\label{sec:falqon}

\begin{figure*}[ht!]
    \centering
    \includegraphics[height=25em]{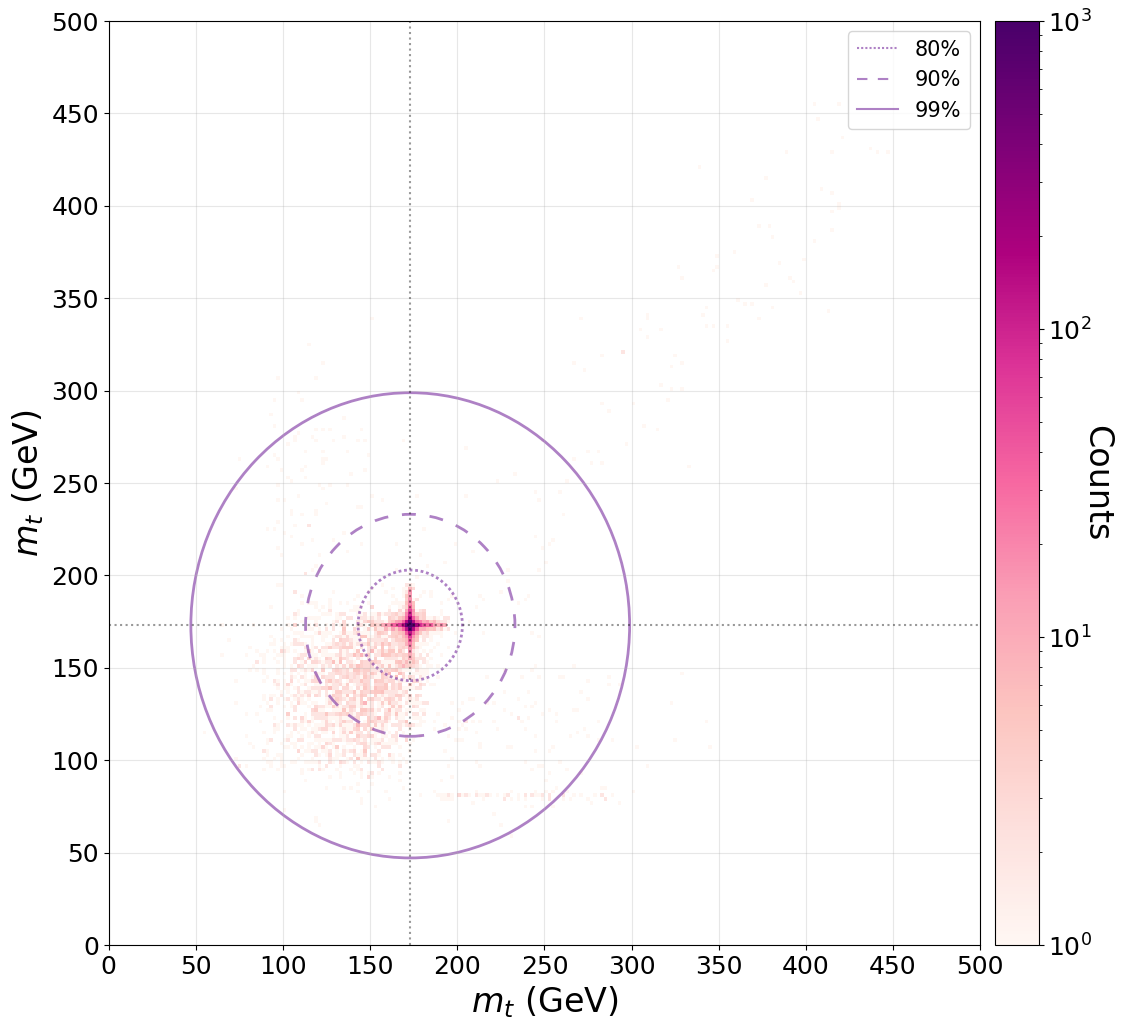}
    \hspace{0.03\textwidth}
    \includegraphics[height=25em]{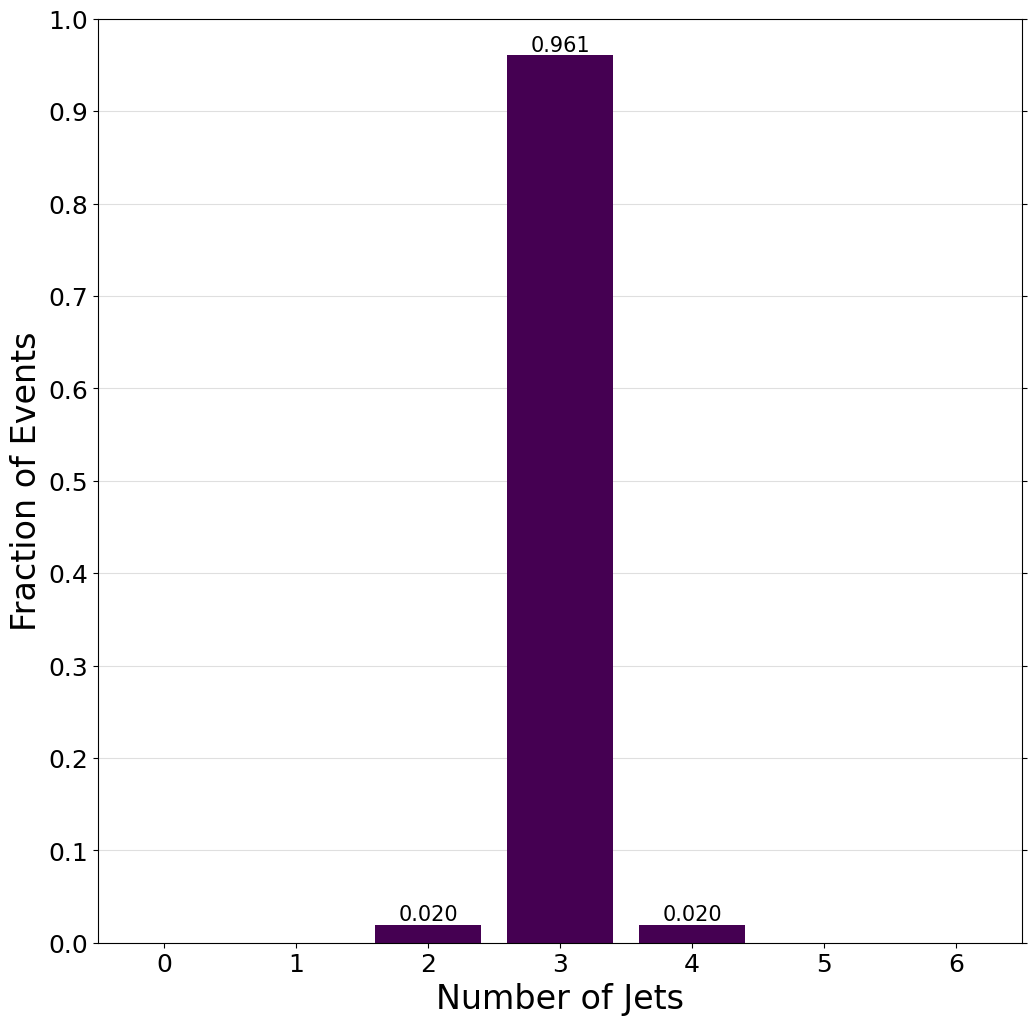}
    \caption{The same as Fig. \ref{fig:parton_qaoa8} but for FALQON with $p=250$. The overall efficiency (matching accuracy) is 75\%. }
    \label{fig:parton_falqon}
\end{figure*}

Feedback-based ALgorithm for Quantum OptimizatioN \cite{Magann_2022} (FALQON) differs from the algorithms above as it eschews the use of the classical optimizer. The circuit layers are built recursively wherein, with each layer being parameterized by a single number, the layer's parameter is determined by the previous layers, inspired by quantum Lyapunov control \cite{Magann_2022_2}. In this algorithm, we begin with a Hamiltonian of the form $H(t)=H_P+\beta(t)H_M$ where $\beta(t)$ is called the control function. As before, we wish to find the minimum energy eigenstate of $H_P$ and thus, for a state $\ket{\psi(t)}$, we require that
\begin{align}
    & \dv{t}\expval{H_P}{\psi(t)}\le0\qquad \text{or}\\
    \qquad& \beta(t)\expval{i[H_M,H_P]}{\psi(t)}\le0.
\end{align}
While the functional form of $\beta(t)$ is arbitrary, it can be dynamically determined via a feedback loop defined by 
\begin{align}
    \beta(t)&=-A(t-\tau) \, , \\
    A(t)&\equiv\expval{i[H_M,H_P]}{\psi(t)} \, ,
\end{align}
where $\tau$ is a small time delay. Following a similar derivation as above and discretizing time into time steps $\Delta t$, we obtain the recursive relationship $\beta_{k+1}=-A_k$ where $A_k=\expval{i[H_M,H_P]}{\psi_k}$ and $\ket{\psi_k}$ is the quantum state of the circuit after the $k$th layer:
\begin{equation}
    \ket{\psi_k}=U_M(\beta_k)U_P\ket{\psi_{k-1}} \, ,
\end{equation}
where 
\begin{equation}
 U_M(\beta_k)=e^{-i\beta_k\Delta tH_M},\ ~~{\rm and}~~\  U_P=e^{-i\Delta tH_P}.
\end{equation}
Unlike the alternative versions (ma-QAOA or XQAOA) of QAOA which try to leverage the classical optimizer more, FALQON removes the optimizer entirely instead relying on expectation values of the operator $i[H_M,H_P]$ to build the circuit. Once the initial value of $\beta(t)$, $\beta(0)=\beta_1$, is chosen, the only parameters are $\Delta t$ and the total number of layers $p$. 
(See Fig. \ref{fig:falqon} for a circuit diagram.)

Fig. \ref{fig:parton_falqon} summarizes our results on the reconstructed masses and the jet assignment, which are comparable to those using ma-QAOA, outperforming conventional QAOA. 
We obtain the results with $\Delta t= 0.08$ and initial parameter $\beta_1=0$ for $p=250$. 
We initialize the algorithm with $\beta_1=0$ to start from an unbiased state. A quick sensitivity scans across different initial values confirm that the feedback loop is robust, with final state convergence remaining stable across the test set. The circuit depth was set to $p=250$ based on an empirical convergence study, where we found that matching efficiency asymptotically plateaus at this depth, with further increases (up to $p=2000$) yielding marginal improvements that do not justify the additional computational overhead. 
We observe that our results are insensitive to the initial value.
A smaller $\Delta t$ converges slower, while a larger $\Delta t$ did not converge.
Feedback-based quantum optimization algorithms are advancing rapidly; for recent developments and additional references, see Refs. \cite{Magann:2021uer, Brady:2024qwv, Malla:2024zbj, Rahman:2024rbn,Brady:2024qwv,Malla:2024zbj}.

\subsection{Variational Quantum Imaginary Time Evolution (VarQITE)}
\label{sec:varqite}

The variational imaginary time evolution (VarQITE) algorithm is designed to efficiently prepare the ground state of a Hamiltonian $H_P$, which encodes a combinatorial optimization problem (though it is not limited to this class of problems) \cite{Motta:2019yya,Yuan2019-oj}. Imaginary time evolution, a concept originating in quantum many-body physics, systematically projects a given initial state $\ket{\psi(0)}$ onto the ground state of $H_P$ by evolving it in imaginary time $\tau = it$, where $t$ is the real time parameter. This process is governed by the imaginary time Schr\"odinger equation
\begin{align}
    \pdv{\tau}\ket{\psi(\tau)} = -(H_P - E_\tau) \ket{\psi(\tau)},
\end{align}
where $E_\tau = \ev{H}{\psi(\tau)}$ is the instantaneous energy resulting from enforcing normalization. For a time-independent Hamiltonian $H_P$, the solution is given by
\begin{align}
    \ket{\psi(\tau)} = \frac{e^{-\tau H_P}}{\sqrt{\ev{e^{-2\tau H_P}}{\psi(0)}}}\ket{\psi(0)}. 
\end{align}

As $\tau \to \infty$, and assuming that $\ket{\psi(0)}$ has a nonzero overlap with the ground state $\ket{\psi_{\text{GS}}}$, this evolution exponentially suppresses excited-state components, thus converging to $\ket{\psi_\text{GS}}$.

Since directly implementing the operator $\exp(-H_P \tau)$ on a quantum computer is challenging, we can instead approximately enforce imaginary time evolution using a variational method. We introduce a parameterized wavefunction
\begin{align}
    \ket{\btheta} = U(\btheta)\ket{+}^{\otimes n},
\end{align}
where $U(\btheta)$ is a unitary operator constructed from a carefully chosen ansatz, and $\btheta = (\theta_1,\theta_2,\ldots,\theta_m)$ is a set of $m$ real parameters. The goal is to adjust these parameters $\btheta(\tau)$ such that the state $\ket{\btheta(\tau)}$ closely approximates the true imaginary time evolved state $\ket{\psi(\tau)}$.

To achieve this, a common approach is to apply the McLachlan variational principle to project the imaginary time evolution onto the variational manifold defined by $\ket{\btheta}$ \cite{Yuan2019-oj,McArdle2019-zp}, which would require $\mathcal{O}(m^2)$ circuit evaluations per time step. Instead, we adopt the approach presented in \cite{Morris2024-gj}, which enforces the imaginary time evolution of each $P_\alpha$, where $H_P=\sum_\alpha P_\alpha$, by
\begin{align}
\nonumber
    \pdv{\expval{P_\alpha}}{\tau} &= \sum_j 2\Re\left[ \bra{\btheta} P_{\alpha}\, \pdv{\theta_j}\ket{\btheta}\right] \dot{\theta}_j\\
    &= -\mel{\btheta}{\{P_{\alpha},H_P-E_{\tau}\}}{\btheta},
\end{align}
with $\dot{\theta}_j = \partial \theta_j / \partial \tau$. Applying this condition to all $P_\alpha$ terms in the Hamiltonian $H_P$ yields the linear system 
\begin{align}
    G \cdot \dot{\btheta} = D,
    \label{eq:varqite_soe}
\end{align}
where
\begin{align}
    G_{\alpha,j} &= \Re\left[ \bra{\btheta} P_{\alpha}\, \pdv{\theta_j}\ket{\btheta}\right], \\
    D_{\alpha} &= -\frac{1}{2}\mel{\btheta}{\{P_{\alpha},H_P-E_{\tau}\}}{\btheta}.
\end{align}

Using the parameter-shift rule \cite{Mitarai2018-mu,Schuld2019-cq}, each column of $G$ can be determined from just two circuit evaluations. Moreover, if the Hamiltonian $H_P$ consists solely of Pauli-$Z$ operators (as those in QUBO formulations do), all $P_\alpha$ terms commute, allowing the calculation of $D$ from a single circuit evaluation. Thus, at each time step, the parameters $\btheta$ can be updated using a forward Euler method, $\btheta \to \btheta + \Delta\tau \dot{\btheta}$, where $\dot{\btheta}$ is obtained by inverting \cref{eq:varqite_soe}. This procedure requires only $2m + 1$ circuit evaluations per time step, offering a substantial improvement in efficiency compared to previous VarQITE approaches.

\begin{figure*}[t!]
    \centering
    \includegraphics[height=25em]{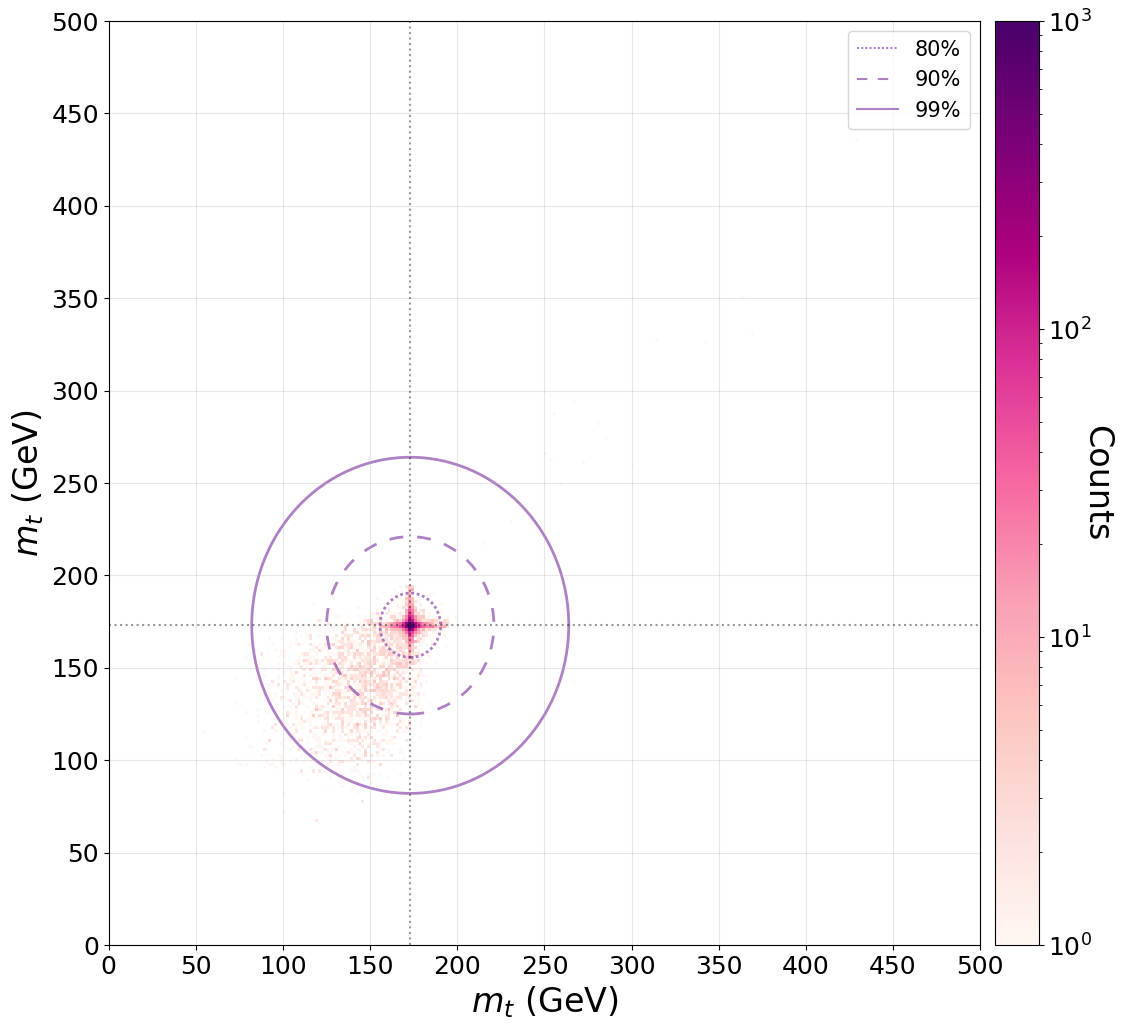}
    \hspace{0.03\textwidth}
    \includegraphics[height=25em]{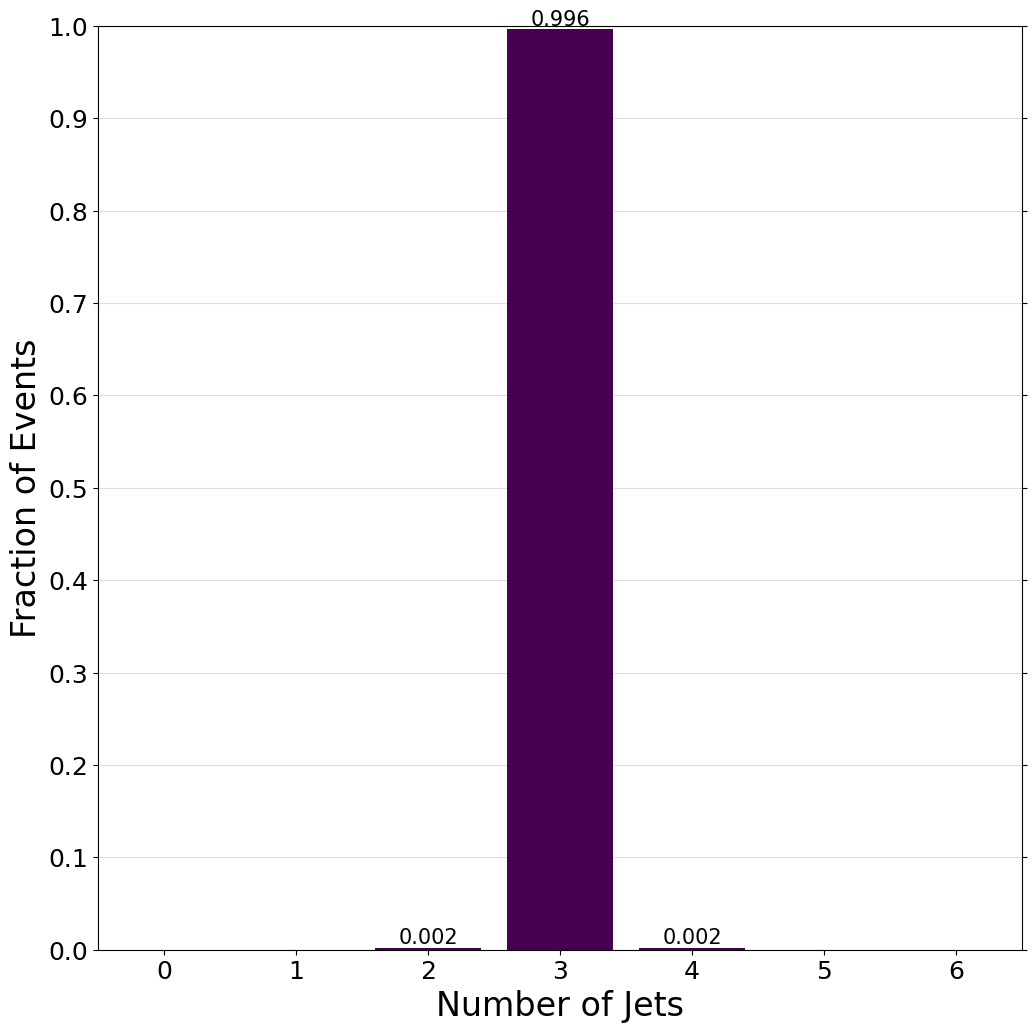}
    \caption{The same as Fig. \ref{fig:parton_qaoa8} but for VarQITE. The overall efficiency (matching accuracy) is 79\%. }
    \label{fig:parton_varqite}
\end{figure*}

Fig. \ref{fig:parton_varqite} summarizes our results on the reconstructed masses and the jet assignment, which are comparable to those using ma-QAOA, outperforming conventional FALQON or QAOA. A value of $\Delta\tau=0.5$ was used as the learning rate with a maximum of 500 optimization steps, and all initial $m$ parameters are set to $\theta_i=\pi$. The average number of evaluations was 150 for $H_0$ (129 for $H_2$). For $H_0$, only two events reached 500 evaluations, and 460 (approximately 3.8\%) exceeded 300. For $H_2$, there were no events that took all 500 evaluations, and 245 events above 300 (approximately 2\%). 
Additionally, this choice of $\Delta\tau$ was determined empirically by evaluating the performance of VarQITE for a few selected values of step size. Smaller increments, such as 0.1 and 0.05, resulted in significantly slower convergence rates and an increased susceptibility to local minima. Conversely, larger step sizes (e.g., $\Delta\tau=1$) introduced numerical instabilities, leading to oscillations in the loss function. A value of 0.5 was found to provide the optimal balance between convergence speed and algorithmic stability for the systems studied. The algorithm was ended early if, for the $i$th step, a precision of $\abs{E_i-E_{i-1}}/E_i\le 10^{-5}$ was obtained, where $E_i$ is the value of the cost function, \ie the expectation value of the Hamiltonian.

As a proof-of-principle demonstration of the practical implementation of VarQITE, we also perform full optimization runs for a representative six-qubit event on the IonQ Forte quantum processor. These real-hardware results are presented in Appendix \ref{sec:quantum_hardware}.

\section{Comparison of different algorithms}
\label{sec:comparison}

\begin{figure*}[th]
    \centering
    \includegraphics[width=\textwidth]{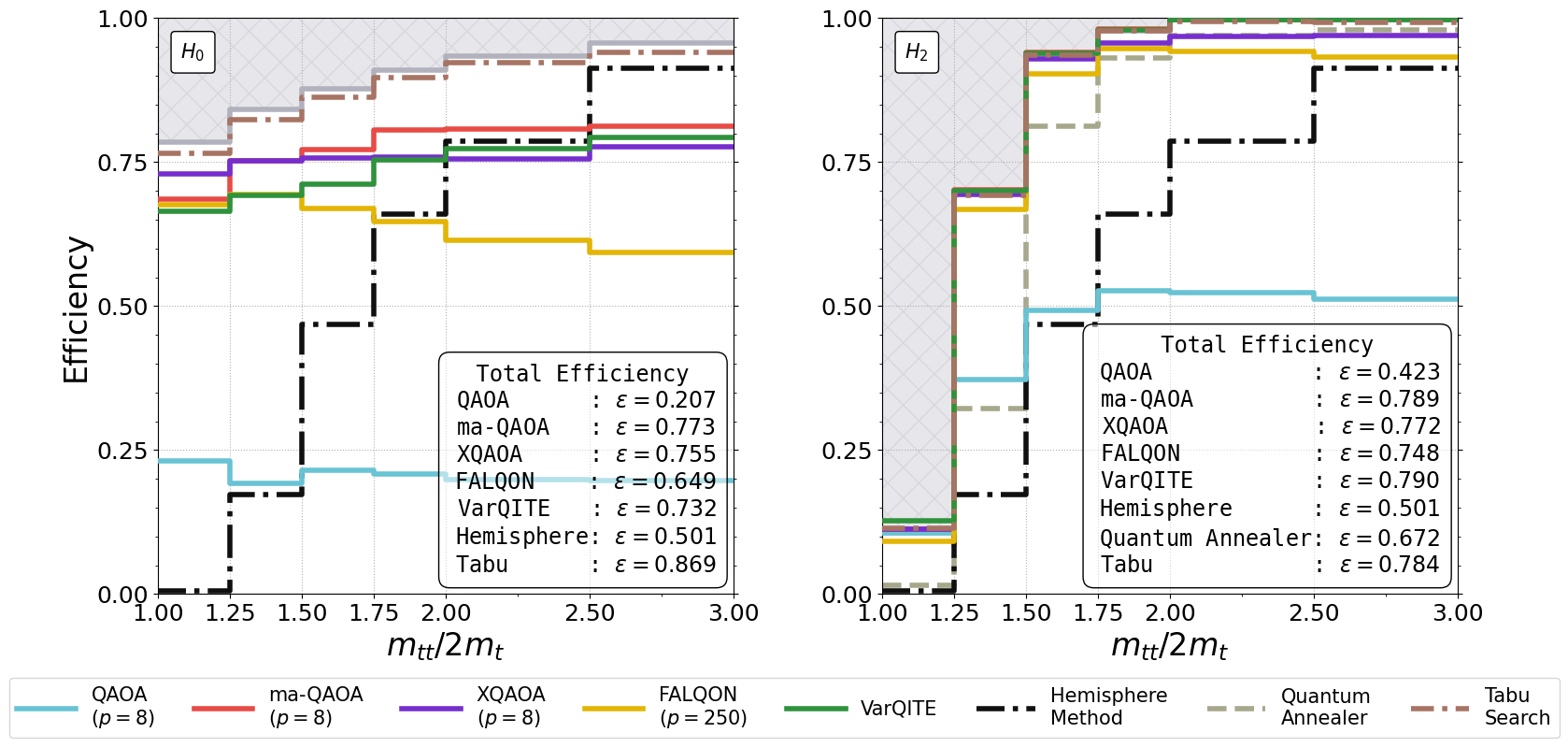}
    \caption{
        The matching accuracy or efficiency of QAOA (light blue), XQAOA (purple), ma-QAOA (red), FALQON (yellow), VarQITE (green), hemisphere (black, dot-dashed), Tabu search (brown, dot-dashed) and quantum annealing from \cite{Kim:2021wrr} (gray, dashed) for parton-level events. We show the results using $H_0$, Eq. (\ref{eq:cost}), on the left and $H_2$, Eq. (\ref{eq:fullH}), on the right. Total efficiency refers to the mean efficiency calculated across all invariant mass bins. The shaded gray region indicates the theoretical maximum efficiency for the given system, which serves as a benchmark for the chosen Hamiltonian rather than a reflection of the algorithms themselves. The $H_0$ Hamiltonian performs better in the low boost regime, while the $H_2$ shows significantly better performance in the boosted regime.
    }
    \label{fig:parton_efficiency_h0_h0h1}
\end{figure*}

We have briefly reviewed VQAs (QAOA and its variants, ma-QAOA and XQAOA) as well as FALQON and VarQITE. In this section, we present all results as a function of the boostness of the mother particles (two top quarks) \cite{Kim:2021wrr}, since boosted particles lead to the collimated daughter particles, which are easier to resolve\footnote{Note that even in the absence of ISR/FSR, the top quarks can be boosted because the high center-of-mass energy at the LHC often produces partonic collisions well above the $2m_t$ mass threshold, converting the excess energy into high transverse momentum and causing the decay products to become collimated.}. 

In Fig. \ref{fig:parton_efficiency_h0_h0h1}, we show the matching accuracy of QAOA (light blue), ma-QAOA (red), XQAOA (purple), FALQON (yellow), VarQITE (green), the hemisphere method (black, dot-dashed) and Tabu search (brown, dot-dashed) for parton-level events using $H_0$, Eq. (\ref{eq:cost}), on the left and $H_2$, Eq. (\ref{eq:h1}), on the right. 
As stated at the end of the introduction, the matching accuracy represents the efficiency of each method, indicating the fraction of events that are successfully resolved by each approach. Despite the identical parent masses in top quark pair production, which would suggest that $H_0$ is a sufficient descriptor, our results show that $H_2$ (right panel) yields better performance than $H_0$ (left panel). This is due to the off-shell effects associated with the top quark and $W$ boson. Additionally, $H_2$ helps mitigate combinatorial issues in the boosted region. When jets are collimated, interchanging decay products between the top and anti-top quarks can result in degenerate mass values and incorporating the sum of the masses provides the additional information necessary to resolve these configurations.

The boostness is illustrated by $m_{t \bar t}/(2 m_t)$, where $m_{t\bar t}$ is the invariant mass of the two top quarks and $m_t$ is the top quark mass. 
The shaded region represents the matching accuracy for a theoretical algorithm that chose the minimum eigenstate for every event, indicating the effectiveness of the Hamiltonian. 
Any quantum algorithms that we consider can not lead to results that are better than the minimum of the shaded region. 
The region around $m_{t\bar t}/(2m_t) \sim 1$ represents the threshold production, where the decay products of the top quarks are approximately isotropically distributed. In this case, the combinatorial problem is harder to resolve and the matching accuracy is low for all methods. In particular, the hemisphere method is significantly affected by the boostness, while quantum algorithms using $H_0$ are less sensitive, as shown in the left panel. 

However, when the two top quarks are very boosted ($m_{tt}/2 m_t \geq  2.5$), the hemisphere method outperforms all other methods (for $H_0$ Hamiltonian) although overall accuracy is still much higher with quantum algorithms. We see that when using the full Hamiltonian ($H_2$), there are substantial enhancements in the region with $m_{t\bar t}/(2m_t) \geq 1.5$. Although the performance near the threshold is quite suppressed, the total efficiency of the full Hamiltonian is much better than just $H_0$. Additionally, near the threshold, ma-QAOA, FALQON, and VarQITE outperform the hemisphere method. 
We note that since the hemisphere method is independent of the cost function or Hamiltonian, the two curves in the left and right are identical. 
To benchmark against classical solvers for QUBO problems, we incorporated results from a Tabu search algorithm utilizing 100 iterations and a Tabu tenure of 25 with aspiration criteria \cite{glover1986future,glover1989tabu,glover1990tabu2}. While the hemisphere method is custom-tailored for combinatorial challenges in particle physics, Tabu search serves as a robust, metaheuristic solution for general QUBO problems. Notably, our results indicate that the Tabu search performs more effectively on $H_0$ than on $H_2$. 
It is important to emphasize that the present six-qubit benchmark is not intended to demonstrate a quantum advantage over classical optimization methods. For a problem of this size, classical heuristics such as Tabu search can provide competitive solutions with modest computational resources. Rather, our objective is to formulate the particle-assignment problem as a QUBO and investigate how different quantum optimization algorithms perform within this framework. As the number of final-state particles increases, the corresponding combinatorial search space grows rapidly, making larger and more realistic event topologies an important direction for future studies. The present results should therefore be viewed as a first step toward assessing the potential of quantum optimization for more complex particle-reconstruction problems.

\begin{figure}[t]
    \centering
    \includegraphics[width=0.49\textwidth,,height=0.23\textheight]{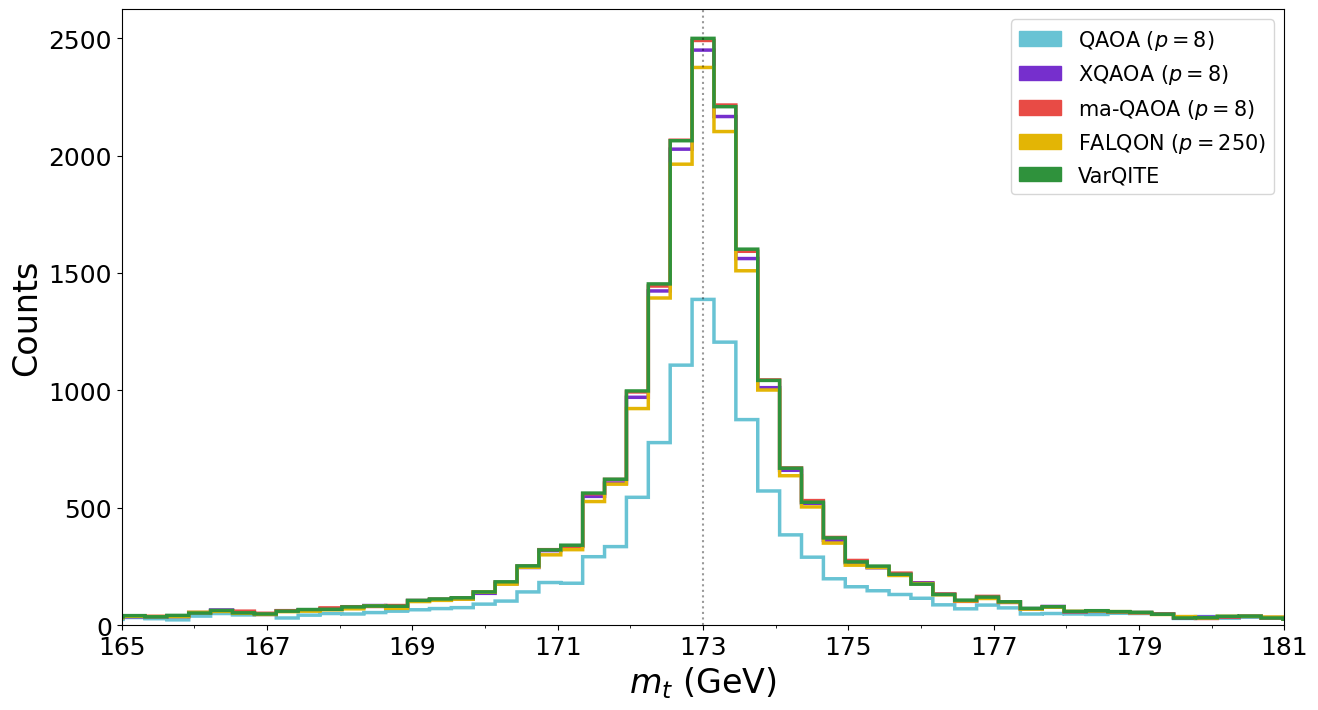}
    \caption{The reconstructed masses of top and anti-top quarks for the different algorithms using the full Hamiltonian. The dotted gray line denotes the top quark mass used for the event generation.
    \label{fig:recomassall}}
\end{figure}

Figure \ref{fig:recomassall} compares the precision of reconstructed top and anti-top quark masses across different algorithms. A narrower distribution indicates a more effective method for top mass reconstruction. The dotted gray line denotes the top quark mass used for the event generation.

\begin{figure}[t!]
    \centering
    \includegraphics[width=0.49\textwidth,height=0.23\textheight]{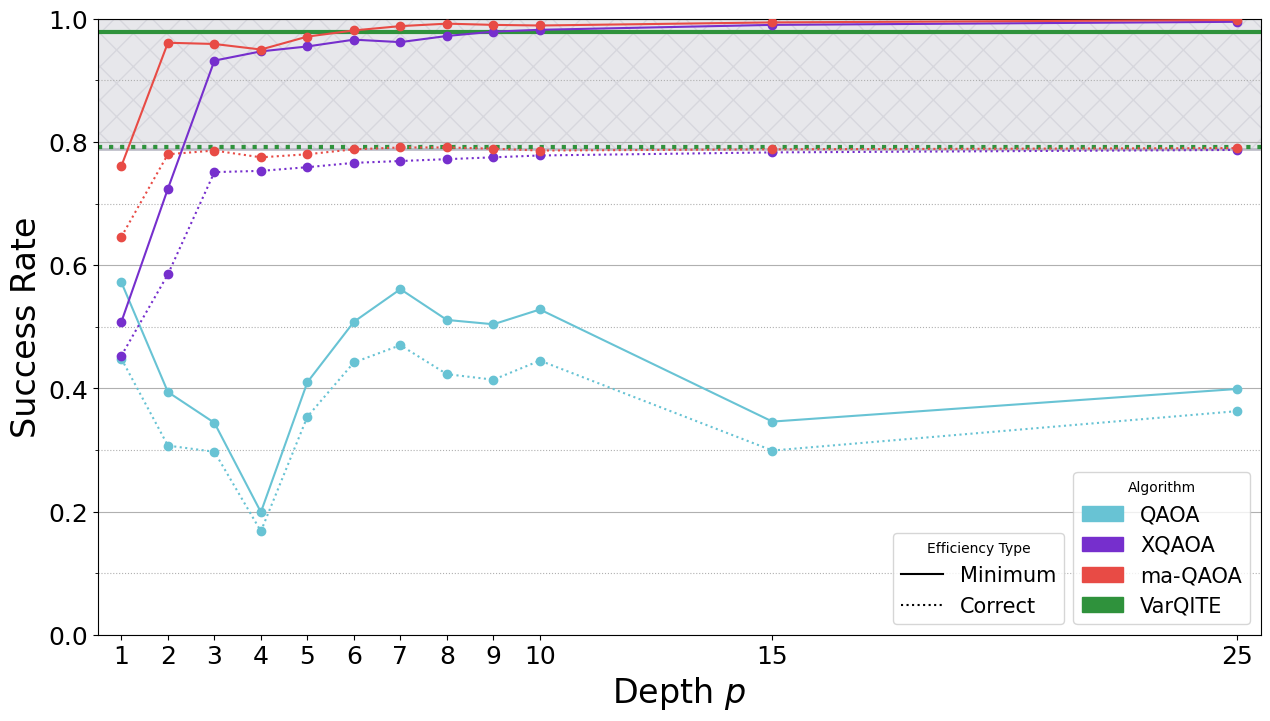}
    \caption{The success rate of QAOA (light blue), ma-QAOA (red), and XQAOA (purple) as a function of the number of iterations (depth, $p$) for parton-level events using $H_2$. Since VarQITE (green) was not considered to have more than a depth of 1, \ie more than 1 copy of the ansatz, it is shown as a horizontal line. The solid curves represent how well the Hamiltonian finds the ground state correctly, while the dotted curves represent how well the quantum algorithm resolves the combinatorial problem. 
    Analogous to the shaded region in Figure \ref{fig:parton_efficiency_h0_h0h1}, the shaded gray region here illustrates the maximum achievable efficiency of 79\% for solving the combinatorial problem via the $H_2$ Hamiltonian. This value represents a fundamental performance ceiling that no quantum algorithm can exceed. 
    } 
    \label{fig:parton_h_full_iteration}
\end{figure}

One important question is how to choose the depth of quantum circuits. Fig. \ref{fig:parton_h_full_iteration} shows the success rates of QAOA (light blue), ma-QAOA (red) and XQAOA (purple) as a function of the number of iterations (depth) for parton-level events using the full Hamiltonian.
The solid curves labeled as ``Minimum'' represent the success rate of the algorithm in finding the ground state of the Hamiltonian, while the dotted curves labeled as ``Correct'' represent the matching accuracy (efficiency) in resolving the combinatorial problem. 
We see that ma-QAOA and XQAOA quickly reaches the maximum efficiency at $p=2$ and VarQITE (represented by the green line) additionally sits at this limit, while the results of QAOA fluctuate significantly. 

It is important to distinguish the reconstruction efficiency from the ``confidence'' of the quantum state. While efficiency is a binary metric (identifying whether the most probable bitstring matches the physical truth), it does not account for the statistical weight of that prediction. A system that yields the correct bitstring with 99.9\% probability and one that yields it with 20\% (where all other states are $<20\%$) share the same efficiency, despite vastly different signal-to-noise ratios.

We define ``confidence'' as the mean probability of measuring the correct bitstring upon sampling. As shown in Fig. \ref{fig:parton_h_full_iteration}, reconstruction efficiency does not scale monotonically with circuit depth ($p$) for all algorithms. For highly expressive architectures like ma-QAOA, efficiency saturates to the theoretical maximum at a shallow depth of $p=2$. At this depth, both ma-QAOA and VarQITE exhibit near 100\% confidence, effectively mitigating the non-deterministic nature of the quantum measurement; the correct state is resolved with high fidelity in very few shots. In contrast, standard QAOA at $p=8$ achieves a confidence of only $\sim 10\%$, indicating a failure to significantly amplify the ground state amplitude. Similarly, FALQON requires substantial complexity, plateauing at $\sim 25\%$ confidence for $p=250$, and requiring $\mathcal{O}(10^3)$ layers to approach 75\%.

\begin{figure}[t!]
    \centering
    \includegraphics[width=0.48\textwidth]{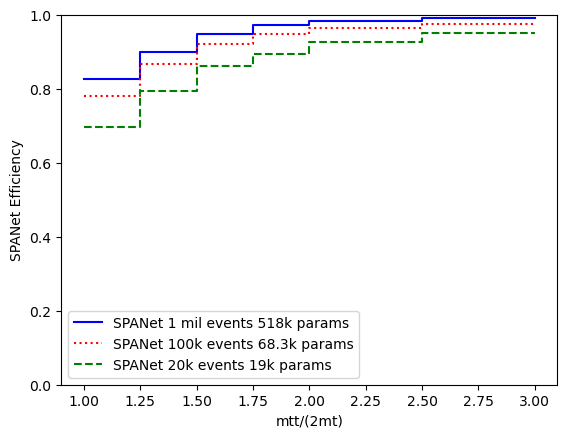}
    \caption{
        Efficiency plot for parton-level events using SPANet \cite{Fenton:2020woz,Shmakov:2021qdz} for several different number of training events and number of the network parameters. The corresponding overall efficiencies are 90.7\% (blue, solid), 87.4\% (red, dotted) and 80.7\% (green, dashed), respectively.}
    \label{fig:spanet}
\end{figure}

Finally we compare quantum algorithms against machine learning (ML) methods. We chose to use SPANet \cite{Fenton:2020woz,Shmakov:2021qdz}, which is a symmetry-preserving attention network reflecting the problem's natural invariance to efficiently find assignments without evaluating all permutations. Refs. \cite{Fenton:2020woz,Shmakov:2021qdz} showed that this general approach is applicable to arbitrarily complex configurations and significantly outperforms current methods. Fig. \ref{fig:spanet} shows the efficiency for parton-level events using SPANet for several different number of training events ($n_{\rm train}$) and number of learnable parameters ($n_{\rm params}$), 
blue for $(n_{\rm train}, n_{\rm params})$ = (1 mil, 518k), 
red for $(n_{\rm train}, n_{\rm params})$ = (100k, 68.3k) and green for $(n_{\rm train}, n_{\rm params})$ = (20k, 19k). The corresponding overall efficiencies are 90.7\% (blue, solid), 87.4\% (red, dotted) and 80.7\% (green, dashed), respectively. These results are obtained with the default settings in SPANet with AdamW optimizer \cite{loshchilov2017decoupled}.

\begin{table*}[ht!]
    \centering
    \renewcommand\arraystretch{1.9}
\resizebox{\textwidth}{!}{%
    \begin{tabular}{c||c|c|c|c|c|c|c|c|c}
            \hline\hline
        \multirow{2}{*}{Methods}      &  \multicolumn{2}{c|}{~~matching accuracy (efficiency)~~} &  ~~\multirow{2}{*}{$n_{\rm train}$}~~ & ~~\multirow{2}{*}{$n_{\rm parameters}$}~~ & ~~~\multirow{2}{*}{depth ($p$)}~~~ & ~~\multirow{2}{*}{$n_{\rm CNOT}$}~~ & ~~~\multirow{2}{*}{$n_{\rm R_Z}$}~~~ & ~~~\multirow{2}{*}{$n_{\rm R_X}$}~~~ & ~~~\multirow{2}{*}{$n_{\rm R_Y}$}~~~ \\
            \cline{2-3}
        & ~parton-level~ & ~smeared events~ & & & & & & \\   
            \hline\hline
       ~Hemisphere \cite{Matsumoto:2006ws,CMS:2007sch}~   & 50\%  & 48\%  &  \multicolumn{7}{c}{N/A} \\
            \hline 
       Tabu \cite{glover1986future,glover1989tabu,glover1990tabu2}   & 78\%  & 76\%  & N/A & 2 & \multicolumn{5}{c}{N/A}  \\
            \hline\hline
        QAOA \cite{Farhi:2014ych} & 43\% & 41\% & \multirow{4}{*}{N/A} & 16 &  \multirow{3}{*}{8} & \multirow{3}{*}{240} & \multirow{3}{*}{120} & \multirow{3}{*}{48} & \multirow{2}{*}{0} \\
        ma-QAOA \cite{herrman2021multiangle} & 79\% & 77\%  & & 168 & & & & & \\
            \cline{10-10}
        XQAOA \cite{vijendran2023expressive}& 77\% & 75\% & & 216 & & & & & 48 \\
            \cline{1-3}
            \cline{5-10}
        FALQON \cite{Magann_2022}& 75\% & 72\% & & 2 & 250 & 7,500 & 3,750 & 1,500 & 0 \\
            \cline{1-3}
            \cline{5-10}
        VarQITE \cite{Motta:2019yya,Yuan2019-oj,Morris2024-gj}& 79\% & 77\% & & 16 & 1 & 30 & 15 & 30 & 0 \\
            \hline\hline 
        \multirow{2}{*}{SPANet} \cite{Fenton:2020woz,Shmakov:2021qdz} & 91\% & 70\% & $5\times 10^5$ & $10^6$ & \multicolumn{5}{c}{\multirow{2}{*}{N/A}} \\
            \cline{2-5}
        & 81\%  & 62\% & $2\times 10^4$ & $1.9\times 10^3$ \\
            \hline\hline
    \end{tabular}
    }
    \caption{Summary of the performance of various methods and the corresponding parameters for the full Hamiltonian $H_2$.}
    \label{tab:summary_table}
\end{table*}

There are several notable differences between quantum algorithms and ML methods. 
First, quantum algorithms are independent of the particle masses, while the top mass information is indirectly learned in ML methods. 
If the top mass changes, ML needs new training. 
Second, quantum algorithms do not require the training step at all and the Hamiltonian is computed on an event-by-event basis, while ML methods learn properties of all training samples. However, QAOA and its variants require an optimization process, while FALQON does not. 
Third, SPANet is a supervised learning algorithm, which assumes the event-topology (2-step 2-body in this study). On the other hand, the quantum algorithms advertised in this paper do not require any training nor event topology. 
Finally, SPANet is the state-of-art-ML method for resolving the combinatorial problem in the multi-jet final state \cite{Fenton:2020woz,Shmakov:2021qdz}, and quantum algorithms that we have investigated already give comparable results. Considering the current era is only the beginning of quantum algorithms, we anticipate substantial improvement with new quantum algorithms in the near future.

We summarize some of the important parameters in Table \ref{tab:summary_table} for the various methods we investigated in resolving the combinatorial ambiguities in the fully hadronic production of top pairs. 
The corresponding accuracies or efficiencies for both parton-level and smeared events are also presented (see Appendix \ref{sec:smearing} for description of smeared events). QAOA, ma-QAOA and FALQON all have the same number of gates per layer. For $n$ qubits, each layer contains $\binom{n}{2}$ $R_{ZZ}$ gates and $n$ $R_X$ gates, and each $R_{ZZ}$ gate includes 2 CNOT gates and 1 $R_Z$ gate, which leads to 
\begin{align}
    n_{\rm CNOT} &= 2p\binom{n}{2}\\
    n_{R_Z} &= p\binom{n}{2}\\
    n_{R_X} &= pn
\end{align}
CNOT, $R_Z$ and $R_X$ gates, respectively. 
Consequently, the computational resources in terms of gate complexity scale quadratically with the number of particles per circuit layer, $\mathcal{O}(p \cdot n^2)$. This scaling is a direct result of the dense all-to-all connectivity required by the problem Hamiltonians $H_0$ and $H_2$. The classical optimization resources also scale with the number of parameters. For instance, ma-QAOA requires $p(n + \binom{n}{2})$ adjustable parameters, transferring a significant computational burden to the classical optimizer as $n$ or $p$ increases.  
For $p=8$, the circuit contains 240 CNOT gates, 120 $R_Z$ gates and 48 $R_X$ gates. FALQON with depth $p=250$ uses 7500 CNOT gates, 3750 $R_Z$ gates and 1500 $R_X$. The number of parameters ($n_{\rm parameters}$) is $2p=16$ for QAOA and $p(n+\binom{n}{2})=168$ for ma-QAOA. FALQON uses only two free parameters $\Delta t$ and $\beta_1$, while VarQITE has $1+\binom{n}{2}=16$ including the time step $\Delta\tau$ and a free parameter for every pair of qubits. More details can be found in Appendix \ref{app:circuits}.
ML methods such as SPANet needs the training procedure, which requires the sufficient number of training samples ($n_{\rm train}$) and a large number of parameters ($n_{\rm parameters}$).
We note that in Table \ref{tab:summary_table}, the $n_{\rm parameters}$ represents two distinct concepts: either the count of learnable parameters for SPANet or the number of adjustable parameters for quantum algorithms, respectively.

When discussing the performance of a quantum algorithm, the number of CNOT gates is often counted because CNOT gates are a key building block for creating entanglement between qubits, which is a crucial aspect of many quantum algorithms, and therefore, the number of CNOT gates directly impacts the complexity and execution time of the algorithm on a quantum computer \cite{Liu:2023aht, patel2003efficient}.

\section{Summary and outlook}
\label{sec:summary}

Combinatorial problems have become a major challenge in particle physics experiments. Unlike the simpler environment of lepton colliders, hadron colliders involve numerous jets - some from heavy particle decays and others from initial state radiation - making event reconstruction complex. Finite detector resolution and the inability to distinguish quarks from anti-quarks add to this difficulty. Solving the combinatorial challenge of associating jets with specific particle decays improves the precision of particle measurements and advances the discovery of new physics via better identification of event topology and determination of masses and spins \cite{Franceschini:2022vck}. Several ideas have been proposed including kinematic methods and ML approaches. In this paper, we have proposed the use of quantum algorithms and compared them against existing methods, taking the fully hadronic channel of the top quark production at the LHC as a concrete example. We have investigated the performance of three types of quantum algorithms: variational quantum algorithms (QAOA and its variants), a feedback-based algorithm (FALQON), and VarQITE, an imaginary time evolution algorithm. 
We have used 6 qubits corresponding to 6 particles in the final state. Our results have shown that the efficiency for selecting the correct pairing can be improved by utilizing quantum algorithms over conventional kinematic methods. 
At the same time, the present six-qubit study does not establish a quantum advantage over general-purpose classical optimization methods such as Tabu search; rather, it provides a proof-of-principle study of the QUBO formulation and quantum algorithms for this class of particle-physics combinatorial problems.
Additionally, the performance of quantum algorithms is comparable to ML methods. 
While our results indicate that the gate counts for these VQAs scale as $\mathcal{O}(p \cdot n^2)$, maintaining high reconstruction efficiency as $n$ grows will likely require a commensurate increase in circuit depth $p$. In principle, these quantum algorithms can be generalized to $n$ particles \cite{Lotshaw:2022wkm,Shaydulin:2023fpr,Weidenfeller:2022hkn,Montanez-Barrera:2024tos}. However, demonstrating empirical scaling advantages for larger problem sizes requires overcoming significant hurdles, particularly the presence of Barren Plateaus and local minima in the classical optimization landscape. A rigorous asymptotic scaling analysis of the reconstruction efficiency versus problem size $n$ is beyond the scope of the current study, and fundamental scalability of VQAs such as QAOA remains an open question in the field.

Moreover we must acknowledge that the practical implementation of these strategies on current quantum hardware remains a significant challenge. The circuit depths explored in this study—exemplified by our 8-layer, 6-qubit configurations—already face substantial hurdles due to quantum noise and decoherence. In practice, the transpilation of these circuits into native hardware gates further increases their depth, often pushing them toward or beyond the noise floor of current NISQ devices, even when employing advanced error suppression. While our findings demonstrate a clear theoretical advantage in matching efficiency, the transition from simulation to large-scale, real-world application will depend on continued advancements in hardware stability, gate fidelity, and error-mitigation techniques. A formal study of these scalability issues is beyond the scope of the current paper. However, relevant discussions regarding scaling in quantum annealing frameworks can be found in Refs. \cite{Kim:2021wrr, Wei:2019rqy}. 

In particular, the challenges of Barren Plateaus (BPs) and local minima are central to the scalability of variational quantum simulations. While detailed study is necessary, one could address these concerns through the following strategies:

\begin{enumerate}
    \item Ansatz-Dependent Mitigation: The susceptibility to BPs is highly dependent on the choice of ansatz and the circuit depth \cite{McClean:2018jps}. While BPs represent a fundamental challenge for deep, hardware-efficient circuits with random initializations \cite{Cerezo:2020mtn}, the use of physically-motivated initializations \cite{Stamatopoulos:2020xez} and layer-wise training \cite{Skolik:2020jvn} could significantly mitigate these effects by keeping the optimization within a ``favorable" region of the Hilbert space where gradients remain non-vanishing.
    \item Algorithmic Resilience (FALQON \& VarQITE): The choice of the classical-quantum feedback loop is critical. FALQON \cite{Magann:2021evo} is inherently distinct from standard gradient-based VQAs, since it utilizes a feedback-based parameter update designed to achieve monotonic energy decrease without requiring the explicit calculation of global gradients, potentially offering higher resilience to BPs. Similarly, VarQITE \cite{McArdle:2019zrh} simulates the natural gradient flow in the imaginary time domain. By following the geometry of the state space (the Fubini-Study metric), VarQITE often identifies more direct paths to the ground state, bypassing many of the ``trap" structures and local minima encountered by standard Euclidean gradient descent \cite{Stokes:2019pmg}.
    \item Convergence and Verification: To ensure the reconstruction is not contaminated by local minima, one could  employ multiple random restarts/initialization. Furthermore, the physical consistency of the reconstructed state serves as an internal diagnostic tool. While a global minimum cannot be strictly guaranteed for non-convex landscapes, these verification steps ensure the reliability of the results presented. Furthermore, warm-starting \cite{Egger:2020dbb} mitigates barren plateaus and local minima by replacing random initialization with physically informed parameters from classical approximations. This strategy positions the optimizer within a favorable "basin of attraction" where gradients are non-vanishing, bypassing the flat, high-energy regions of the cost landscape to ensure more reliable convergence toward the global minimum.
\end{enumerate}

Hybrid quantum-classical algorithms such as VQAs are well-suited for current noisy quantum devices, leveraging a hybrid approach that combines quantum circuits with classical optimization to solve complex problems \cite{Ge:2022vmm,Callison:2022zfe,Campos:2024wvc}. Their flexibility and resource efficiency make them ideal for various tasks in optimization and machine learning problems, while also offering some resilience to noise. However, they face challenges such as the potential inefficiency of classical optimizers, difficulty in selecting the right ansatz, and the risk of barren plateaus, where optimization stagnates. Additionally, while VQAs are more noise-tolerant than some algorithms, scalability remains limited due to noise and classical bottlenecks \cite{Cerezo:2020jpv}.
On the other hand, feedback-based quantum algorithms offer real-time adaptability, allowing quantum operations to be adjusted based on intermediate measurements, which helps mitigate noise and improve precision, especially in quantum sensing and control. They are crucial for certain quantum error correction techniques and can optimize the use of available resources by reducing the need for deep circuits. However, frequent measurements risk collapsing quantum states, increasing complexity and overhead due to the need for fast classical processing. In some cases, feedback can amplify noise if not properly designed, and the approach may not be broadly applicable across all quantum algorithms \cite{Magann:2021evo}. 
The Variational Quantum Imaginary Time Evolution (VarQITE) algorithm offers a practical approach for approximating ground states on near-term quantum devices by mimicking imaginary time evolution through a variational framework. It is hardware-friendly, avoids non-unitary operations, and provides some resilience to noise. However, its performance heavily depends on the quality of the ansatz, and it incurs significant classical and measurement overhead due to the need to compute and invert the overlap matrix \cite{Motta:2019yya,Yuan2019-oj,McArdle2019-zp,Morris2024-gj}.

In general, kinematic methods and quantum algorithms do not require a training process, whereas ML methods heavily depend on training a network with numerous events. VQAs involve parameter optimization using classical optimizers, while feedback-based algorithms do not. It is important to note that each method has its own strengths and weaknesses. The performance of different methods can vary depending on the underlying event topology and the study point, meaning no single method is consistently superior. Therefore it is prudent to keep a broad range of methods available in the analysis toolbox. 

The interplay between these methods highlights the tension between optimality and generalizability. While the straightforward kinematic methods are robust and universally applicable (model-independent), their performance is not the best. In contrast, the ML methods offer greater sensitivity and improved physics performance, but they are not easily generalizable to other signal processes, \ie they need to be retrained. Quantum algorithms provide an alternative solution to the combinatorial problems but they have their own challenges such as 
\begin{itemize}
    \item the effect of quantum noise (appendix \ref{sec:noise}), 
    \item statistical uncertainty due to finite shot count (appendix \ref{sec:finiteshots}),
    \item detector effects (appendix \ref{sec:smearing}), and 
    \item the performance of quantum algorithms for different processes (appendix \ref{sec:diff_processes}).
\end{itemize}
In the Appendix, we will make short comments on these points as well as on quantum algorithms based on a non-adiabatic path (appendix \ref{sec:adapt}).
Ultimately, to make quantum algorithms more practical, a more in-depth investigation is necessary. 

While this study establishes a baseline for the use of quantum algorithms to resolve the combinatorial problem using parton-level events, several questions must be addressed to bridge the gap toward experimental data. In particular, the inclusion of initial and final state radiation (ISR/FSR), parton shower, and hadronization will introduce additional radiation and jet activity, potentially increasing combinatorial complexity. 
It is critical to address cases where missing transverse energy arises, whether from undetected soft jets or from invisible particles in decay chains, such as neutrinos in the dileptonic $t\bar t$ events. Accounting for such missing information or additional jets may require a fundamental change in the problem Hamiltonian to reconstruct the full kinematics.
Future work could extend beyond the fully resolved topology to the boosted regime, where jet merging fundamentally alters the combinatorial landscape. Refining these algorithms to account for realistic detector effects and showering physics remains a critical step toward a comprehensive quantum-enhanced reconstruction framework at the LHC. 

\medskip
\textbf{Data Availability: } The datasets generated and analyzed in current study are available on \href{https://github.com/crumpstrr33/collider_combinatorics_with_QAs/tree/main/data}{GitHub}, \url{https://github.com/crumpstrr33/collider_combinatorics_with_QAs/tree/main/data}.

\medskip
\textbf{Code Availability: } The codes used to generate the results in this study is publicly available on \href{https://github.com/crumpstrr33/collider_combinatorics_with_QAs}{GitHub}, \url{https://github.com/crumpstrr33/collider_combinatorics_with_QAs}.

\medskip
\textbf{Author Contributions: }
MP, KK, and TK conceptualized the core idea of this study. ZD, JS, and HY prepared the simulation datasets and conducted the primary analysis using various algorithms. Classical machine learning implementations were executed by ZD, KK, and TK. WA, AK, and MR performed the full optimization runs of the VarQITE algorithm on the 36-qubit IonQ Forte system, and drafted Section IV.E, Appendix G, alongside other sections of the text. All authors participated in discussing the results and contributed to the preparation of the final manuscript.

\medskip
\textbf{Additional Information: }
Correspondence and requests for materials should be addressed to Myeonghun Park, parc.seoultech@seoultech.ac.kr.

\bigskip
\emph{Acknowledgements:} 
We thank Konstantin Matchev, Sergei Gleyzer, Andrea Delgado and Talal Chowdhury for useful discussion, and  Ahamed Hammad for introducing us the FALQON algorithm. We also thank Quantum Information Research Support Center and IonQ for their support through the mentoring program.
KK is supported in part by the National Science Foundation through Grant PHY-2609759. JS is supported by US DOE AI-HEP grant DE-SC0024673. JS and CD is supported in part by College of Liberal Arts and Sciences Research Fund at the University of Kansas. 
TK is supported in part by the National Science Foundation (NSF) under Grant CNS2212565, Grant CNS2225577, and Grant CNS1955561, and the Office of Naval Research (ONR) under Grant N000142112472. MP was supported by the National Research Foundation of Korea (NRF) under grant NRF-2021R1A2C4002551 during the early stages of this work.
MP and KK would like to thank the Aspen Center for Physics and the organizers of Summer 2024 workshop, ``Fundamental Physics in the Era of Big Data and Machine Learning'' (supported by National Science Foundation grant PHY-2210452) for hospitality during the completion of this manuscript.
This research used resources of the National Energy Research
Scientific Computing Center, a DOE Office of Science User Facility supported by the Office of Science of the U.S. Department of Energy under Contract No. DE-AC02-05CH11231 using NERSC award HEP-ERCAP0037550.

\quad
\appendix

\section{Quantum circuits}
\label{app:circuits}

\begin{figure}[t]
    \centering
    \includegraphics[width=0.49\textwidth]{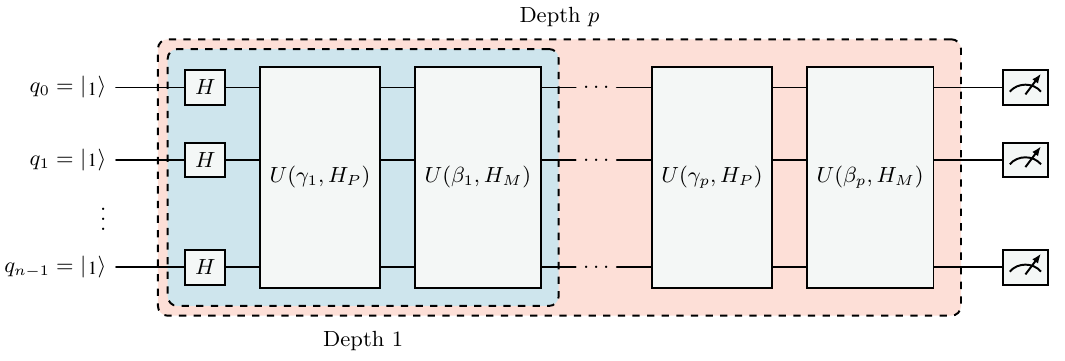}   
    \caption{QAOA of depth $p$ for $n$ qubits.}
    \label{fig:qaoa}
\end{figure}

\begin{figure}[t]
    \centering
    \includegraphics[width=0.3\textwidth]{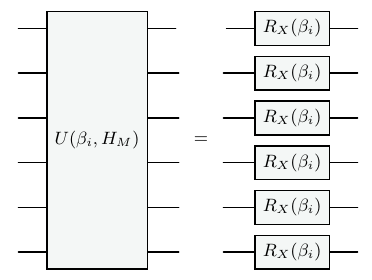}
    \caption{The $i$th mixer layer of QAOA for $n=6$ qubits.}
    \label{fig:mixer_layer_qaoa}
\end{figure}

\begin{figure}[t]
    \centering
    \includegraphics[width=0.49\textwidth]{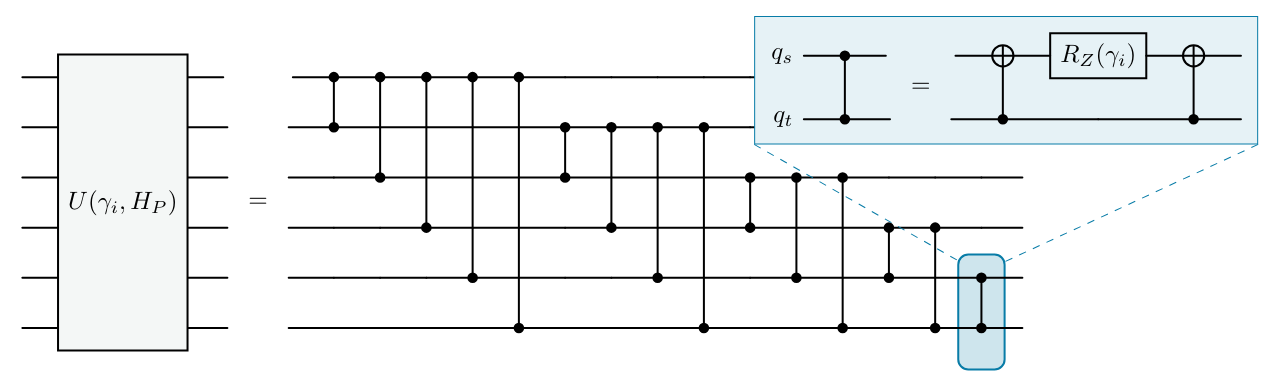}    
    \caption{The $i$th cost or problem layer of QAOA.}
    \label{fig:cost_layer}
\end{figure}

\begin{figure}[t]
    \centering
    \includegraphics[width=0.3\textwidth]{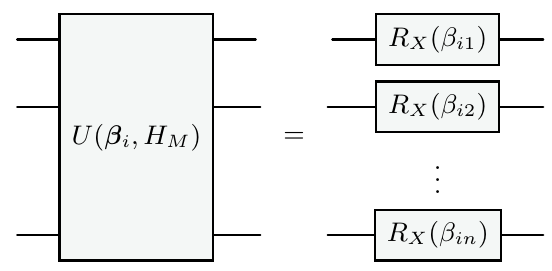}
    \caption{The $i$th mixer layer of ma-QAOA for $n$ qubits.}
    \label{fig:mixer_layer-maqaoa}
\end{figure}

\noindent\textbf{QAOA:} The schematic QAOA diagram of depth $p$ for $n$ qubits is shown in Fig. \ref{fig:qaoa}, where $U(\beta, H_M)$ and $U(\gamma_i, H_P)$ are shown in Fig. \ref{fig:mixer_layer_qaoa} and Fig. \ref{fig:cost_layer}, respectively. $H$ is the Hadamard gate, $R_X$ and $R_Z$ are Pauli rotations of $X$ and $Z$.
In the cost layer as shown in Fig. \ref{fig:cost_layer}, each vertical bar represents the $R_{ZZ}(\gamma_i)$ gate between the two qubits. As shown decomposed into common basis gates in the inlay. The gate applied to qubits $q_s$ and $q_t$ is defined by $R_{ZZ}(\gamma_i)\equiv\exp[-Z_s Z_t\gamma_i]$ where $Z_s$ ($Z_t$) is the Pauli-Z gate acting on qubit $q_s$ ($q_t$), which in practice are decomposed into CNOT gates and single-qubit Pauli-Z rotation.
Fig. \ref{fig:mixer_layer-maqaoa} shows the mixer layer in the $i$th layer of ma-QAOA for $n$ qubits, which is similar to the mixer layer of QAOA in Fig. \ref{fig:mixer_layer_qaoa}. However the Mixer layer of ma-QAOA has a vector $\bm{\beta}_i$ of parameters such that each qubit receives its own rotation that the optimizer can freely vary. \\

\begin{figure}[t]
    \centering
    \includegraphics[width=0.49\textwidth]{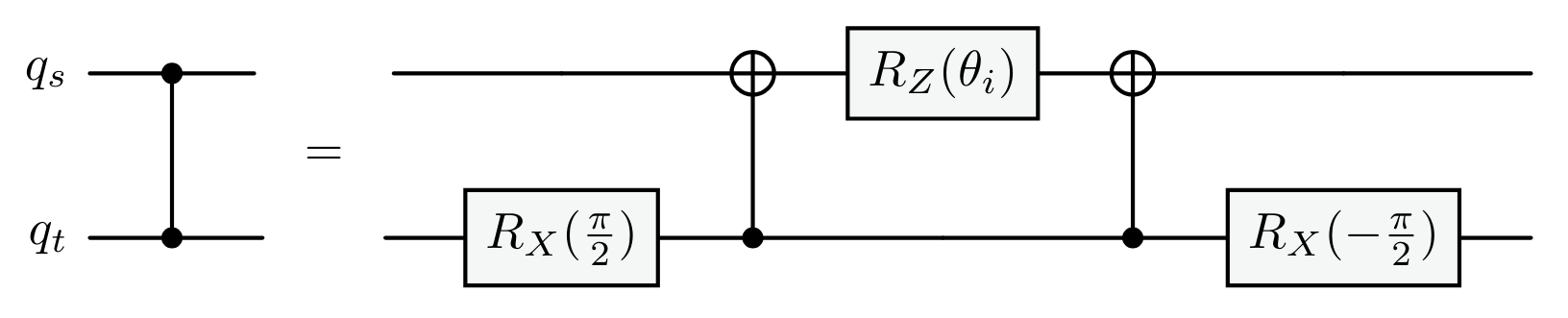}
    \caption{The decomposition of an $R_{ZY}$ gate into common basis gates. It is a $\pi/2$ rotation from an $R_{ZZ}$ gate.}
    \label{fig:rzy-gate}
\end{figure}

\begin{figure*}[ht!]
    \centering
    \includegraphics[width=0.95\textwidth]{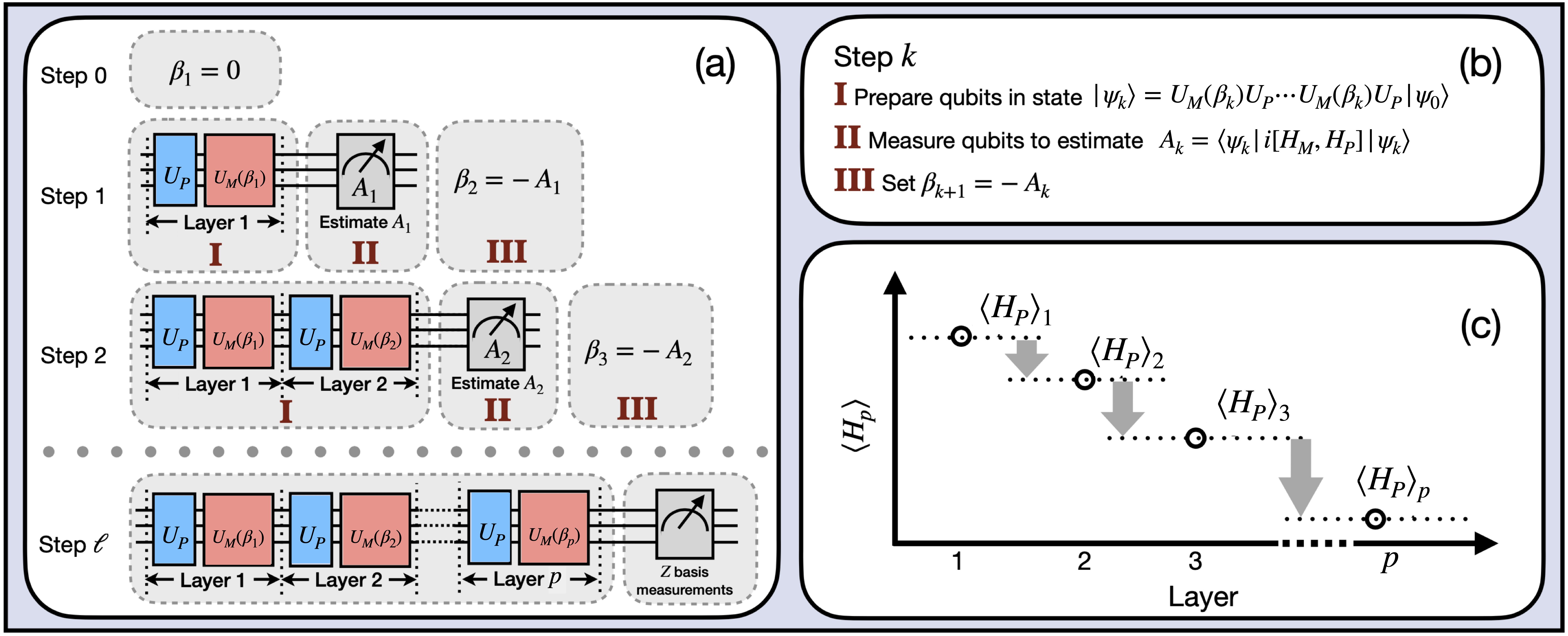}
    \caption{Illustration of the FALQON algorithm. Taken from Ref. \cite{Magann_2022}.}
\label{fig:falqon}
\end{figure*}

\noindent\textbf{FALQON:} The basic idea of FALQON is summarized in Fig. \ref{fig:falqon} (taken from Ref. \cite{Magann_2022}). The procedure for implementing FALQON begins by setting the initial condition $\beta_1 = 0$, as shown in Fig. \ref{fig:falqon}(a). The qubits are first prepared in the state $|\psi_0\rangle$, followed by applying a single FALQON layer to generate $|\psi_1\rangle = U_{{M}}(\beta_1)U_{{P}}|\psi_0\rangle$. After this, the qubits are measured to estimate $A_1$, which is then used to set $\beta_2 = -A_1$. In subsequent steps, for $k = 2, \dots, p$, the process is repeated as described in Fig. \ref{fig:falqon}(b): the qubits are initialized in $|\psi_0\rangle$, and $k$ layers are applied to produce $|\psi_k\rangle = U_{{M}}(\beta_{k}) U_{{P}}\cdots U_{M}(\beta_1) U_{{P}} |\psi_0\rangle$. 
The qubits are then measured to estimate $A_k$, and the result is used to update $\beta_{k+1}$. This iterative process ensures that $\langle H_{\textrm{p}}\rangle$ decreases with each layer, as shown in Fig. \ref{fig:falqon}(c): $\langle\psi_1| H_{{P}} |\psi_1\rangle \geq \langle\psi_2| H_{{P}} |\psi_2\rangle \geq \cdots \geq \langle\psi_p| H_{{P}} |\psi_p\rangle$. This layer-by-layer reduction improves the quality of the solution to the combinatorial optimization problem as the circuit depth increases. The procedure can be halted when $\langle H_{{P}}\rangle$ converges or when a predefined number of layers $p$ is reached.\\

\noindent\textbf{VarQITE:} We adopt the ansatz in Ref. \cite{Morris2024-gj} wherein it is explicitly fleshed-out in the supplemental materials. The circuit consists of $R_{ZY}$ gates, as shown in Fig. \ref{fig:rzy-gate}, between every pair of qubits. For better convergence, the gates are ordered by the parameter $$\rho_j=\sum_{k}\abs{\omega_{jk}}$$ for qubit $j$ where $\omega_{jk}$ is the coefficient for the $Z_jZ_k$ term of the Hamiltonian, {\it i.e.,} for $H_0$, $\omega_{ij}=J_{ij}$ and for $H_2$, $\omega_{ij}=J_{ij}+\lambda P_{ij}/2$. The gates are then ordered in descending order such that our final state is $$\ket{\btheta}=R_{ZY}^{n-1,n}(\theta_{n_{ZY}}) \cdots R_{ZY}^{0,2}(\theta_2)R_{ZY}^{0,1}(\theta_1)\ket{+}^{\otimes n}$$ where $n_{ZY}=\binom{n}{2}$ is the number of parameters and $R_{ZY}^{i,j}$ is the $R_{ZY}$ gate between qubits in the $i$-th and $j$-th place as ordered by $\rho_j$. For example, the qubits of the first rotation have the largest and second-largest value of $\rho_j$, respectively.

%

\section{Effects of quantum noise}
\label{sec:noise}

Quantum noise can be modeled as a set of operators $\{E_k\}$ that obey the completeness relation,
\begin{align}
    \sum_kE_k^\dagger E_k=I,
\end{align}
and act on the density matrix of the quantum system as $\mathcal{E}(\rho)=\sum_kE_k\rho E_k^\dagger$. In practice, for a single qubit, these operators can often be written as a linear combination of Pauli matrices,
\begin{align}
    E_k=\alpha_kI+\sum_{i=1}^3\beta_{ki}\sigma_i.
\end{align}
Using this form, we can define a few commonly used noise channels. The {\it bit flip} channel is given by
\begin{align}
    E_0=\sqrt{1-p}I,\hspace{3mm}E_1=\sqrt{p}X,
\end{align}
wherein, with probability $p$, a bit flip occurs, \ie an X gate. Likewise, a {\it phase flip} channel is given by
\begin{align}
    E_0=\sqrt{1-p}I,\hspace{3mm}E_1=\sqrt{p}Z.
\end{align}
The {\it depolarizing channel} represents a uniform contraction of the Bloch sphere. Physically, there is a probability $p$ that the qubit is depolarized. It is given by
\begin{align}
    \mathcal{E}(\rho)=\frac{p}{2}I+(1-p)\rho
\end{align}
or, the operator representation,
\begin{align}
    E_0=\sqrt{1-p}I,\hspace{3mm}E_i=\sqrt{p/3}P_i
\end{align}
where $P_1=X$, $P_2=Y$, and $P_3=Z$. Lastly, {\it amplitude damping} encodes potential energy dissipation of the qubit, say the spontaneous emission of a photon. Its operators are
\begin{align}
    E_0=\begin{pmatrix}
        1 & 0 \\ 0 & \sqrt{1-\gamma}
    \end{pmatrix},\hspace{3mm}
    E_1=\begin{pmatrix}
        0 & \sqrt{\gamma} \\ 0 & 0
    \end{pmatrix}
\end{align}
where $\gamma\in[0,1]$ is the amplitude damping probability, e.g. a representation of the probability to emit a photon.

There are additional, commonly used, operator representations of noise, such as generalized amplitude damping and phase damping, which represent additional modes of quantum information loss, but they are not discussed here. More information can be found in \cite{Nielsen_Chuang_2010}.

\begin{figure}[t]
    \centering
    \includegraphics[width=0.49\textwidth]{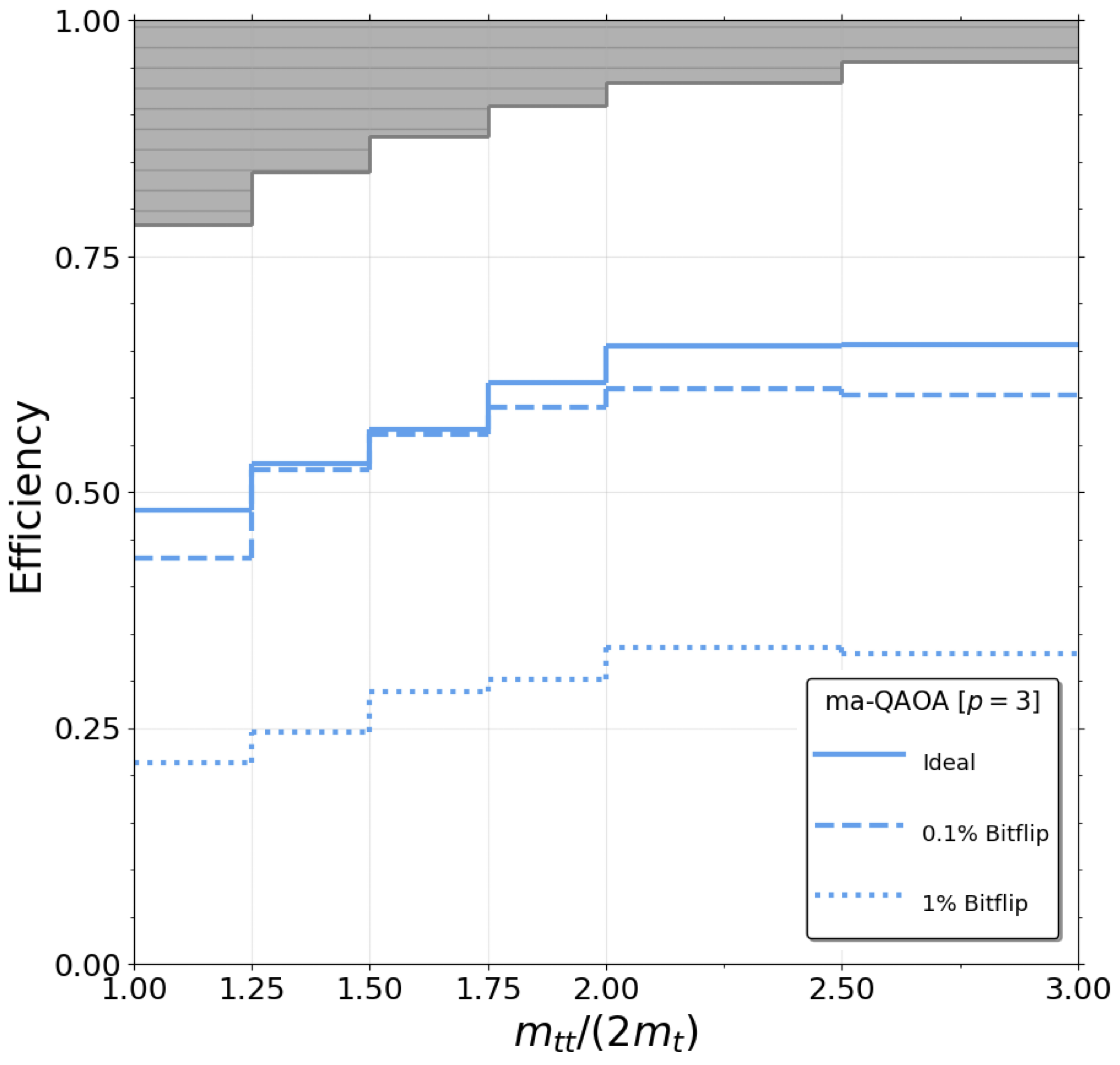}
    \caption{Efficiency plot for ma-QAOA with bitflip noise on every gate for the ideal case (solid), 0.1\% chance (dashed) and 1\% chance (dotted). Shown are for depths of $p=3$.}
    \label{fig:eff_bitflip}
\end{figure}

Error rates for hardware will different, depending on quantum computing architecture. 
Nevertheless, the two-qubit gate error rate is about 1\% for most current quantum computers and some of them are already close to $\sim$0.1\% \cite{Ezratty:2023fxe}, which we take to be our benchmark points.
Fig. \ref{fig:eff_bitflip} shows the efficiency of ma-QAOA with bitflip noise on every gate for 0.1\% (dashed) and 1\% error rate (dotted). The solid curve represent the ideal case (no noise). Results are obtained for depths of $p=3$. We anticipate that future improvements will significantly reduce the error rate, potentially minimizing the impact of quantum noise. However, exploring error correction in this combinatorial optimization problem remains valuable.

\section{Effects of Finite Shots}
\label{sec:finiteshots}

\begin{figure}
    \centering
    \includegraphics[width=0.99\linewidth]{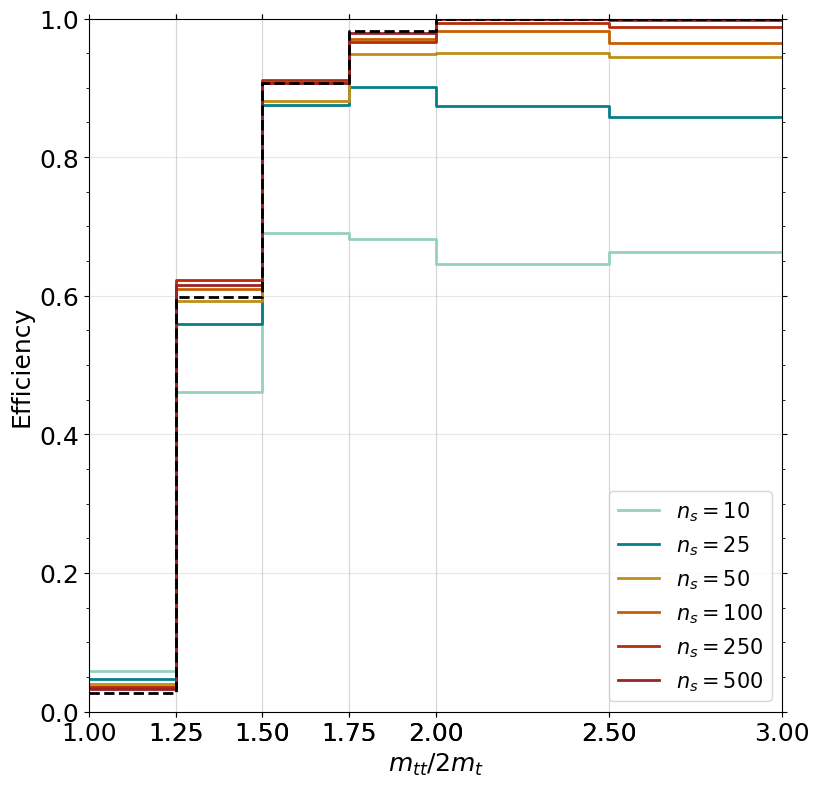}
    \caption{Efficiency plot for ma-QAOA at $p=5$ for a varying number of shots. The dashed black line is the exact, or infinite shot, result as would be used in the main body of this paper. By $n_s$ shots, we are referring to the number of shots per step of the optimization process. That is, the classical optimization is based on an approximation of the energy expectation value.}
    \label{fig:eff_shots}
\end{figure}

On real hardware, quantum states and expectation values are approximated based on a finite number of circuit executions or ``shots'', below labeled $n_s$. In the above analysis, exact analytic state vectors are used in the simulations, \ie $n_s=\infty$, and thus it is of interest to understand the effects of a finite number of shots. In Fig. \ref{fig:eff_shots}, ma-QAOA is run at a depth of $p=5$ for different values of $n_s$. The black dashed line represents the ideal case, $n_s=\infty$. With $n_s\sim\mathcal{O}(10^3)$, we see no noticeable difference from the ideal case. Even with only $n_s\sim\mathcal{O}(10^2)$, there is only a difference on the order of a percent.

\section{Detector effects}
\label{sec:smearing}

\begin{figure*}
    \centering
    \includegraphics[width=\textwidth]{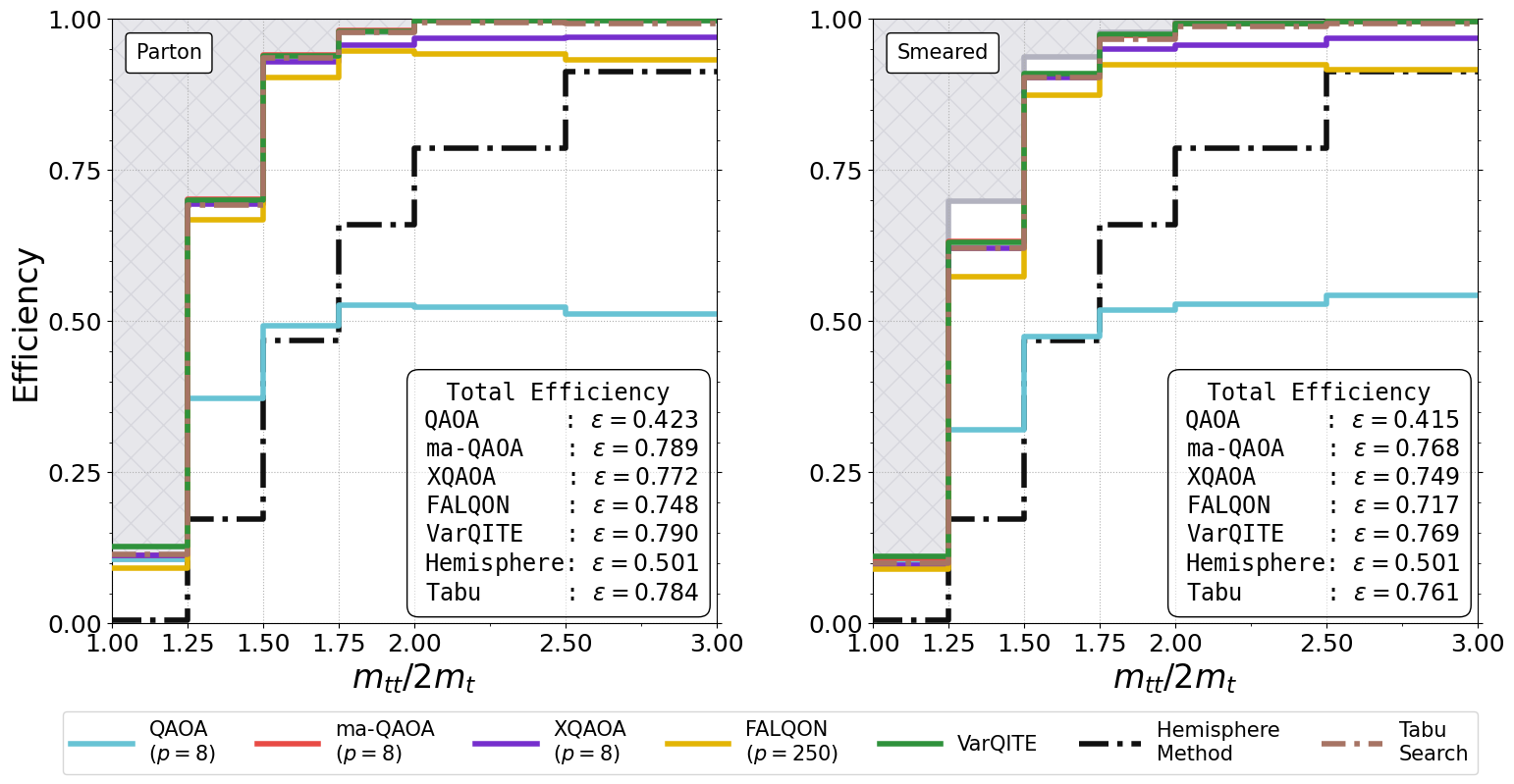}
    \caption{Akin to Fig. \ref{fig:parton_efficiency_h0_h0h1}, shown is the efficiency for parton-level events (left) and smeared events (right) using the Hamiltonian $H_2$ for QAOA (light blue), XQAOA (purple), ma-QAOA (red), FALQON (yellow), VarQITE (green), and hemisphere (black, dot-dashed). 
    }
    \label{fig:efficiency_h0_h1}
\end{figure*}

To simulate detector effects, we follow the  parametrization used in ATLAS detector performances report for the HL-LHC \cite{TheATLAScollaboration:2013sgb}\footnote{We study the effects of detector resolution as described in this appendix. We chose not to use existing detector simulation packages such as Delphes \cite{deFavereau:2013fsa}, since they do not retain the correspondence between parton-level truth information and reconstructed objects. In practice, one typically identifies the parton that best matches a reconstructed jet using the angular distance $\Delta R$ \cite{Fenton:2020woz,Shmakov:2021qdz}. However, this procedure provides only an approximate association and does not preserve the true MC-level correspondence.}. 
The energy resolution is parameterized by three terms; noise ($N$), stochastic ($S$), and constant ($C$) terms
\begin{align}
\frac{\sigma}{E} =  \sqrt{
\bigg( \frac{N}{E} \bigg)^2 +\bigg( \frac{S}{\sqrt{E}}\bigg)^2  +C^2~,
}
\end{align}
where in our analysis we use $N=5.3$, $S=0.74$ and $C=0.05$ for jets, and the energy $E$ is in GeV. We smear particle momenta of parton-level events following the above parametrization and refer to this dataset as ``Smeared events''. Basic cuts of $p_T > 25$ GeV, rapidity $|\eta |< 2.5$ and $\Delta R(j,j) > 0.4 $ are applied to the smeared particles. Fig. \ref{fig:efficiency_h0_h1} shows the matching accuracy of QAOA (light blue), XQAOA (purple), ma-QAOA (red), FALQON (yellow), VqrQITE (green) and hemisphere (dot-dashed)for parton-level events (left) and smeared events (right) using both Eq. (\ref{eq:cost}) and Eq. (\ref{eq:h1}) for the full Hamiltonian, $H_2$. The performance degrades slightly but remains very similar, showing the robustness of the quantum algorithms. Table \ref{tab:summary_table} presents the matching accuracy for various methods across both parton-level and smeared events, showing that quantum algorithms exhibit only slight changes, while ML methods undergo a noticeable degradation. While these results are quite promising, a more thorough investigation is needed including ISR and parton shower / hadronization and more realistic detector simulation.

\section{Different processes with the same final state}
\label{sec:diff_processes}

\begin{figure*}[t!]
    \centering
    \includegraphics[height=24em]{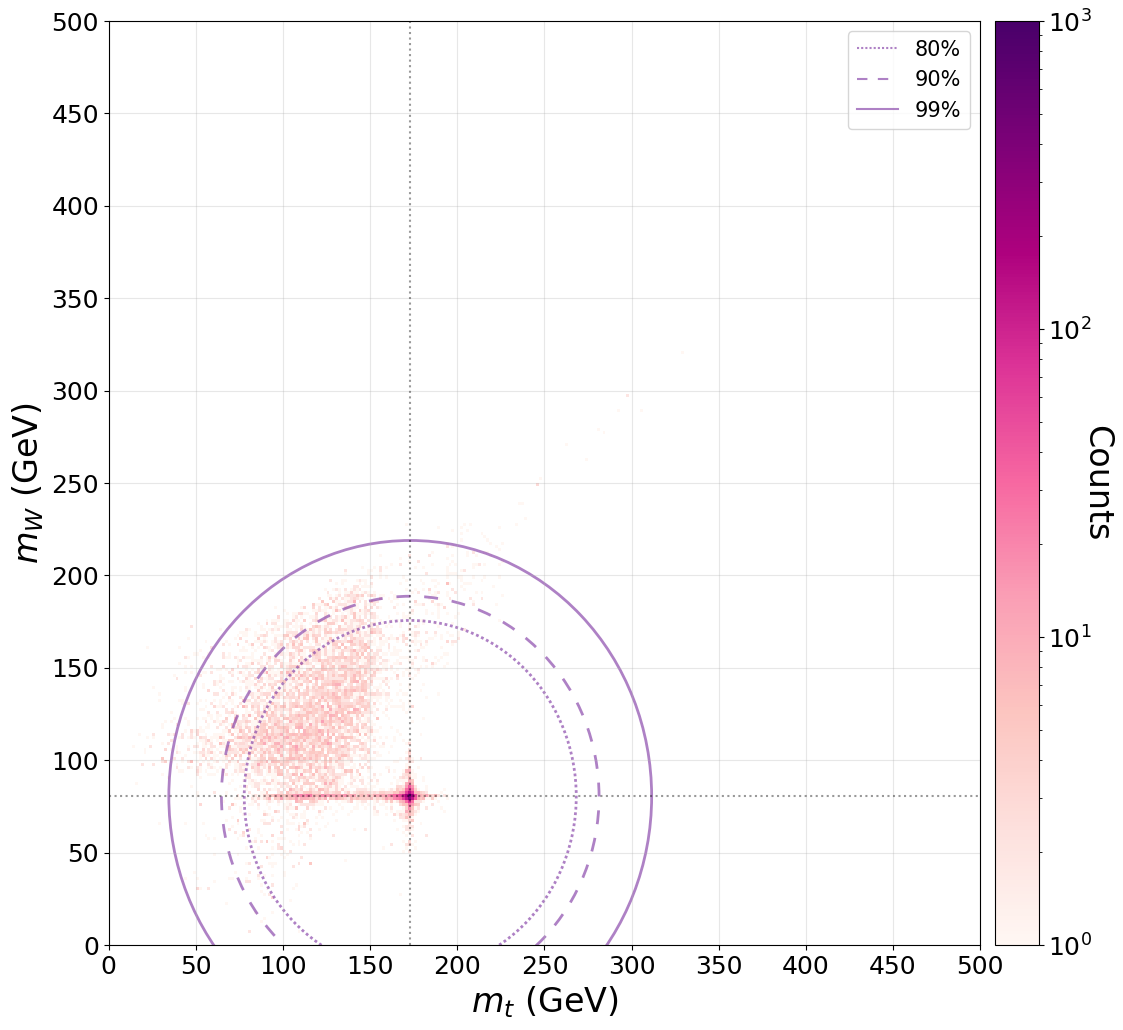} 
    \hspace*{0.03\textwidth}
    \includegraphics[height=24em]{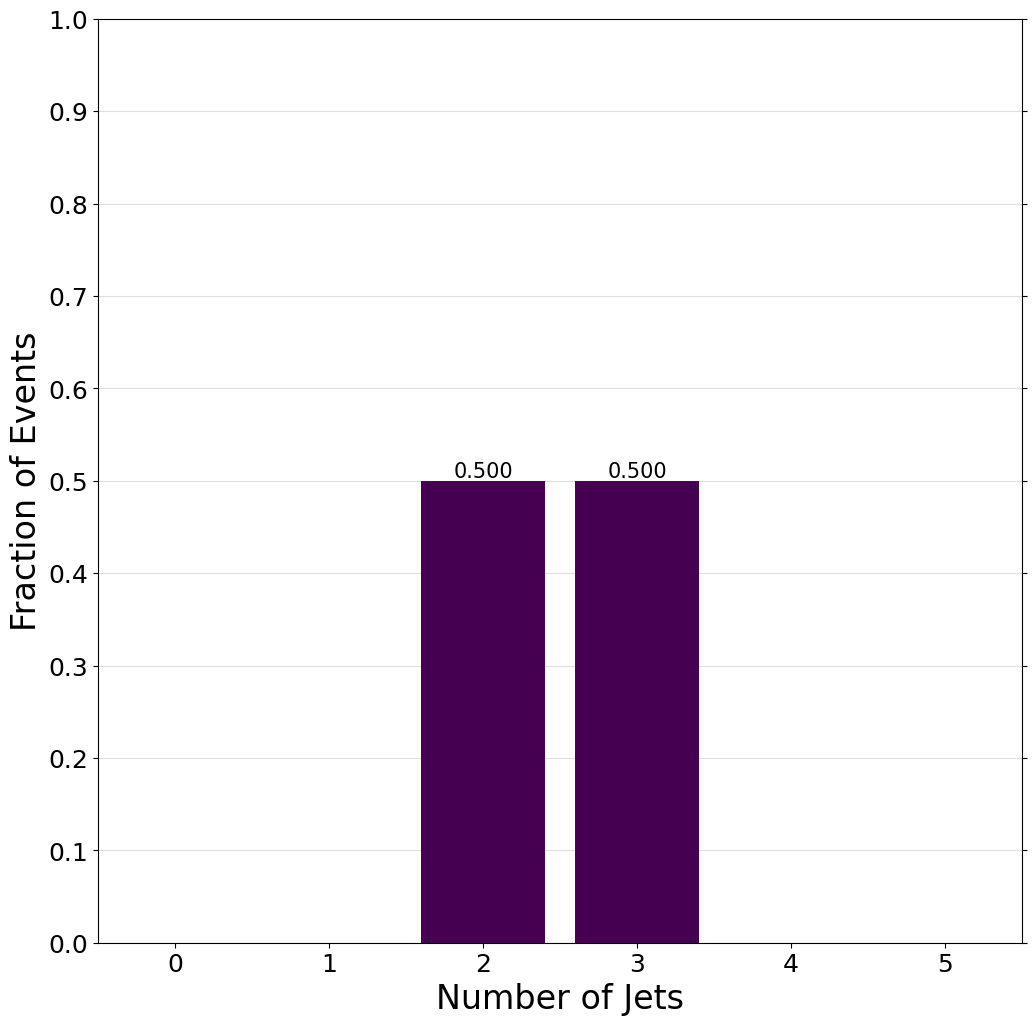} 
    \\
    \includegraphics[height=24em]{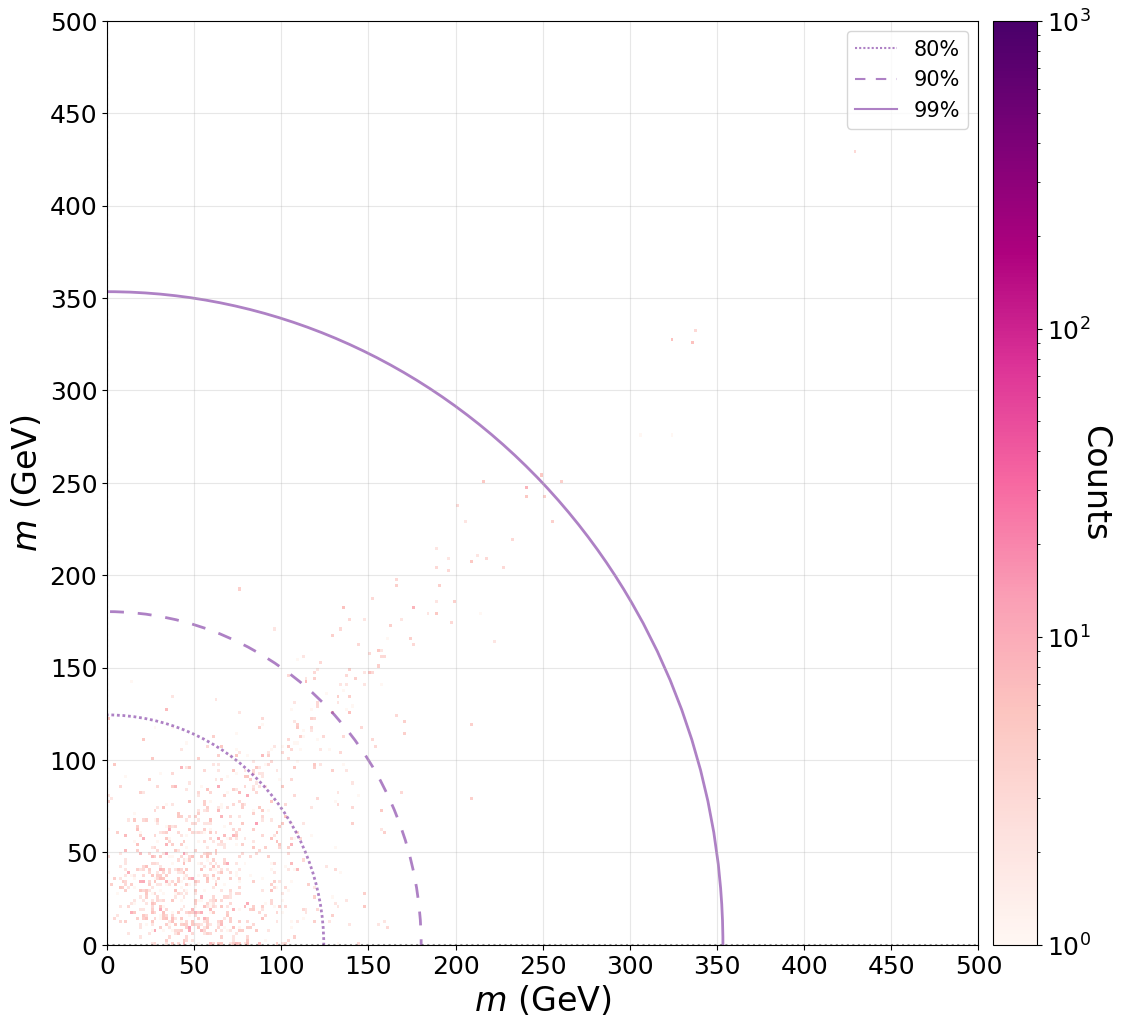} 
    \hspace*{0.03\textwidth}
    \includegraphics[height=24em]{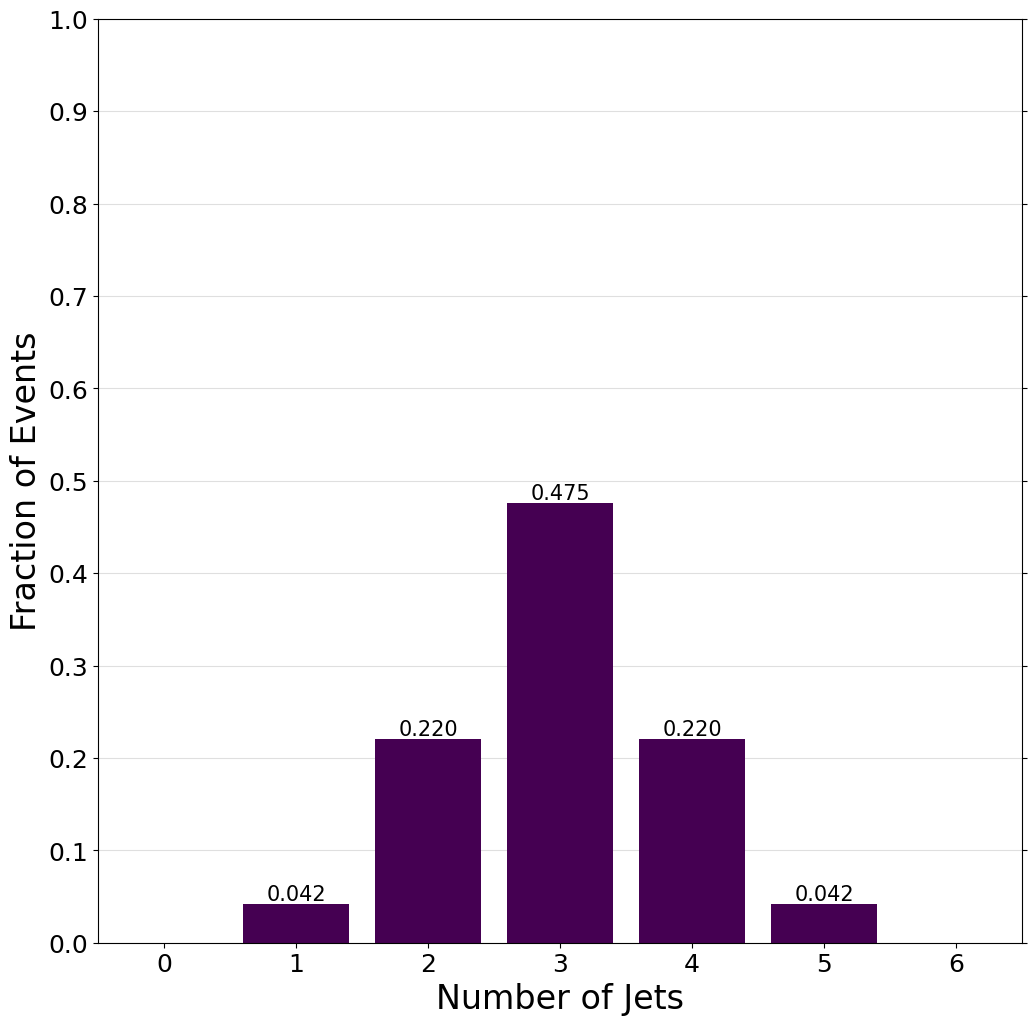} 
    \caption{The same as Fig. \ref{fig:parton_qaoa8} but for $tW$ production (top) and multi-jet production (bottom). The reconstructed masses (left) and jet-assignment (right) for parton-level events using ma-QAOA with depth $p=8$.  The ground state is found for the full Hamiltonian.  The overall efficiency for $tW$ is 22\%. Results using FALQON (not shown) are similar to those using ma-QAOA. \label{fig:parton_tW_6jets_qaoa8}}
\end{figure*}

In this section, we apply the QAOA algorithm to $tW$ production with 5 jets and QCD production with 6 jets. Fig. \ref{fig:parton_tW_6jets_qaoa8} shows the reconstructed masses (left) and jet-assignment (right) with parton-level events using QAOA with depth $p=8$ for $tW$ production (top) and multi-jet production (bottom). The ground state is found for the full Hamiltonian.  As expected, the top panel shows a large concentration near the correct masses ($m_t$, $m_W$) = (173, 80) GeV with the correct jet assignment for top and $W$ boson, even if the mass information is not used in QAOA (quantum algorithms in general). The overall efficiency for $tW$ is 22\%.
Similarly, since the multi-jet production does not have a specific mass scale, the bottom panel shows that most events are populated near zero mass with random jet assignment. It is important to note that the same Hamiltonians in Eqs. (\ref{eq:H0}-\ref{eq:fullH}) lead to these results. We used 12k $tW$ and 3k 6 jet events. 

\section{ADAPT-QAOA}
\label{sec:adapt}

As an example of quantum algorithms based on a non-adiabatic path, we consider the Adaptive Derivative Assembled Problem Tailored QAOA \cite{zhu2022adaptive} (ADAPT-QAOA). In this algorithm, the choice of the mixer Hamiltonian is not fixed but instead is drawn from a pool of potential Hamiltonians layer by layer. The $k$th layer will have $H_M=A_k$ where $A_k$ is one of the single-qubit mixers (\ref{eq:adapt-single}) or multi-qubit mixers (\ref{eq:adapt-multi}): 
\begin{widetext}
\begin{align}
    \text{Single-qubit mixers}
    &\begin{cases}
        X_i&\quad\text{--}\quad\text{$X$ gate on the $i$th qubit}\\
        Y_i&\quad\text{--}\quad\text{$Y$ gate on the $i$th qubit}\\
        \sum_{i=1}^nX_i&\quad\text{--}\quad\text{$X$ gate on every qubit, \ie QAOA}\\
        \sum_{i=1}^nY_i&\quad\text{--}\quad\text{$Y$ gate on every qubit}        
    \end{cases}\label{eq:adapt-single} \\
    \text{Multi-qubit mixers}
    &\begin{cases}
        X_iX_j&\quad\text{--}\quad\text{$X$ gates on the $i$th and $j$th qubits}\\
        X_iY_j&\quad\text{--}\quad\text{$X$ and $Y$ gates on the $i$th and $j$th qubits}\\
        X_iZ_j&\quad\text{--}\quad\text{$X$ and $Z$ gates on the $i$th and $j$th qubits}\\
        Y_iZ_j&\quad\text{--}\quad\text{$Y$ and $Z$ gates on the $i$th and $j$th qubits}\\
        Y_iY_j&\quad\text{--}\quad\text{$Y$ gates on the $i$th and $j$th qubits}\\
        Z_iZ_j&\quad\text{--}\quad\text{$Z$ gates on the $i$th and $j$th qubits}
        \label{eq:adapt-multi}
    \end{cases}
\end{align}
\end{widetext}

Namely, the two QAOA-like mixers (the third and fourth above) and all length one and two Pauli strings.
As such, ADAPT-QAOA shares similarities with FALQON as the circuit is iteratively built. For layer $k$, we aim to find the mixer that maximizes the energy gradient. That is, if $\mathcal{A}$ is the mixer pool, then $\forall A_j\in\mathcal{A}$, we calculate
\begin{widetext}
\begin{align}   
    \Delta E_k^j\equiv\pdv{\beta_k}\expval{H_P}{\psi_k}\bigg|_{\beta_k=0}&=\pdv{\beta_k}\expval{e^{i\gamma_kH_P}e^{i\beta_kA_j}H_Pe^{-i\beta_kA_j}e^{-i\gamma_kH_P}}{\psi_{k-1}}\bigg|_{\beta_k=0}\\
    &=\expval{e^{i\gamma_kH_P}e^{i\beta_kA_j}\left(iA_jH_P-iH_PA_j\right)e^{-i\beta_kA_j}e^{-i\gamma_kH_P}}{\psi_{k-1}}\bigg|_{\beta_k=0}\\
    &=-i\expval{e^{i\gamma_kH_P}[H_P,A_j]e^{-i\gamma_kH_P}}{\psi_{k-1}}
\end{align}
\end{widetext}
and choose $\displaystyle A_k=\argmax_{A_j\in\mathcal{A}}\Delta E_k^j$. Since we don't know $\gamma_k$ a priori, we set it to some predefined value $\gamma_0$. We set both initial parameters and $\gamma_0$ to 0.01.  Once the mixer for the $k$th layer, $A_k$, has been chosen, the parameter of the $k$th layer is optimized as with normal QAOA. This optimization procedure is then repeated layer by layer while keeping all previous layers fixed. The algorithm can be stopped when a specific circuit depth has been reached or $\Delta E_k^j$ is below some threshold. The improvement provided by ADAPT-QAOA comes taking a shortcut non-adiabatic path\cite{Chai_2022}. One downside to this method is the large mixer pool could potentially lead to many measurements. We attempted to implement this algorithm but ADAPT-QAOA did not improve the matching efficiency. 
While a comprehensive investigation is needed, we observed that ADAPT-QAOA consistently converges to an equal superposition of two or three states. Although the global minimum is typically present within this superposition, the algorithm fails to isolate it uniquely. This behavior remained unchanged despite testing various initial parameter configurations. We attribute this phenomenon to the challenges identified in Ref. \cite{Sridhar:2023our}, where ADAPT-QAOA is shown to frequently converge to excited states, representing local minima, rather than the true ground state. Our findings, specifically the frequent 50/50 split between the ground state and a nearby excited state, align closely with the issues that motivated the warm-start methodology proposed in that study.

There are a few other studies on manipulating the adiabatic path of QAOA-type algorithms \cite{Chandarana_2022, Wurtz_2022}. Other physics-inspired QAOA variants include Mean-Field Approximate Optimization Algorithm \cite{Misra_Spieldenner_2023}, Improving Quantum Approximate Optimization by Noise-Directed Adaptive Remapping
\cite{Maciejewski:2024aaf}, ADAPT-QAOA with a classically inspired initial state\cite{Sridhar:2023our} and Dynamic-ADAPT-QAOA with shallow and noise-resilient circuits \cite{Yanakiev:2023ibw}.

\section{Quantum hardware results}
\label{sec:quantum_hardware}

\begin{figure}[t!]
    \centering
    \includegraphics[width=0.99\linewidth]{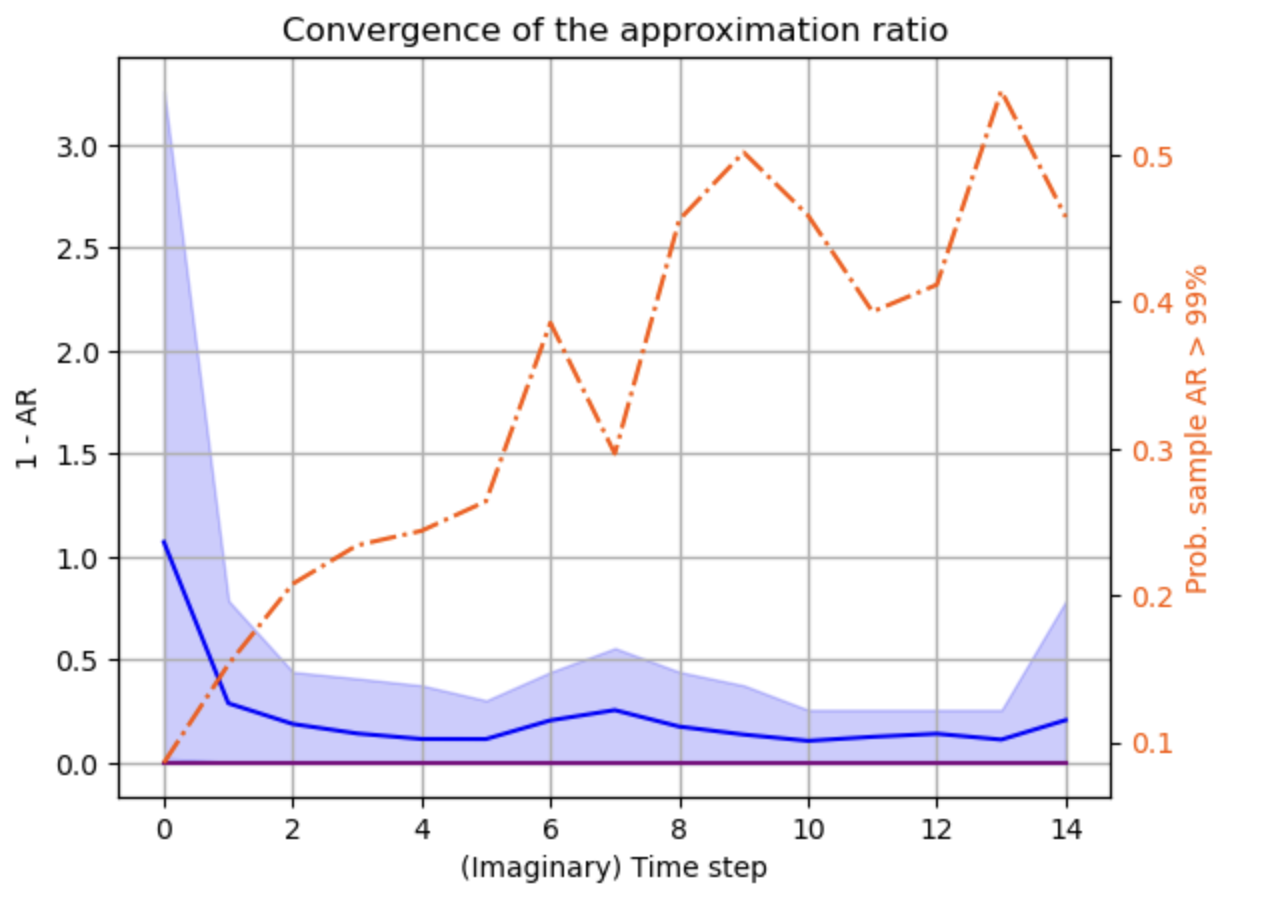}
    \includegraphics[width=0.99\linewidth]{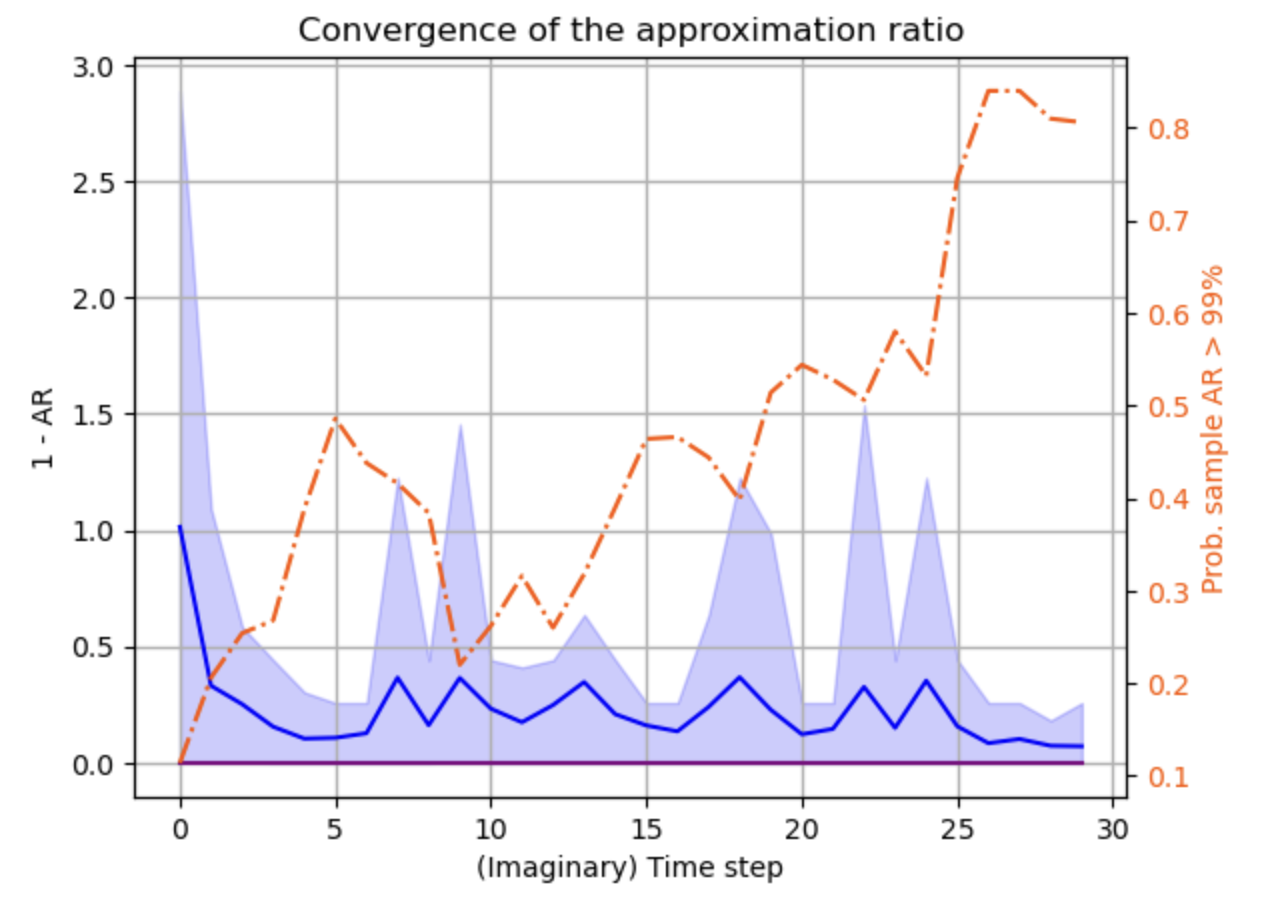}
    \caption{(top) Convergence of the average sampled energy over the course of a full VarQITE optimization on IonQ Forte. The VarQITE procedure is used to find the ground state of an $H_0$-type Hamiltonian for a particular event. We used $128$ shots to evaluate each of the $2(15) + 1 = 31$ circuits per VarQITE iteration. (bottom) Convergence of the average sampled energy over the course of a full VarQITE optimization, obtained by executing six gradient circuits simultaneously on the $36$-qubit IonQ Forte system. We again use $128$ shots per iteration. In this case, we need only run five $36$-qubit circuits and an additional $6$-qubit on IonQ Forte per iteration, instead of running $31$ circuits sequentially.}
    \label{fig:forte-varqite}    
\end{figure}
We executed a select number of hardware experiments to demonstrate the feasibility of certain optimization algorithms on existing quantum processors. In particular, we used the IonQ Forte system to conduct two full optimization runs using the VarQITE algorithm explained in Section~\ref{sec:varqite}.

The top panel in Fig. \ref{fig:forte-varqite} illustrates the convergence of our hybrid quantum-classical VarQITE optimizer as a function of the iteration count. 
The Approximation Ratio (AR) is defined as the ratio of the observed objective function value to the optimal value. The blue shaded band illustrates the variance within the observed samples, spanning the 10th to the 90th percentile, while the red curve serves as a metric for the quality of the solutions generated by the device. 
The figure illustrates how the parametrized ansatz converges onto the ground state of the underlying problem Hamiltonian as we tweak the ansatz parameters upon each iteration. These experimental results demonstrate the viability of real systems for implementing optimization protocols useful in high energy particle physics.

In particular, the VarQITE procedure is used to find the ground state of an $H0$-type Hamiltonian for a randomly selected event (with event index $115$). We executed $2(15) + 1 = 31$ $6$-qubit circuits on IonQ Forte at each iteration, taking $128$ shots for each circuit. Figure \ref{fig:forte-varqite} illustrates the convergence of our procedure across $14$ imaginary time steps, making for a total of $434$ circuit executed on IonQ Forte. This VarQITE procedure has proven to be quite robust to shot noise in previous empirical studies like \cite{Morris2024-gj} and \cite{Aboumrad:2025accelerating}. This means we can keep the shot count relatively low at $128$ and still observe convergence to the optimal solution. 

To reduce the resource requirements of our algorithm and improve the time-to-solution of our experimental VarQITE runs, we introduced a quantum circuit stacking procedure that enables parallel execution of multiple narrow circuits on the $36$-qubit IonQ Forte system. Explicitly, we exploit all of IonQ Forte's $36$ all-to-all connected qubits by stacking six $6$-qubit circuits one above another on disjoint sets of qubits and executing them simultaneously as a single $36$-qubit circuit. Each circuit occupies only a subset of the hardware wires, while the remaining wires belong to other stacked circuits. 

To recover the statistics of any one circuit upon measurement, we simply trace out the qubits belonging to other subcircuits. Operationally, this is equivalent to having measured only those wires in the first place: by the law of total probability, marginalizing the full-system samples yields exactly the same subsystem probability distribution as directly sampling the subsystem alone.

This procedure resulted in a $5$-$6$x speed up in the total time-to-solution of our VarQITE runs, since it allowed us to run six gradient circuits at a time in each VarQITE iteration. What's more, the bottom panel in Fig. \ref{fig:forte-varqite} shows that performance degradation due to cross-talk between wires in different sub-circuits was negligible on IonQ Forte.

\bibliography{bibliography.bib}

@misc{Aboumrad:2025accelerating,
      title={Accelerating large-scale linear algebra using variational quantum imaginary time evolution}, 
      author={Willie Aboumrad and Daiwei Zhu and Claudio Girotto and François-Henry Rouet and Jezer Jojo and Robert Lucas and Jay Pathak and Ananth Kaushik and Martin Roetteler},
      year={2025},
      eprint={2503.13128},
      archivePrefix={arXiv},
      primaryClass={quant-ph},
      url={https://arxiv.org/abs/2503.13128}, 
}

@article{Barr:2011xt,
    author = "Barr, A. J. and Khoo, T. J. and Konar, P. and Kong, K. and Lester, C. G. and Matchev, K. T. and Park, M.",
    title = "{Guide to transverse projections and mass-constraining variables}",
    eprint = "1105.2977",
    archivePrefix = "arXiv",
    primaryClass = "hep-ph",
    reportNumber = "CAV-HEP-11-08",
    doi = "10.1103/PhysRevD.84.095031",
    journal = "Phys. Rev. D",
    volume = "84",
    pages = "095031",
    year = "2011"
}

@article{Barr:2010zj,
    author = "Barr, Alan J. and Lester, Christopher G.",
    title = "{A Review of the Mass Measurement Techniques proposed for the Large Hadron Collider}",
    eprint = "1004.2732",
    archivePrefix = "arXiv",
    primaryClass = "hep-ph",
    doi = "10.1088/0954-3899/37/12/123001",
    journal = "J. Phys. G",
    volume = "37",
    pages = "123001",
    year = "2010"
}

@article{glover1986future,
  title={Future paths for integer programming and links to artificial intelligence},
  author={Glover, Fred},
  journal={Computers \& Operations Research},
  volume={13},
  number={5},
  pages={533--549},
  year={1986},
  publisher={Elsevier},
  url={https://doi.org}
}

@article{Barzen:2025nfo,
    author = "Barzen, Johanna and Leymann, Frank",
    title = "{On the Differential Topology of Expressivity of Parameterized Quantum Circuits}",
    eprint = "2507.16401",
    archivePrefix = "arXiv",
    primaryClass = "quant-ph",
    doi = "10.3390/appliedmath5030121",
    month = "7",
    year = "2025"
}

@article{Holmes:2021qjw,
    author = {Holmes, Zo{\"e} and Sharma, Kunal and Cerezo, M. and Coles, Patrick J.},
    title = "{Connecting Ansatz Expressibility to Gradient Magnitudes and Barren Plateaus}",
    eprint = "2101.02138",
    archivePrefix = "arXiv",
    primaryClass = "quant-ph",
    reportNumber = "LA-UR-21-20034",
    doi = "10.1103/PRXQuantum.3.010313",
    journal = "PRX Quantum",
    volume = "3",
    number = "1",
    pages = "010313",
    year = "2022"
}

@article{glover1989tabu,
  title={Tabu Search—Part I},
  author={Glover, Fred},
  journal={ORSA Journal on Computing},
  volume={1},
  number={3},
  pages={190--206},
  year={1989},
  publisher={INFORMS},
  doi={10.1287/ijoc.1.3.190},
  url={https://pubsonline.informs.org/doi/10.1287/ijoc.1.3.190}
}

@article{glover1990tabu2,
  title={Tabu Search—Part II},
  author={Glover, Fred},
  journal={ORSA Journal on Computing},
  volume={2},
  number={1},
  pages={4--32},
  year={1990},
  publisher={INFORMS},
  doi={10.1287/ijoc.2.1.4}
}

@article{Franceschini:2022vck,
    author = "Franceschini, Roberto and Kim, Doojin and Kong, Kyoungchul and Matchev, Konstantin T. and Park, Myeonghun and Shyamsundar, Prasanth",
    title = "{Kinematic variables and feature engineering for particle phenomenology}",
    eprint = "2206.13431",
    archivePrefix = "arXiv",
    primaryClass = "hep-ph",
    reportNumber = "MI-HET-780, FERMILAB-PUB-22-481-QIS",
    doi = "10.1103/RevModPhys.95.045004",
    journal = "Rev. Mod. Phys.",
    volume = "95",
    number = "4",
    pages = "045004",
    year = "2023"
}

@misc{rehfeldt2022fasterexactsolutionsparse,
      title={Faster exact solution of sparse MaxCut and QUBO problems}, 
      author={Daniel Rehfeldt and Thorsten Koch and Yuji Shinano},
      year={2022},
      eprint={2202.02305},
      archivePrefix={arXiv},
      primaryClass={math.OC},
      url={https://arxiv.org/abs/2202.02305}, 
}

@article{Malla:2024zbj,
    author = "Malla, Rajesh K. and Sukeno, Hiroki and Yu, Hongye and Wei, Tzu-Chieh and Weichselbaum, Andreas and Konik, Robert M.",
    title = "{Feedback-based quantum algorithm inspired by counterdiabatic driving}",
    eprint = "2401.15303",
    archivePrefix = "arXiv",
    primaryClass = "quant-ph",
    doi = "10.1103/PhysRevResearch.6.043068",
    journal = "Phys. Rev. Res.",
    volume = "6",
    number = "4",
    pages = "043068",
    year = "2024"
}

@article{Motta:2019yya,
    author = "Motta, Mario and Sun, Chong and Tan, Adrian Teck Keng and Rourke, Matthew J. O' and Ye, Erika and Minnich, Austin J. and Brand\~ao, Fernando G. S. L. and Chan, Garnet Kin-Lic",
    title = "{Determining eigenstates and thermal states on a quantum computer using quantum imaginary time evolution}",
    eprint = "1901.07653",
    archivePrefix = "arXiv",
    primaryClass = "quant-ph",
    doi = "10.1038/s41567-019-0704-4",
    journal = "Nature Phys.",
    volume = "16",
    pages = "205--210",
    year = "2019"
}

@article{Yuan2019-oj,
  title     = "Theory of variational quantum simulation",
  author    = "Yuan, Xiao and Endo, Suguru and Zhao, Qi and Li, Ying and
               Benjamin, Simon C",
  journal   = "Quantum",
  publisher = "Verein zur Forderung des Open Access Publizierens in den
               Quantenwissenschaften",
  volume    =  3,
  number    =  191,
  pages     =  191,
  month     =  oct,
  year      =  2019,
  url       = "https://quantum-journal.org/papers/q-2019-10-07-191/pdf/",
  eprint    = "1812.08767v4",
  doi       = "10.22331/q-2019-10-07-191",
  issn      = "2521-327X"
}

@article{Morris2024-gj,
  title         = "Performant near-term quantum combinatorial optimization",
  author        = "Morris, Titus D and Kaushik, Ananth and Roetteler, Martin and
                   Lotshaw, Phillip C",
  journal       = "arXiv [quant-ph]",
  month         =  "24~" # apr,
  year          =  2024,
  url           = "http://arxiv.org/abs/2404.16135",
  archivePrefix = "arXiv",
  primaryClass  = "quant-ph",
  eprint        = "2404.16135"
}

@article{McArdle2019-zp,
  title     = "Variational ansatz-based quantum simulation of imaginary time
               evolution",
  author    = "McArdle, Sam and Jones, Tyson and Endo, Suguru and Li, Ying and
               Benjamin, Simon C and Yuan, Xiao",
  journal   = "Npj Quantum Inf.",
  publisher = "Springer Science and Business Media LLC",
  volume    =  5,
  number    =  1,
  pages     = "1--6",
  month     =  sep,
  year      =  2019,
  url       = "https://www.nature.com/articles/s41534-019-0187-2",
  doi       = "10.1038/s41534-019-0187-2",
  issn      = "2056-6387,2056-6387"
}

@article{Schuld2019-cq,
  title     = "Evaluating analytic gradients on quantum hardware",
  author    = "Schuld, Maria and Bergholm, Ville and Gogolin, Christian and
               Izaac, Josh and Killoran, Nathan",
  journal   = "Phys. Rev. A",
  publisher = "American Physical Society",
  volume    =  99,
  number    =  3,
  pages     =  032331,
  month     =  mar,
  year      =  2019,
  url       = "http://link.aps.org/pdf/10.1103/PhysRevA.99.032331",
  doi       = "10.1103/PhysRevA.99.032331"
}

@article{Mitarai2018-mu,
  title     = "Quantum circuit learning",
  author    = "Mitarai, K and Negoro, M and Kitagawa, M and Fujii, K",
  journal   = "Phys. Rev. A",
  publisher = "American Physical Society (APS)",
  volume    =  98,
  number    =  3,
  pages     =  032309,
  month     =  sep,
  year      =  2018,
  url       = "http://link.aps.org/pdf/10.1103/PhysRevA.98.032309",
  doi       = "10.1103/physreva.98.032309",
  issn      = "2469-9926,2469-9934"
}

@article{Brady:2024qwv,
    author = "Brady, Lucas T. and Hadfield, Stuart",
    title = "{Feedback-based optimally controlled quantum states}",
    eprint = "2409.15426",
    archivePrefix = "arXiv",
    primaryClass = "quant-ph",
    doi = "10.1103/PhysRevA.111.062406",
    journal = "Phys. Rev. A",
    volume = "111",
    number = "6",
    pages = "062406",
    year = "2025"
}

@article{Rahman:2024rbn,
    author = {Rahman, Salahuddin Abdul and Karabacak, {\"O}zkan and Wisniewski, Rafal},
    title = "{Feedback-Based Quantum Algorithm for~Constrained Optimization Problems}",
    eprint = "2406.08169",
    archivePrefix = "arXiv",
    primaryClass = "quant-ph",
    doi = "10.1007/978-3-031-85700-3_20",
    journal = "Lect. Notes Comput. Sci.",
    volume = "15580",
    pages = "277--289",
    year = "2025"
}

@article{Magann:2021uer,
    author = "Magann, Alicia B. and Rudinger, Kenneth M. and Grace, Matthew D. and Sarovar, Mohan",
    title = "{Lyapunov-control-inspired strategies for quantum combinatorial optimization}",
    eprint = "2108.05945",
    archivePrefix = "arXiv",
    primaryClass = "quant-ph",
    doi = "10.1103/PhysRevA.106.062414",
    journal = "Phys. Rev. A",
    volume = "106",
    number = "6",
    pages = "062414",
    year = "2022"
}

@article{Okawa:2024goh,
    author = "Okawa, Hideki and Tao, Xian-Zhe and Zeng, Qing-Guo and Yung, Man-Hong",
    title = "{Quantum-annealing-inspired algorithms for multijet clustering}",
    eprint = "2410.14233",
    archivePrefix = "arXiv",
    primaryClass = "quant-ph",
    doi = "10.1016/j.physletb.2025.139393",
    journal = "Phys. Lett. B",
    volume = "864",
    pages = "139393",
    year = "2025"
}

@article{Shmakov:2021qdz,
    author = "Shmakov, Alexander and Fenton, Michael James and Ho, Ta-Wei and Hsu, Shih-Chieh and Whiteson, Daniel and Baldi, Pierre",
    title = "{SPANet: Generalized permutationless set assignment for particle physics using symmetry preserving attention}",
    eprint = "2106.03898",
    archivePrefix = "arXiv",
    primaryClass = "hep-ex",
    doi = "10.21468/SciPostPhys.12.5.178",
    journal = "SciPost Phys.",
    volume = "12",
    number = "5",
    pages = "178",
    year = "2022"
}

@article{Fenton:2020woz,
    author = "Fenton, Michael James and Shmakov, Alexander and Ho, Ta-Wei and Hsu, Shih-Chieh and Whiteson, Daniel and Baldi, Pierre",
    title = "{Permutationless many-jet event reconstruction with symmetry preserving attention networks}",
    eprint = "2010.09206",
    archivePrefix = "arXiv",
    primaryClass = "hep-ex",
    doi = "10.1103/PhysRevD.105.112008",
    journal = "Phys. Rev. D",
    volume = "105",
    number = "11",
    pages = "112008",
    year = "2022"
}

@article{Yanakiev:2023ibw,
    author = "Yanakiev, Nikola and Mertig, Normann and Long, Christopher K. and Arvidsson-Shukur, David R. M.",
    title = "{Dynamic adaptive quantum approximate optimization algorithm for shallow, noise-resilient circuits}",
    eprint = "2309.00047",
    archivePrefix = "arXiv",
    primaryClass = "quant-ph",
    doi = "10.1103/PhysRevA.109.032420",
    journal = "Phys. Rev. A",
    volume = "109",
    number = "3",
    pages = "032420",
    year = "2024"
}

@article{Misra_Spieldenner_2023,
   title={Mean-Field Approximate Optimization Algorithm},
   volume={4},
   ISSN={2691-3399},
   url={http://dx.doi.org/10.1103/PRXQuantum.4.030335},
   DOI={10.1103/prxquantum.4.030335},
   number={3},
   journal={PRX Quantum},
   publisher={American Physical Society (APS)},
   author={Misra-Spieldenner, Aditi and Bode, Tim and Schuhmacher, Peter K. and Stollenwerk, Tobias and Bagrets, Dmitry and Wilhelm, Frank K.},
   year={2023},
   month=sep }

@article{Meyer:2022fjx,
    author = "Meyer, Johannes Jakob and Mularski, Marian and Gil-Fuster, Elies and Mele, Antonio Anna and Arzani, Francesco and Wilms, Alissa and Eisert, Jens",
    title = "{Exploiting Symmetry in Variational Quantum Machine Learning}",
    eprint = "2205.06217",
    archivePrefix = "arXiv",
    primaryClass = "quant-ph",
    doi = "10.1103/PRXQuantum.4.010328",
    journal = "PRX Quantum",
    volume = "4",
    number = "1",
    pages = "010328",
    year = "2023"
}

@article{Wei:2019rqy,
    author = "Wei, Annie Y. and Naik, Preksha and Harrow, Aram W. and Thaler, Jesse",
    title = "{Quantum Algorithms for Jet Clustering}",
    eprint = "1908.08949",
    archivePrefix = "arXiv",
    primaryClass = "hep-ph",
    reportNumber = "MIT-CTP 5137",
    doi = "10.1103/PhysRevD.101.094015",
    journal = "Phys. Rev. D",
    volume = "101",
    number = "9",
    pages = "094015",
    year = "2020"
}

@article{Skolik:2022qwn,
    author = {Skolik, Andrea and Cattelan, Michele and Yarkoni, Sheir and B\"ack, Thomas and Dunjko, Vedran},
    title = "{Equivariant quantum circuits for learning on weighted graphs}",
    eprint = "2205.06109",
    archivePrefix = "arXiv",
    primaryClass = "quant-ph",
    doi = "10.1038/s41534-023-00710-y",
    journal = "npj Quantum Inf.",
    volume = "9",
    number = "1",
    pages = "47",
    year = "2023"
}

@article{CMS:2007sch,
    author = "Bayatian, G. L. and others",
    collaboration = "CMS",
    title = "{CMS technical design report, volume II: Physics performance}",
    reportNumber = "CERN-LHCC-2006-021, CMS-TDR-008-2",
    doi = "10.1088/0954-3899/34/6/S01",
    journal = "J. Phys. G",
    volume = "34",
    number = "6",
    pages = "995--1579",
    year = "2007"
}

@article{Matsumoto:2006ws,
    author = "Matsumoto, Shigeki and Nojiri, Mihoko M. and Nomura, Daisuke",
    title = "{Hunting for the Top Partner in the Littlest Higgs Model with T-parity at the CERN LHC}",
    eprint = "hep-ph/0612249",
    archivePrefix = "arXiv",
    reportNumber = "KEK-TH-1120",
    doi = "10.1103/PhysRevD.75.055006",
    journal = "Phys. Rev. D",
    volume = "75",
    pages = "055006",
    year = "2007"
}

@article{Delgado:2022snu,
    author = "Delgado, Andrea and Thaler, Jesse",
    title = "{Quantum annealing for jet clustering with thrust}",
    eprint = "2205.02814",
    archivePrefix = "arXiv",
    primaryClass = "quant-ph",
    reportNumber = "MIT-CTP 5428",
    doi = "10.1103/PhysRevD.106.094016",
    journal = "Phys. Rev. D",
    volume = "106",
    number = "9",
    pages = "094016",
    year = "2022"
}

@article{deLejarza:2022bwc,
    author = "de Lejarza, Jorge J. Mart\'\i{}nez and Cieri, Leandro and Rodrigo, Germ\'an",
    title = "{Quantum clustering and jet reconstruction at the LHC}",
    eprint = "2204.06496",
    archivePrefix = "arXiv",
    primaryClass = "hep-ph",
    reportNumber = "IFIC/22-14 FTUV-22-0413.2034",
    doi = "10.1103/PhysRevD.106.036021",
    journal = "Phys. Rev. D",
    volume = "106",
    number = "3",
    pages = "036021",
    year = "2022"
}

@article{Kim:2024jte,
    author = "Kim, Seongmin and Luo, Tengfei and Lee, Eungkyu and Suh, In-Saeng",
    title = "{Distributed Quantum Approximate Optimization Algorithm on Integrated High-Performance Computing and Quantum Computing Systems for Large-Scale Optimization}",
    eprint = "2407.20212",
    archivePrefix = "arXiv",
    primaryClass = "cs.DC",
    month = "7",
    year = "2024"
}

@article{Pires:2021fka,
    author = "Pires, Diogo and Bargassa, Pedrame and Seixas, Jo\~ao and Omar, Yasser",
    title = "{A Digital Quantum Algorithm for Jet Clustering in High-Energy Physics}",
    eprint = "2101.05618",
    archivePrefix = "arXiv",
    primaryClass = "physics.data-an",
    month = "1",
    year = "2021"
}

@article{Pires:2020urc,
    author = "Pires, Diogo and Omar, Yasser and Seixas, Jo\~ao",
    title = "{Adiabatic quantum algorithm for multijet clustering in high energy physics}",
    eprint = "2012.14514",
    archivePrefix = "arXiv",
    primaryClass = "hep-ex",
    doi = "10.1016/j.physletb.2023.138000",
    journal = "Phys. Lett. B",
    volume = "843",
    pages = "138000",
    year = "2023"
}

@article{Gustafson:2022dsq,
    author = {Gustafson, G\"osta and Prestel, Stefan and Spannowsky, Michael and Williams, Simon},
    title = "{Collider events on a quantum computer}",
    eprint = "2207.10694",
    archivePrefix = "arXiv",
    primaryClass = "hep-ph",
    doi = "10.1007/JHEP11(2022)035",
    journal = "JHEP",
    volume = "11",
    pages = "035",
    year = "2022"
}

@article{Bauer:2021gup,
    author = "Bauer, Christian W. and Freytsis, Marat and Nachman, Benjamin",
    title = "{Simulating Collider Physics on Quantum Computers Using Effective Field Theories}",
    eprint = "2102.05044",
    archivePrefix = "arXiv",
    primaryClass = "hep-ph",
    doi = "10.1103/PhysRevLett.127.212001",
    journal = "Phys. Rev. Lett.",
    volume = "127",
    number = "21",
    pages = "212001",
    year = "2021"
}

@article{Bauer:2019qxa,
    author = "Bauer, Christian W. and de Jong, Wibe A. and Nachman, Benjamin and Provasoli, Davide",
    title = "{Quantum Algorithm for High Energy Physics Simulations}",
    eprint = "1904.03196",
    archivePrefix = "arXiv",
    primaryClass = "hep-ph",
    doi = "10.1103/PhysRevLett.126.062001",
    journal = "Phys. Rev. Lett.",
    volume = "126",
    number = "6",
    pages = "062001",
    year = "2021"
}

@article{Hadfield_2019,
	doi = {10.3390/a12020034},
  
	url = {https://doi.org/10.3390%2Fa12020034},
  
	year = 2019,
	month = {feb},
  
	publisher = {{MDPI} {AG}
},
  
	volume = {12},
  
	number = {2},
  
	pages = {34},
  
	author = {Stuart Hadfield and Zhihui Wang and Bryan O{\textquotesingle}Gorman and Eleanor Rieffel and Davide Venturelli and Rupak Biswas},
  
	title = {From the Quantum Approximate Optimization Algorithm to a Quantum Alternating Operator Ansatz},
  
	journal = {Algorithms}
}

@inproceedings{Funcke_2023,
	doi = {10.22323/1.430.0228},
  
	url = {https://doi.org/10.22323%2F1.430.0228},
  
	year = 2023,
	month = {feb},
  
	publisher = {Sissa Medialab},
  
	author = {Lena Funcke and Tobias Hartung and Karl Jansen and Stefan Kühn},
  
	title = {Review on Quantum Computing for Lattice Field Theory},
  
	booktitle = {Proceedings of The 39th International Symposium on Lattice Field Theory {\textemdash} {PoS}({LATTICE}2022)}
}

@misc{chen2020QCNN,
      title={Quantum Long Short-Term Memory}, 
      author={Samuel Yen-Chi Chen and Shinjae Yoo and Yao-Lung L. Fang},
      year={2020},
      eprint={2009.01783},
      archivePrefix={arXiv},
      primaryClass={quant-ph}
}

@article{Bergholm:2018cyq,
    author = "Bergholm, Ville and others",
    title = "{PennyLane: Automatic differentiation of hybrid quantum-classical computations}",
    eprint = "1811.04968",
    archivePrefix = "arXiv",
    primaryClass = "quant-ph",
    month = "11",
    year = "2018"
}

@article{TheATLAScollaboration:2013sgb,
    author={ATLAS Collaboration},
    title = "{Performance assumptions for an upgraded ATLAS detector at a High-Luminosity LHC}",
    reportNumber = "ATL-PHYS-PUB-2013-004",
    year = "2013"
}

@article{deFavereau:2013fsa,
    author = "de Favereau, J. and Delaere, C. and Demin, P. and Giammanco, A. and Lema\^\i{}tre, V. and Mertens, A. and Selvaggi, M.",
    collaboration = "DELPHES 3",
    title = "{DELPHES 3, A modular framework for fast simulation of a generic collider experiment}",
    eprint = "1307.6346",
    archivePrefix = "arXiv",
    primaryClass = "hep-ex",
    doi = "10.1007/JHEP02(2014)057",
    journal = "JHEP",
    volume = "02",
    pages = "057",
    year = "2014"
}

@misc{chen2020quantum,
      title={Quantum Convolutional Neural Networks for High Energy Physics Data Analysis}, 
      author={Samuel Yen-Chi Chen and Tzu-Chieh Wei and Chao Zhang and Haiwang Yu and Shinjae Yoo},
      year={2020},
      eprint={2012.12177},
      archivePrefix={arXiv},
      primaryClass={cs.LG}
}

@misc{chen2021hybrid,
      title={Hybrid Quantum-Classical Graph Convolutional Network}, 
      author={Samuel Yen-Chi Chen and Tzu-Chieh Wei and Chao Zhang and Haiwang Yu and Shinjae Yoo},
      year={2021},
      eprint={2101.06189},
      archivePrefix={arXiv},
      primaryClass={cs.LG}
}

@article{Abbas_2021,
	doi = {10.1038/s43588-021-00084-1},
  
	url = {https://doi.org/10.1038%2Fs43588-021-00084-1},
  
	year = 2021,
	month = {jun},
  
	publisher = {Springer Science and Business Media {LLC}
},
  
	volume = {1},
  
	number = {6},
  
	pages = {403--409},
  
	author = {Amira Abbas and David Sutter and Christa Zoufal and Aurelien Lucchi and Alessio Figalli and Stefan Woerner},
  
	title = {The power of quantum neural networks},
  
	journal = {Nature Computational Science}
}

@article{Du_2020,
	doi = {10.1103/physrevresearch.2.033125},
  
	url = {https://doi.org/10.1103%2Fphysrevresearch.2.033125},
  
	year = 2020,
	month = {jul},
  
	publisher = {American Physical Society ({APS})},
  
	volume = {2},
  
	number = {3},
  
	author = {Yuxuan Du and Min-Hsiu Hsieh and Tongliang Liu and Dacheng Tao},
  
	title = {Expressive power of parametrized quantum circuits},
  
	journal = {Physical Review Research}
}

@article{Sim_2019,
	doi = {10.1002/qute.201900070},
  
	url = {https://doi.org/10.1002%2Fqute.201900070},
  
	year = 2019,
	month = {oct},
  
	publisher = {Wiley},
  
	volume = {2},
  
	number = {12},
  
	author = {Sukin Sim and Peter D. Johnson and Al{\'{a}
}n Aspuru-Guzik},
  
	title = {Expressibility and Entangling Capability of Parameterized Quantum Circuits for Hybrid Quantum-Classical Algorithms},
  
	journal = {Advanced Quantum Technologies}
}

@article{PhysRevX.4.021041,
  title = {Entanglement in a Quantum Annealing Processor},
  author = {Lanting, T. and Przybysz, A. J. and Smirnov, A. Yu. and Spedalieri, F. M. and Amin, M. H. and Berkley, A. J. and Harris, R. and Altomare, F. and Boixo, S. and Bunyk, P. and Dickson, N. and Enderud, C. and Hilton, J. P. and Hoskinson, E. and Johnson, M. W. and Ladizinsky, E. and Ladizinsky, N. and Neufeld, R. and Oh, T. and Perminov, I. and Rich, C. and Thom, M. C. and Tolkacheva, E. and Uchaikin, S. and Wilson, A. B. and Rose, G.},
  journal = {Phys. Rev. X},
  volume = {4},
  issue = {2},
  pages = {021041},
  numpages = {14},
  year = {2014},
  month = {May},
  publisher = {American Physical Society},
  doi = {10.1103/PhysRevX.4.021041},
  url = {https://link.aps.org/doi/10.1103/PhysRevX.4.021041}
}

@article{Bepari_2021,
	doi = {10.1103/physrevd.103.076020},
  
	url = {https://doi.org/10.1103%2Fphysrevd.103.076020},
  
	year = 2021,
	month = {apr},
  
	publisher = {American Physical Society ({APS})},
  
	volume = {103},
  
	number = {7},
  
	author = {Khadeejah Bepari and Sarah Malik and Michael Spannowsky and Simon Williams},
  
	title = {Towards a quantum computing algorithm for helicity amplitudes and parton showers},
  
	journal = {Physical Review D}
}

@ARTICLE{2012NatPh...8..264C,
       author = {{Cirac}, J. Ignacio and {Zoller}, Peter},
        title = "{Goals and opportunities in quantum simulation}",
      journal = {Nature Physics},
         year = 2012,
        month = apr,
       volume = {8},
       number = {4},
        pages = {264-266},
          doi = {10.1038/nphys2275},
       adsurl = {https://ui.adsabs.harvard.edu/abs/2012NatPh...8..264C}
}

@article{Georgescu_2014,
	doi = {10.1103/revmodphys.86.153},
  
	url = {https://doi.org/10.1103%2Frevmodphys.86.153},
  
	year = 2014,
	month = {mar},
  
	publisher = {American Physical Society ({APS})},
  
	volume = {86},
  
	number = {1},
  
	pages = {153--185},
  
	author = {I.{\hspace{0.167em}
}M. Georgescu and S. Ashhab and Franco Nori},
  
	title = {Quantum simulation},
  
	journal = {Reviews of Modern Physics}
}

@misc{herrmann2023quantum,
      title={Quantum utility -- definition and assessment of a practical quantum advantage}, 
      author={Nils Herrmann and Daanish Arya and Marcus W. Doherty and Angus Mingare and Jason C. Pillay and Florian Preis and Stefan Prestel},
      year={2023},
      eprint={2303.02138},
      archivePrefix={arXiv},
      primaryClass={quant-ph}
}

@article{Daley2022PracticalQA,
  title={Practical quantum advantage in quantum simulation},
  author={Andrew J. Daley and Immanuel Bloch and C. Kokail and Stuart Flannigan and N Pearson and Matthias Troyer and Peter Zoller},
  journal={Nature},
  year={2022},
  volume={607},
  pages={667 - 676},
  url={https://api.semanticscholar.org/CorpusID:251132664}
}

@article{Bapst:2019llh,
    author = "Bapst, Fr{\'e}d{\'e}ric and Bhimji, Wahid and Calafiura, Paolo and Gray, Heather and Lavrijsen, Wim and Linder, Lucy",
    title = "{A pattern recognition algorithm for quantum annealers}",
    eprint = "1902.08324",
    archivePrefix = "arXiv",
    primaryClass = "quant-ph",
    doi = "10.1007/s41781-019-0032-5",
    journal = "Comput. Softw. Big Sci.",
    volume = "4",
    number = "1",
    pages = "1",
    year = "2020"
}

@article{McArdle:2019zrh,
    author = "McArdle, Sam and Jones, Tyson and Endo, Suguru and Li, Ying and Benjamin, Simon C. and Yuan, Xiao",
    title = "{Variational ansatz-based quantum simulation of imaginary time evolution}",
    eprint = "1804.03023",
    archivePrefix = "arXiv",
    primaryClass = "quant-ph",
    doi = "10.1038/s41534-019-0187-2",
    journal = "npj Quantum Inf.",
    volume = "5",
    pages = "75",
    year = "2019"
}

@article{Stokes:2019pmg,
    author = "Stokes, James and Izaac, Josh and Killoran, Nathan and Carleo, Giuseppe",
    title = "{Quantum Natural Gradient}",
    eprint = "1909.02108",
    archivePrefix = "arXiv",
    primaryClass = "quant-ph",
    doi = "10.22331/q-2020-05-25-269",
    journal = "Quantum",
    volume = "4",
    pages = "269",
    year = "2020"
}

@article{Erdmann:2013rxa,
    author = "Erdmann, Johannes and Guindon, Stefan and Kroeninger, Kevin and Lemmer, Boris and Nackenhorst, Olaf and Quadt, Arnulf and Stolte, Philipp",
    title = "{A likelihood-based reconstruction algorithm for top-quark pairs and the KLFitter framework}",
    eprint = "1312.5595",
    archivePrefix = "arXiv",
    primaryClass = "hep-ex",
    doi = "10.1016/j.nima.2014.02.029",
    journal = "Nucl. Instrum. Meth. A",
    volume = "748",
    pages = "18--25",
    year = "2014"
}

@article{Clemente:2022nll,
    author = "Clemente, Giuseppe and Crippa, Arianna and Jansen, Karl and Ram{\'\i}rez-Uribe, Selomit and Renter{\'\i}a-Olivo, Andr{\'e}s E. and Rodrigo, Germ{\'a}n and Sborlini, German F. R. and Vale Silva, Luiz",
    title = "{Variational quantum eigensolver for causal loop Feynman diagrams and directed acyclic graphs}",
    eprint = "2210.13240",
    archivePrefix = "arXiv",
    primaryClass = "hep-ph",
    reportNumber = "IFIC/22-28",
    doi = "10.1103/PhysRevD.108.096035",
    journal = "Phys. Rev. D",
    volume = "108",
    number = "9",
    pages = "096035",
    year = "2023"
}

@article{Afik:2025ejh,
    author = "Afik, Yoav and others",
    title = "{Quantum information meets high-energy physics: input to the update of the European strategy for particle physics}",
    eprint = "2504.00086",
    archivePrefix = "arXiv",
    primaryClass = "hep-ph",
    doi = "10.1140/epjp/s13360-025-06752-9",
    journal = "Eur. Phys. J. Plus",
    volume = "140",
    number = "9",
    pages = "855",
    year = "2025"
}

@article{Tuysuz:2024hyl,
    author = {T{\"u}ys{\"u}z, Cenk and Demidik, Maria and Coopmans, Luuk and Rinaldi, Enrico and Croft, Vincent and Haddad, Yacine and Rosenkranz, Matthias and Jansen, Karl},
    title = "{Learning to generate high-dimensional distributions with low-dimensional quantum Boltzmann machines}",
    eprint = "2410.16363",
    archivePrefix = "arXiv",
    primaryClass = "quant-ph",
    month = "10",
    year = "2024"
}

@article{Kiss:2022pjw,
    author = "Kiss, Oriel and Grossi, Michele and Kajomovitz, Enrique and Vallecorsa, Sofia",
    title = "{Conditional Born machine for Monte Carlo event generation}",
    eprint = "2205.07674",
    archivePrefix = "arXiv",
    primaryClass = "quant-ph",
    doi = "10.1103/PhysRevA.106.022612",
    journal = "Phys. Rev. A",
    volume = "106",
    number = "2",
    pages = "022612",
    year = "2022"
}

@article{Bravo-Prieto:2021ehz,
    author = "Bravo-Prieto, Carlos and Baglio, Julien and C{\`e}, Marco and Francis, Anthony and Grabowska, Dorota M. and Carrazza, Stefano",
    title = "{Style-based quantum generative adversarial networks for Monte Carlo events}",
    eprint = "2110.06933",
    archivePrefix = "arXiv",
    primaryClass = "quant-ph",
    reportNumber = "CERN-TH-2021-139, TIF-UNIMI-2021-14",
    doi = "10.22331/q-2022-08-17-777",
    journal = "Quantum",
    volume = "6",
    pages = "777",
    year = "2022"
}

@article{Egger:2020dbb,
    author = "Egger, Daniel J. and Marecek, Jakub and Woerner, Stefan",
    title = "{Warm-starting quantum optimization}",
    eprint = "2009.10095",
    archivePrefix = "arXiv",
    primaryClass = "quant-ph",
    doi = "10.22331/q-2021-06-17-479",
    journal = "Quantum",
    volume = "5",
    pages = "479",
    year = "2021"
}

@article{Skolik:2020jvn,
    author = "Skolik, Andrea and McClean, Jarrod R. and Mohseni, Masoud and van der Smagt, Patrick and Leib, Martin",
    title = "{Layerwise learning for quantum neural networks}",
    eprint = "2006.14904",
    archivePrefix = "arXiv",
    primaryClass = "quant-ph",
    doi = "10.1007/s42484-020-00036-4",
    journal = "Quantum Machine Intelligence",
    volume = "3",
    number = "1",
    pages = "5",
    year = "2021"
}

@article{Stamatopoulos:2020xez,
    author = "Stamatopoulos, Nikitas and Egger, Daniel J. and Sun, Yue and Zoufal, Christa and Iten, Raban and Shen, Ning and Woerner, Stefan",
    title = "{Option Pricing using Quantum Computers}",
    eprint = "1905.02666",
    archivePrefix = "arXiv",
    primaryClass = "quant-ph",
    doi = "10.22331/q-2020-07-06-291",
    journal = "Quantum",
    volume = "4",
    pages = "291",
    year = "2020"
}

@article{Cerezo:2020mtn,
    author = "Cerezo, M. and Sone, Akira and Volkoff, Tyler and Cincio, Lukasz and Coles, Patrick J.",
    title = "{Cost function dependent barren plateaus in shallow parametrized quantum circuits}",
    eprint = "2001.00550",
    archivePrefix = "arXiv",
    primaryClass = "quant-ph",
    reportNumber = "LA-UR-19-32681, LA-UR-19-32681",
    doi = "10.1038/s41467-021-21728-w",
    journal = "Nature Commun.",
    volume = "12",
    number = "1",
    pages = "1791",
    year = "2021"
}

@article{McClean:2018jps,
    author = "McClean, Jarrod R. and Boixo, Sergio and Smelyanskiy, Vadim N. and Babbush, Ryan and Neven, Hartmut",
    title = "{Barren plateaus in quantum neural network training landscapes}",
    eprint = "1803.11173",
    archivePrefix = "arXiv",
    primaryClass = "quant-ph",
    doi = "10.1038/s41467-018-07090-4",
    journal = "Nature Commun.",
    volume = "9",
    pages = "4812",
    year = "2018"
}

@article{Nakanishi:2019rrm,
    author = "Nakanishi, Ken M. and Fujii, Keisuke and Todo, Synge",
    title = "{Sequential minimal optimization for quantum-classical hybrid algorithms}",
    eprint = "1903.12166",
    archivePrefix = "arXiv",
    primaryClass = "quant-ph",
    doi = "10.1103/physrevresearch.2.043158",
    journal = "Phys. Rev. Res.",
    volume = "2",
    number = "4",
    pages = "043158",
    year = "2020"
}

@article{Crippa:2023ieq,
    author = "Crippa, Arianna and others",
    title = "{Quantum Algorithms for Charged Particle Track Reconstruction in the LUXE Experiment}",
    eprint = "2304.01690",
    archivePrefix = "arXiv",
    primaryClass = "quant-ph",
    reportNumber = "DESY-23-045, MIT-CTP/5481",
    doi = "10.1007/s41781-023-00109-6",
    journal = "Comput. Softw. Big Sci.",
    volume = "7",
    number = "1",
    pages = "14",
    year = "2023"
}

@article{Nicotra:2023rmn,
    author = "Nicotra, Davide and Lucio Martinez, Miriam and de Vries, Jacco Andreas and Merk, Marcel and Driessens, Kurt and Westra, Ronald Leonard and Dibenedetto, Domenica and C{\'a}mpora P{\'e}rez, Daniel Hugo",
    title = "{A quantum algorithm for track reconstruction in the LHCb vertex detector}",
    eprint = "2308.00619",
    archivePrefix = "arXiv",
    primaryClass = "quant-ph",
    doi = "10.1088/1748-0221/18/11/P11028",
    journal = "JINST",
    volume = "18",
    number = "11",
    pages = "P11028",
    year = "2023"
}

@inproceedings{Delgado:2022tpc,
    author = "Delgado, Andrea and others",
    title = "{Quantum Computing for Data Analysis in High-Energy Physics}",
    booktitle = "{Snowmass 2021}",
    eprint = "2203.08805",
    archivePrefix = "arXiv",
    primaryClass = "physics.data-an",
    month = "3",
    year = "2022"
}

@article{Tuysuz:2020ocw,
    author = {T\"uys\"uz, Cenk and Carminati, Federico and Demirk\"oz, Bilge and Dobos, Daniel and Fracas, Fabio and Novotny, Kristiane and Potamianos, Karolos and Vallecorsa, Sofia and Vlimant, Jean-Roch},
    editor = "Doglioni, C. and Kim, D. and Stewart, G. A. and Silvestris, L. and Jackson, P. and Kamleh, W.",
    title = "{Particle Track Reconstruction with Quantum Algorithms}",
    eprint = "2003.08126",
    archivePrefix = "arXiv",
    primaryClass = "quant-ph",
    doi = "10.1051/epjconf/202024509013",
    journal = "EPJ Web Conf.",
    volume = "245",
    pages = "09013",
    year = "2020"
}

@inproceedings{Alam:2022crs,
    author = "Alam, M. Sohaib and others",
    title = "{Quantum computing hardware for HEP algorithms and sensing}",
    booktitle = "{Snowmass 2021}",
    eprint = "2204.08605",
    archivePrefix = "arXiv",
    primaryClass = "quant-ph",
    reportNumber = "FERMILAB-PUB-22-260-SQMS",
    month = "4",
    year = "2022"
}

@inproceedings{Catterall:2022wjq,
    author = "Catterall, Simon and others",
    title = "{Report of the Snowmass 2021 Theory Frontier Topical Group on Quantum Information Science}",
    booktitle = "{Snowmass 2021}",
    eprint = "2209.14839",
    archivePrefix = "arXiv",
    primaryClass = "quant-ph",
    reportNumber = "FERMILAB-FN-1199-T",
    month = "9",
    year = "2022"
}

@article{Bauer:2022hpo,
    author = "Bauer, Christian W. and others",
    title = "{Quantum Simulation for High-Energy Physics}",
    eprint = "2204.03381",
    archivePrefix = "arXiv",
    primaryClass = "quant-ph",
    reportNumber = "UMD-PP-022-04, LA-UR-22-22100, RIKEN-iTHEMS-Report-22, IQuS@UW-21-027, MITRE-21-03848-2, FERMILAB-PUB-22-249-SQMS-T",
    doi = "10.1103/PRXQuantum.4.027001",
    journal = "PRX Quantum",
    volume = "4",
    number = "2",
    pages = "027001",
    year = "2023"
}

@misc{humble2022snowmass,
      title={Snowmass White Paper: Quantum Computing Systems and Software for High-energy Physics Research}, 
      author={Travis S. Humble and Andrea Delgado and Raphael Pooser and Christopher Seck and Ryan Bennink and Vicente Leyton-Ortega and C. -C. Joseph Wang and Eugene Dumitrescu and Titus Morris and Kathleen Hamilton and Dmitry Lyakh and Prasanna Date and Yan Wang and Nicholas A. Peters and Katherine J. Evans and Marcel Demarteau and Alex McCaskey and Thien Nguyen and Susan Clark and Melissa Reville and Alberto Di Meglio and Michele Grossi and Sofia Vallecorsa and Kerstin Borras and Karl Jansen and Dirk Krücker},
      year={2022},
      eprint={2203.07091},
      archivePrefix={arXiv},
      primaryClass={quant-ph}
}

@article{Zhou_2020,
	doi = {10.1103/physrevx.10.021067},
  
	url = {https://doi.org/10.1103%2Fphysrevx.10.021067},
  
	year = 2020,
	month = {jun},
  
	publisher = {American Physical Society ({APS})},
  
	volume = {10},
  
	number = {2},
  
	author = {Leo Zhou and Sheng-Tao Wang and Soonwon Choi and Hannes Pichler and Mikhail D. Lukin},
  
	title = {Quantum Approximate Optimization Algorithm: Performance, Mechanism, and Implementation on Near-Term Devices},
  
	journal = {Physical Review X}
}

@article{Lee:2020qil,
    author = "Lee, Jason Sang Hun and Park, Inkyu and Watson, Ian James and Yang, Seungjin",
    title = "{Zero-permutation jet-parton assignment using a self-attention network}",
    eprint = "2012.03542",
    archivePrefix = "arXiv",
    primaryClass = "hep-ex",
    doi = "10.1007/s40042-024-01037-3",
    journal = "J. Korean Phys. Soc.",
    volume = "84",
    number = "6",
    pages = "427--438",
    year = "2024"
}

@article{Crooks:2018vud,
    author = "Crooks, Gavin E.",
    title = "{Performance of the Quantum Approximate Optimization Algorithm on the Maximum Cut Problem}",
    eprint = "1811.08419",
    archivePrefix = "arXiv",
    primaryClass = "quant-ph",
    month = "11",
    year = "2018"
}

@article{Kim:2021wrr,
    author = "Kim, Minho and Ko, Pyungwon and Park, Jae-hyeon and Park, Myeonghun",
    title = "{Leveraging Quantum Annealer to identify an Event-topology at High Energy Colliders}",
    eprint = "2111.07806",
    archivePrefix = "arXiv",
    primaryClass = "hep-ph",
    reportNumber = "KIAS--P21050",
    month = "11",
    year = "2021"
}

@article{Ezratty:2023fxe,
    author = "Ezratty, Olivier",
    title = "{Is there a Moore's law for quantum computing?}",
    eprint = "2303.15547",
    archivePrefix = "arXiv",
    primaryClass = "quant-ph",
    month = "3",
    year = "2023"
}

@article{Cerezo:2020jpv,
    author = "Cerezo, M. and others",
    title = "{Variational quantum algorithms}",
    eprint = "2012.09265",
    archivePrefix = "arXiv",
    primaryClass = "quant-ph",
    reportNumber = "LA-UR-20-30142",
    doi = "10.1038/s42254-021-00348-9",
    journal = "Nature Rev. Phys.",
    volume = "3",
    number = "9",
    pages = "625--644",
    year = "2021"
}

@article{Magann:2021evo,
    author = "Magann, Alicia B. and Rudinger, Kenneth M. and Grace, Matthew D. and Sarovar, Mohan",
    title = "{Feedback-Based Quantum Optimization}",
    eprint = "2103.08619",
    archivePrefix = "arXiv",
    primaryClass = "quant-ph",
    doi = "10.1103/PhysRevLett.129.250502",
    journal = "Phys. Rev. Lett.",
    volume = "129",
    number = "25",
    pages = "250502",
    year = "2022"
}

@ARTICLE{QUBO-survey,
title = {The unconstrained binary quadratic programming problem: a survey},
author = {Kochenberger, Gary and Hao, Jin-Kao and Glover, Fred and Lewis, Mark and Lü, Zhipeng and Wang, Haibo and Wang, Yang},
year = {2014},
journal = {Journal of Combinatorial Optimization},
volume = {28},
number = {1},
pages = {58-81},
url = {https://EconPapers.repec.org/RePEc:spr:jcomop:v:28:y:2014:i:1:d:10.1007_s10878-014-9734-0}
}

@article{Debnath:2017ktz,
    author = "Debnath, Dipsikha and Kim, Doojin and Kim, Jeong Han and Kong, Kyoungchul and Matchev, Konstantin T.",
    title = "{Resolving Combinatorial Ambiguities in Dilepton $t\bar t$ Event Topologies with Constrained $M_2$ Variables}",
    eprint = "1706.04995",
    archivePrefix = "arXiv",
    primaryClass = "hep-ph",
    reportNumber = "CERN-TH-2017-130",
    doi = "10.1103/PhysRevD.96.076005",
    journal = "Phys. Rev. D",
    volume = "96",
    number = "7",
    pages = "076005",
    year = "2017"
}

@article{Forestano:2023lnb,
    author = "Forestano, Roy T. and others",
    title = "{A Comparison between Invariant and Equivariant Classical and Quantum Graph Neural Networks}",
    eprint = "2311.18672",
    archivePrefix = "arXiv",
    primaryClass = "quant-ph",
    doi = "10.3390/axioms13030160",
    journal = "Axioms",
    volume = "13",
    number = "3",
    pages = "160",
    year = "2024"
}

@article{Dong:2023oqb,
    author = "Dong, Zhongtian and others",
    title = "{$\mathbb{Z}_2\times \mathbb{Z}_2$~Equivariant Quantum Neural Networks: Benchmarking against Classical Neural Networks}",
    eprint = "2311.18744",
    archivePrefix = "arXiv",
    primaryClass = "quant-ph",
    doi = "10.3390/axioms13030188",
    journal = "Axioms",
    volume = "13",
    number = "3",
    pages = "188",
    year = "2024"
}

@article{Alwall:2014hca,
    author = "Alwall, J. and Frederix, R. and Frixione, S. and Hirschi, V. and Maltoni, F. and Mattelaer, O. and Shao, H. -S. and Stelzer, T. and Torrielli, P. and Zaro, M.",
    title = "{The automated computation of tree-level and next-to-leading order differential cross sections, and their matching to parton shower simulations}",
    eprint = "1405.0301",
    archivePrefix = "arXiv",
    primaryClass = "hep-ph",
    reportNumber = "CERN-PH-TH-2014-064, CP3-14-18, LPN14-066, MCNET-14-09, ZU-TH-14-14",
    doi = "10.1007/JHEP07(2014)079",
    journal = "JHEP",
    volume = "07",
    pages = "079",
    year = "2014"
}

@article{Yarkoni:2021zvu,
    author = {Yarkoni, Sheir and Raponi, Elena and B\"ack, Thomas and Schmitt, Sebastian},
    title = "{Quantum annealing for industry applications: introduction and review}",
    eprint = "2112.07491",
    archivePrefix = "arXiv",
    primaryClass = "quant-ph",
    doi = "10.1088/1361-6633/ac8c54",
    journal = "Rept. Prog. Phys.",
    volume = "85",
    number = "10",
    pages = "104001",
    year = "2022"
}

@article{Yang:2024bqw,
    author = "Yang, Ji-Chong and Zhang, Shuai and Yue, Chong-Xing",
    title = "{A quantum machine learning classifier to search for new physics}",
    eprint = "2410.18847",
    archivePrefix = "arXiv",
    primaryClass = "hep-ph",
    doi = "10.1007/JHEP01(2026)023",
    journal = "JHEP",
    volume = "01",
    pages = "023",
    year = "2026"
}

@article{Shaydulin:2023fpr,
    author = "Shaydulin, Ruslan and others",
    title = "{Evidence of scaling advantage for the quantum approximate optimization algorithm on a classically intractable problem}",
    eprint = "2308.02342",
    archivePrefix = "arXiv",
    primaryClass = "quant-ph",
    doi = "10.1126/sciadv.adm6761",
    journal = "Sci. Adv.",
    volume = "10",
    number = "22",
    pages = "adm6761",
    year = "2024"
}

@article{Montanez-Barrera:2024tos,
    author = "Montanez-Barrera, J. A. and Michielsen, Kristel",
    title = "{Toward a linear-ramp QAOA protocol: evidence of a scaling advantage in solving some combinatorial optimization problems}",
    eprint = "2405.09169",
    archivePrefix = "arXiv",
    primaryClass = "quant-ph",
    doi = "10.1038/s41534-025-01082-1",
    journal = "npj Quantum Inf.",
    volume = "11",
    number = "1",
    pages = "131",
    year = "2025"
}

@article{Weidenfeller:2022hkn,
    author = "Weidenfeller, Johannes and Valor, Lucia C. and Gacon, Julien and Tornow, Caroline and Bello, Luciano and Woerner, Stefan and Egger, Daniel J.",
    title = "{Scaling of the quantum approximate optimization algorithm on superconducting qubit based hardware}",
    eprint = "2202.03459",
    archivePrefix = "arXiv",
    primaryClass = "quant-ph",
    doi = "10.22331/q-2022-12-07-870",
    journal = "Quantum",
    volume = "6",
    pages = "870",
    year = "2022"
}

@article{Lotshaw:2022wkm,
    author = "Lotshaw, Phillip C. and Nguyen, Thien and Santana, Anthony and McCaskey, Alexander and Herrman, Rebekah and Ostrowski, James and Siopsis, George and Humble, Travis S.",
    title = "{Scaling quantum approximate optimization on near-term hardware}",
    eprint = "2201.02247",
    archivePrefix = "arXiv",
    primaryClass = "quant-ph",
    doi = "10.1038/s41598-022-14767-w",
    journal = "Sci. Rep.",
    volume = "12",
    number = "1",
    pages = "12388",
    year = "2022"
}

@article{Rajak:2022tgo,
   title={Quantum annealing: an overview},
   volume={381},
   ISSN={1471-2962},
   url={http://dx.doi.org/10.1098/rsta.2021.0417},
   DOI={10.1098/rsta.2021.0417},
   number={2241},
   journal={Philosophical Transactions of the Royal Society A: Mathematical, Physical and Engineering Sciences},
   publisher={The Royal Society},
   author={Rajak, Atanu and Suzuki, Sei and Dutta, Amit and Chakrabarti, Bikas K.},
   year={2022},
   month=dec }

@inproceedings{Kingma:2014vow,
    author = "Kingma, Diederik P. and Ba, Jimmy",
    title = "{Adam: A Method for Stochastic Optimization}",
    eprint = "1412.6980",
    archivePrefix = "arXiv",
    primaryClass = "cs.LG",
    month = "12",
    year = "2014"
}

@article{Liu:2023aht,
    author = "Liu, Xia and Yang, Huan and Yang, Li",
    title = "{Minimizing CNOT-count in quantum circuit of the extended Shor{\textquoteright}s algorithm for ECDLP}",
    eprint = "2305.11410",
    archivePrefix = "arXiv",
    primaryClass = "quant-ph",
    doi = "10.1186/s42400-023-00181-w",
    journal = "Cybersecurity",
    volume = "6",
    number = "1",
    pages = "48",
    year = "2023"
}

@phdthesis{Campos:2024wvc,
    author = "Campos, Roberto",
    title = "{Hybrid Quantum-Classical Algorithms}",
    eprint = "2406.12371",
    archivePrefix = "arXiv",
    primaryClass = "quant-ph",
    school = "Unlisted",
    year = "2024"
}

@article{Callison:2022zfe,
    author = "Callison, Adam and Chancellor, Nicholas",
    title = "{Hybrid quantum-classical algorithms in the noisy intermediate-scale quantum era and beyond}",
    eprint = "2207.06850",
    archivePrefix = "arXiv",
    primaryClass = "quant-ph",
    doi = "10.1103/PhysRevA.106.010101",
    journal = "Phys. Rev. A",
    volume = "106",
    number = "1",
    pages = "010101",
    year = "2022"
}

@article{Ge:2022vmm,
    author = "Ge, Xiaozhen and Wu, Re-Bing and Rabitz, Herschel",
    title = "{The optimization landscape of hybrid quantum\textendash{}classical algorithms: From quantum control to NISQ applications}",
    eprint = "2201.07448",
    archivePrefix = "arXiv",
    primaryClass = "quant-ph",
    doi = "10.1016/j.arcontrol.2022.06.001",
    journal = "Annual Reviews in Control",
    volume = "54",
    pages = "314--323",
    year = "2022"
}

@article{Hammad:2023wme,
    author = "Hammad, A. and Kong, Kyoungchul and Park, Myeonghun and Shim, Soyoung",
    title = "{Quantum Metric Learning for New Physics Searches at the LHC}",
    eprint = "2311.16866",
    archivePrefix = "arXiv",
    primaryClass = "hep-ph",
    month = "11",
    year = "2023"
}

@article{Unlu:2024nvo,
    author = "Unlu, Eyup B. and others",
    title = "{Hybrid Quantum Vision Transformers for Event Classification in High Energy Physics}",
    eprint = "2402.00776",
    archivePrefix = "arXiv",
    primaryClass = "quant-ph",
    doi = "10.3390/axioms13030187",
    journal = "Axioms",
    volume = "13",
    number = "3",
    pages = "187",
    year = "2024"
}

@article{Cara:2024spj,
    author = "Cara, Mar\c{c}al Comajoan and others",
    title = "{Quantum Vision Transformers for Quark\textendash{}Gluon Classification}",
    eprint = "2405.10284",
    archivePrefix = "arXiv",
    primaryClass = "quant-ph",
    doi = "10.3390/axioms13050323",
    journal = "Axioms",
    volume = "13",
    number = "5",
    pages = "323",
    year = "2024"
}

@article{Lucas:2013ahy,
    author = "Lucas, Andrew",
    title = "{Ising formulations of many NP problems}",
    eprint = "1302.5843",
    archivePrefix = "arXiv",
    primaryClass = "cond-mat.stat-mech",
    doi = "10.3389/fphy.2014.00005",
    journal = "Front. in Phys.",
    volume = "2",
    pages = "5",
    year = "2014"
}

@article{Glover:2018ikr,
    author = "Glover, Fred and Kochenberger, Gary and Du, Yu",
    title = "{Quantum Bridge Analytics I: a tutorial on formulating and using QUBO models}",
    eprint = "1811.11538",
    archivePrefix = "arXiv",
    primaryClass = "cs.DS",
    doi = "10.1007/s10288-019-00424-y",
    journal = "4OR",
    volume = "17",
    number = "4",
    pages = "335--371",
    year = "2019"
}

@article{Farhi:2014ych,
    author = "Farhi, Edward and Goldstone, Jeffrey and Gutmann, Sam",
    title = "{A Quantum Approximate Optimization Algorithm}",
    eprint = "1411.4028",
    archivePrefix = "arXiv",
    primaryClass = "quant-ph",
    reportNumber = "MIT-CTP/4610",
    month = "11",
    year = "2014"
}

@article{Blekos:2023nil,
    author = "Blekos, Kostas and Brand, Dean and Ceschini, Andrea and Chou, Chiao-Hui and Li, Rui-Hao and Pandya, Komal and Summer, Alessandro",
    title = "{A review on Quantum Approximate Optimization Algorithm and its variants}",
    eprint = "2306.09198",
    archivePrefix = "arXiv",
    primaryClass = "quant-ph",
    doi = "10.1016/j.physrep.2024.03.002",
    journal = "Phys. Rept.",
    volume = "1068",
    pages = "1--66",
    year = "2024"
}

@article{DiMeglio:2023nsa,
    author = "Di Meglio, Alberto and others",
    title = "{Quantum Computing for High-Energy Physics: State of the Art and Challenges}",
    eprint = "2307.03236",
    archivePrefix = "arXiv",
    primaryClass = "quant-ph",
    reportNumber = "FERMILAB-PUB-23-468-ETD",
    doi = "10.1103/PRXQuantum.5.037001",
    journal = "PRX Quantum",
    volume = "5",
    number = "3",
    pages = "037001",
    year = "2024"
}

@article{Abbas:2023agz,
    author = "Abbas, Amira and others",
    title = "{Challenges and opportunities in quantum optimization}",
    eprint = "2312.02279",
    archivePrefix = "arXiv",
    primaryClass = "quant-ph",
    doi = "10.1038/s42254-024-00770-9",
    journal = "Nature Rev. Phys.",
    volume = "6",
    number = "12",
    pages = "718--735",
    year = "2024"
}

@article{P5:2023wyd,
    author = "Asai, Shoji and others",
    collaboration = "P5",
    title = "{Exploring the Quantum Universe: Pathways to Innovation and Discovery in Particle Physics}",
    eprint = "2407.19176",
    archivePrefix = "arXiv",
    primaryClass = "hep-ex",
    reportNumber = "OSTI Technical Report 2368847",
    doi = "10.2172/2368847",
    month = "12",
    year = "2023"
}

@article{Zhu:2024own,
    author = "Zhu, Yongfeng and Zhuang, Weifeng and Qian, Chen and Ma, Yunheng and Liu, Dong E. and Ruan, Manqi and Zhou, Chen",
    title = "{A novel quantum realization of jet clustering in high-energy physics experiments}",
    eprint = "2407.09056",
    archivePrefix = "arXiv",
    primaryClass = "quant-ph",
    doi = "10.1016/j.scib.2024.12.020",
    journal = "Sci. Bull.",
    volume = "70",
    number = "4",
    pages = "460--463",
    year = "2025"
}

@article{Zhou:2018fwi,
    author = "Zhou, Leo and Wang, Sheng-Tao and Choi, Soonwon and Pichler, Hannes and Lukin, Mikhail D.",
    title = "{Quantum Approximate Optimization Algorithm: Performance, Mechanism, and Implementation on Near-Term Devices}",
    eprint = "1812.01041",
    archivePrefix = "arXiv",
    primaryClass = "quant-ph",
    doi = "10.1103/PhysRevX.10.021067",
    journal = "Phys. Rev. X",
    volume = "10",
    number = "2",
    pages = "021067",
    year = "2020"
}

@article{Alhazmi:2022qbf,
    author = "Alhazmi, Haider and Dong, Zhongtian and Huang, Li and Kim, Jeong Han and Kong, Kyoungchul and Shih, David",
    title = "{Resolving combinatorial ambiguities in dilepton tt\textasciimacron{} event topologies with neural networks}",
    eprint = "2202.05849",
    archivePrefix = "arXiv",
    primaryClass = "hep-ph",
    doi = "10.1103/PhysRevD.105.115011",
    journal = "Phys. Rev. D",
    volume = "105",
    number = "11",
    pages = "115011",
    year = "2022"
}

@misc{herrman2021multiangle,
      title={Multi-angle Quantum Approximate Optimization Algorithm}, 
      author={Rebekah Herrman and Phillip C. Lotshaw and James Ostrowski and Travis S. Humble and George Siopsis},
      year={2021},
      eprint={2109.11455},
      archivePrefix={arXiv},
      primaryClass={quant-ph}
}

@misc{vijendran2023expressive,
      title={An Expressive Ansatz for Low-Depth Quantum Optimisation}, 
      author={V. Vijendran and Aritra Das and Dax Enshan Koh and Syed M. Assad and Ping Koy Lam},
      year={2023},
      eprint={2302.04479},
      archivePrefix={arXiv},
      primaryClass={quant-ph}
}

@article{Magann_2022,
	doi = {10.1103/physrevlett.129.250502},
	url = {https://doi.org/10.1103%2Fphysrevlett.129.250502},
	year = 2022,
	month = {dec},
	publisher = {American Physical Society ({APS})},
	volume = {129},
	number = {25},
  	author = {Alicia B. Magann and Kenneth M. Rudinger and Matthew D. Grace and Mohan Sarovar},
	title = {Feedback-Based Quantum Optimization},
	journal = {Physical Review Letters}
}

@misc{zhu2022adaptive,
      title={An adaptive quantum approximate optimization algorithm for solving combinatorial problems on a quantum computer}, 
      author={Linghua Zhu and Ho Lun Tang and George S. Barron and F. A. Calderon-Vargas and Nicholas J. Mayhall and Edwin Barnes and Sophia E. Economou},
      year={2022},
      eprint={2005.10258},
      archivePrefix={arXiv},
      primaryClass={quant-ph}
}

@book{Nielsen_Chuang_2010,
    place={Cambridge},
    title={Quantum Computation and Quantum Information: 10th Anniversary Edition},
    publisher={Cambridge University Press},
    author={Nielsen, Michael A. and Chuang, Isaac L.},
    year={2010}
}

@article{loshchilov2017decoupled,
  title={Decoupled weight decay regularization},
  author={Loshchilov, I},
  journal={arXiv preprint arXiv:1711.05101},
  year={2017}
}

@article{Magann_2022_2,
   title={Lyapunov-control-inspired strategies for quantum combinatorial optimization},
   volume={106},
   ISSN={2469-9934},
   url={http://dx.doi.org/10.1103/PhysRevA.106.062414},
   DOI={10.1103/physreva.106.062414},
   number={6},
   journal={Physical Review A},
   publisher={American Physical Society (APS)},
   author={Magann, Alicia B. and Rudinger, Kenneth M. and Grace, Matthew D. and Sarovar, Mohan},
   year={2022},
   month=dec
}

@article{Sridhar:2023our,
    author = "Sridhar, Vishvesha K. and Chen, Yanzhu and Gard, Bryan and Barnes, Edwin and Economou, Sophia E.",
    title = "{ADAPT-QAOA with a classically inspired initial state}",
    eprint = "2310.09694",
    archivePrefix = "arXiv",
    primaryClass = "quant-ph",
    month = "10",
    year = "2023"
}

@article{patel2003efficient,
  title={Efficient synthesis of linear reversible circuits},
  author={Patel, Ketan N and Markov, Igor L and Hayes, John P},
  journal={arXiv preprint quant-ph/0302002},
  year={2003}
}

@article{Maciejewski:2024aaf,
    author = "Maciejewski, Filip B. and Biamonte, Jacob and Hadfield, Stuart and Venturelli, Davide",
    title = "{Improving Quantum Approximate Optimization by Noise-Directed Adaptive Remapping}",
    eprint = "2404.01412",
    archivePrefix = "arXiv",
    primaryClass = "quant-ph",
    doi = "10.22331/q-2025-11-06-1906",
    journal = "Quantum",
    volume = "9",
    pages = "1906",
    year = "2025"
}

@article{Chai_2022,
   title={Shortcuts to the quantum approximate optimization algorithm},
   volume={105},
   ISSN={2469-9934},
   url={http://dx.doi.org/10.1103/PhysRevA.105.042415},
   DOI={10.1103/physreva.105.042415},
   number={4},
   journal={Physical Review A},
   publisher={American Physical Society (APS)},
   author={Chai, Yahui and Han, Yong-Jian and Wu, Yu-Chun and Li, Ye and Dou, Menghan and Guo, Guo-Ping},
   year={2022},
   month=apr }

@article{Wurtz_2022,
   title={Counterdiabaticity and the quantum approximate optimization algorithm},
   volume={6},
   ISSN={2521-327X},
   url={http://dx.doi.org/10.22331/q-2022-01-27-635},
   DOI={10.22331/q-2022-01-27-635},
   journal={Quantum},
   publisher={Verein zur Forderung des Open Access Publizierens in den Quantenwissenschaften},
   author={Wurtz, Jonathan and Love, Peter J.},
   year={2022},
   month=jan, pages={635} }

@article{Chandarana_2022,
   title={Digitized-counterdiabatic quantum approximate optimization algorithm},
   volume={4},
   ISSN={2643-1564},
   url={http://dx.doi.org/10.1103/PhysRevResearch.4.013141},
   DOI={10.1103/physrevresearch.4.013141},
   number={1},
   journal={Physical Review Research},
   publisher={American Physical Society (APS)},
   author={Chandarana, P. and Hegade, N. N. and Paul, K. and Albarrán-Arriagada, F. and Solano, E. and del Campo, A. and Chen, Xi},
   year={2022},
   month=feb }

\end{document}